%% file: arXiv_ExploringMassDependence_FinalUpdate.tex
\documentclass[twocolumn,aps,prd,groupedaddress,superscriptaddress,amsmath,amssymb,noeprint,nofootinbib,10pt,floatfix]{revtex4-2}
\input{preamble_packages}
\input{preamble_commands}

\makeatletter 

\renewcommand\onecolumngrid{
\do@columngrid{one}{\@ne}%
\def\set@footnotewidth{\onecolumngrid}
\def\footnoterule{\kern-6pt\hrule width 1.5in\kern6pt}%
}

\renewcommand\twocolumngrid{
\def\footnoterule{
	\dimen@\skip\footins\divide\dimen@\thr@@
	\kern-\dimen@\hrule width.5in\kern\dimen@}
\do@columngrid{mlt}{\tw@}
}%

\makeatother    

\begin{document}

\title{Exploring the Effects of Mass Dependence in Spontaneous Collapse Models}

\author{Nicol\`{o} Piccione}
\email{nicolo'.piccione@units.it}
\affiliation{Department of Physics, University of Trieste, Strada Costiera 11, 34151 Trieste, Italy}
\affiliation{Istituto Nazionale di Fisica Nucleare, Trieste Section, Via Valerio 2, 34127 Trieste, Italy}

\author{Angelo Bassi}
\affiliation{Department of Physics, University of Trieste, Strada Costiera 11, 34151 Trieste, Italy}
\affiliation{Istituto Nazionale di Fisica Nucleare, Trieste Section, Via Valerio 2, 34127 Trieste, Italy}

\begin{abstract}

	Spontaneous collapse models aim to solve the long-standing measurement problem in quantum mechanics by modifying the theory’s dynamics to include objective wave function collapses. These collapses occur randomly in space, bridging the gap between quantum and classical behavior. A central feature of these models is their dependence on mass density, which directly influences how and when collapse events occur. In this work, we explore a generalized framework in which the collapse dynamics depend on arbitrary functions of the mass density, extending previous models. We analyze the theoretical consistency of these generalizations, investigate their predictions, and compare them with experimental data. Our findings show that only a limited range of mass-dependence functions are viable, with significant implications for the future development and empirical testability of collapse-based models. Importantly, they also indicate that a well-justified model denoted here as PSL shows much more resilience to experimental falsification than standard collapse models.
\end{abstract}

\maketitle

\section{Introduction}

Despite its enormous empirical success, the meaning and validity of Quantum Mechanics as a fundamental theory of nature remain subjects of debate. The role of the wave function in representing the state of a system is not fully understood and, with it, whether it collapses, and if so, when and how this occurs. These unresolved questions are collectively referred to as the measurement problem in quantum theory~\cite{Book_Bell2004Speakable,Book_Norsen2017foundations,Book_Durr2020understanding,Book_Tumulka2022Foundations}.

Models of spontaneous wave function collapse (collapse models) are a well-studied tentative solution to these problems~\cite{Bassi2003Dynamical,Bassi2013Models,Bassi2023CollapseModels}. Their goal is to replace standard quantum mechanics with a self-consistent theory that coherently explains the quantum-to-classical transition. To achieve this, they modify the Schr\"odinger equation by introducing suitable nonlinear and stochastic terms that encode the collapse of the wave function. Accordingly, collapses occur more or less all the time, alongside standard quantum evolution, but they become dominant only under specific circumstances, such as during measurement processes. 

By modifying the standard quantum dynamics, collapse models become falsifiable, which makes them particularly interesting from an experimental point of view~\cite{Carlesso2022Present}.

The most studied spontaneous collapse models are the Continuous Spontaneous Localization (CSL) model~\cite{Ghirardi1990_CSL}, and the Diósi-Penrose (DP) model~\cite{Diosi1987Universal,Diosi1989Models,Penrose1996gravity,Penrose2014Gravitization}, the latter being based on the assumption that gravity breaks the unitarity of the quantum dynamics. Both models have undergone extensive experimental and theoretical investigations~\cite{Carlesso2019Collapse,Carlesso2022Present}. 

In its original formulation, CSL assumes the existence of a fundamental noise field that couples to the smeared particle number operator. The smearing is essential to prevent energy divergences and defines a characteristic localization length scale commonly denoted by $r_C$. However, it soon became clear that the model would be more resistant to experimental falsification if the coupling was to the (smeared) mass density $\mcM (\bx)$ ~\cite{Pearle1994MassCSL} rather than to the particle number. Moreover, such a coupling leads to an important property named \enquote{composition invariance} or \enquote{compoundation invariance}~\cite{Rimini1997CompoundObjects}. Generally speaking, this property states that, under certain conditions, extended bodies can be effectively treated as point particles characterized by parameters such as total mass, charge, and velocity. In spontaneous collapse dynamics, compoundation invariance means that composite objects much smaller than $r_C$ can be treated as elementary particles, without the need to consider their detailed internal structure. The linear mass-dependent coupling in CSL undoubtedly plays a key role in the effectiveness of the model.

The CSL and DP models assume that the collapse occurs continuously over time. Models based on discrete quantum jumps have also been proposed. For example, the well-known Ghirardi-Rimini-Weber (GRW)~\cite{Ghirardi1986Unified} model is of this kind: every particle undergoes an instantaneous random spontaneous localization in space within a distance $r_C$ and according to a Poissonian distribution in time with rate $\lambda$. Originally, this recipe was not applicable to indistinguishable particles and to quantum fields. Generalized versions of GRW without this limitation now exist~\cite{Tumulka2006spontaneous,Piccione2023Collapse}. In particular, the model of Ref.~\cite{Piccione2023Collapse} posits that spacetime is randomly filled with natural detectors (called collapse points) whose clicking probability is proportional to the average of a certain location-dependent operator. The click of a collapse point corresponds to a spontaneous collapse of the wavefunction. In the limit of large density of collapse points in space and time, and vanishing coupling to quantum matter, one obtains the Poissonian Spontaneous Localization (PSL) model, which is mathematically similar to the model of Ref.~\cite{Tumulka2006spontaneous}. 

\begin{figure*}
	\centering
	\includegraphics[width=0.9\textwidth]{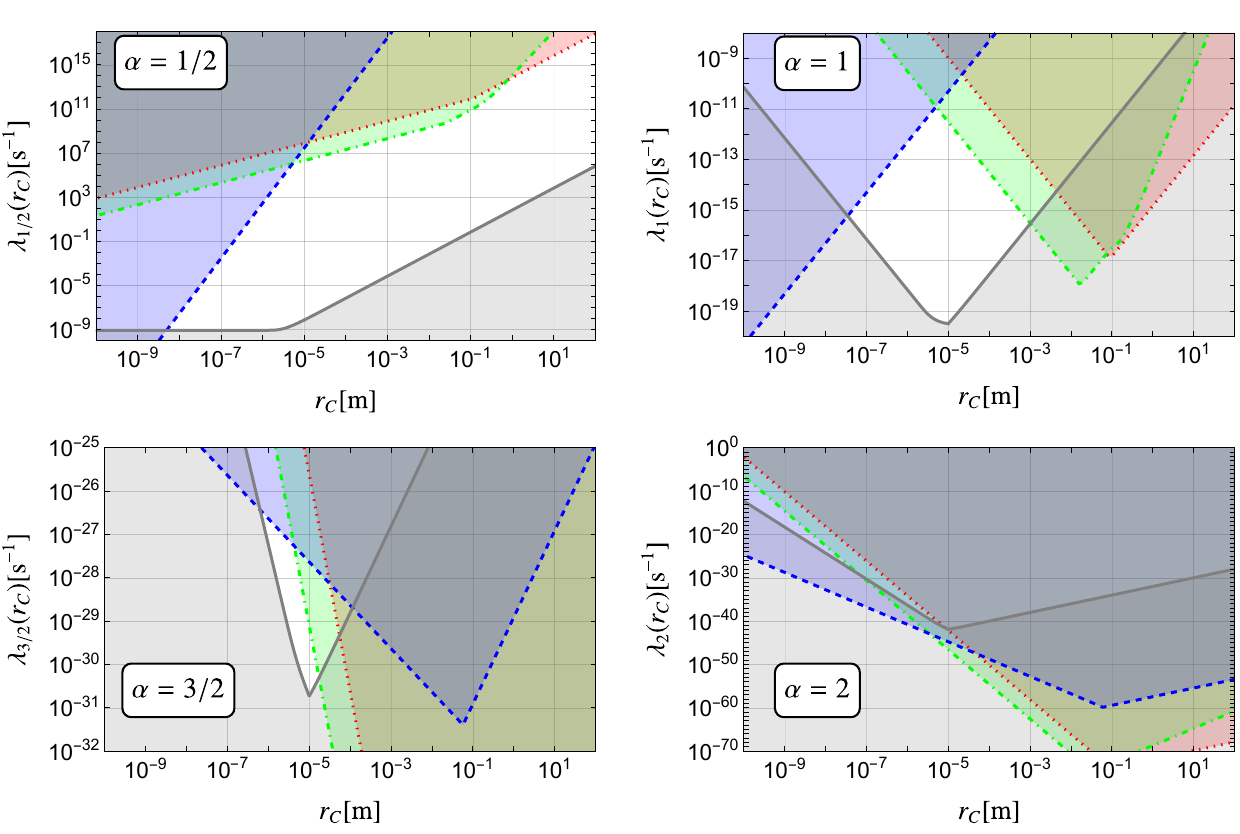}
	\caption{Exclusion plots for different values of $\alpha$ when $f(\mcM(\bx)) = \mcM^\alpha (\bx)/m_0^\alpha$. CSL corresponds to $\alpha=1$ while PSL corresponds to $\alpha=1/2$. In each plot, the experimentally allowed parameters ($\lambda$,$r_C$) have to reside within the white region lower bounded by the continuous gray line and upper bounded by the other lines. The lower bounds (continuous gray lines) are computed by means of Eq.~\eqref{eq:MinimumValueLocalizationRate}, the radiation bounds (dashed blue lines) by employing Eqs.~\eqref{eq:UpperBoundRadiation_Part1} and~\eqref{eq:UpperBoundRadiation_Part2}, and the LIGO and LISA (respectively the dotted red line and the dot-dashed green line) bounds by employing Eq.~\eqref{eq:ForceNoiseDensity} and Table~\ref{tab:GravWaveParameters}.}
	\label{fig:exclusionPlots}
\end{figure*}

Similarly to PSL, CSL can be obtained as a limiting case, assuming that space is filled with natural detectors performing continuous weak measurements of a suitable operator~\cite{Tilloy2016CSLGravity,Tilloy2017LeastDecoherence,GaonaReyes2021GravitationalFeedback}; standard mass-proportional CSL is recovered when the measured operator corresponds to the smeared mass density, which seems a natural choice. However, if the measurements are not continuous over time but instantaneous (as in PSL), the same natural choice leads to a  master equation defined in terms of $\sqrt{\mcM (\bx)}$ rather than $\mcM (\bx)$\footnote{The master equation would be exactly that of CSL if the natural detectors of the PSL model  measure $\mcM^2 (\bx)$, which seems a less natural choice.}. Moreover, assuming that the natural detectors measure the (smeared) mass density becomes mandatory if one wants the spontaneous collapses to source a Newtonian gravitational field in a Markovian way\footnote{This happens because a Markovian feedback must depend linearly on the measured quantity~\cite{Wiseman2002BayesianFeedback,Tilloy2024HybridDynamics}.}~\cite{Tilloy2016CSLGravity,Tilloy2017LeastDecoherence,GaonaReyes2021GravitationalFeedback,Piccione2023Collapse,Piccione2025NewtonianPSL}.

In a nutshell, the above mentioned collapse models differ in how their master equations depend on the mass density. It is therefore relevant to analyze the phenomenology of a CSL-like master equation defined in terms of a general function $\hL(\bx) = f(\mcM(\bx))$ of the mass density, where $f$ is a generic well-behaved function; this is the aim of the present work. Whenever a fully general treatment is not feasible, we focus on arbitrary positive powers of the mass density, with particular attention to the cases of power 1 and $1/2$, which have been previously introduced.

First, we will show that two important features of the CSL model are retained in the generalized version: compoundation invariance, previously introduced, and the decoupling of internal and center-of-mass degrees of freedom\footnote{For the CSL model, this is shown in Ref.~\cite{Bassi2003Dynamical}, sections 8.2 and 8.3.}. Then, specializing to the case $\hL(\bx)\ \propto\ \mcM^\alpha(\bx)$, with $\alpha>0$, we will see how the collapse dynamics affects the motion of single particles, of rigid bodies, and the spontaneous emission of radiation from charged particles. Finally, we will compare the theoretical predictions with available data from experiments in the literature to investigate the experimental upper bounds on these generalized CSL models.

The result of this investigation is
summarized in Fig.~\ref{fig:exclusionPlots}, where we also included the lower bound on the collapse parameters $(\lambda, r_C)$ coming from the theoretical requirement that the collapse should be strong enough to ensure the classical behavior of macroscopic objects~\cite{Toros2018BoundsCalculations}. Models with $\alpha \gtrsim 3/2$ seem highly unlikely to be correct; in particular, the choice $\alpha=2$ is ruled out. In contrast,  the PSL model (corresponding to $\alpha=1/2$) appears to be more robust to experimental falsification than the CSL model (corresponding to $\alpha=1$), presenting a much wider range of allowed parameters. Given that values of $\alpha$ smaller than $1/2$ seem difficult to justify, one arrives at the conclusion that only a limited range of values for $\alpha$ is available.

The paper is organized as follows. In Sec.~\ref{Sec:Models_CSL_PSL}, we introduce the two models and their generalization, also explaining the reasons behind some choices. In Sec.~\ref{Sec:GeneralFeaturesModels}, we prove the compoundation property and the decoupling of internal and center-of-mass motion for rigid bodies. Then, in Sec.~\ref{Sec:Phenomenology}, we investigate the dynamics of single particles, the dynamics of rigid bodies, and the emission of radiation due to spontaneous collapses. In Sec.~\ref{Sec:ExperimentalBounds}, we investigate the theoretical lower bound and the experimental upper bounds of these models. Finally, in Sec.~\ref{Sec:Conclusions} we summarize our results and discuss their implications.

\section{Continuous and Poissonian spontaneous localization models\label{Sec:Models_CSL_PSL}}

In the Continuous Spontaneous Localization (CSL) model, the dynamics are governed by the following stochastic and master equations:
\begin{equations}\label{eq:CSLDynamicsEquations}
	\dd{\ket{\psi_t}}
	&= \prtqB{
		-\frac{i}{\hbar} \hat{H}\dd{t}
		- \frac{\gamma_1}{2 m_0^2}\int \dd[3]{\bx} \prtq{\mcM (\bx)-\ev{\mcM (\bx)}}^2\dd{t}
		\\
		&\quad +\frac{\sqrt{\gamma_1}}{m_0}\int \dd[3]{\bx} \prtq{\mcM (\bx)-\ev{\mcM (\bx)}} \dd{W_t} (\bx)}\ket{\psi_t},
	\\
	\dv{t} \rho_t 
	&= -\frac{i}{\hbar}\comm{\hat{H}}{\rho_t} - \frac{\gamma_1}{2 m_0^2}\int \dd[3]{\bx} \comm{\mcM (\bx)}{\comm{\mcM (\bx)}{\rho_t}}.
\end{equations}
where $\hat{H}$ is the standard Hamiltonian of the system, $m_0$ is the proton mass\footnote{The quantity $1/m_0^{2}$ could be absorbed in $\gamma_1$ and $m_0$ could be changed with another mass by changing $\gamma_1$ accordingly. It is common practice in the spontaneous collapse models literature to use and define $m_0$ as we did.}, $\dd{W_t} (\bx)$ is the infinitesimal Wiener increment such that $\dd{W_t} (\bx)\dd{W_t} (\by)=\delta(\bx-\by)\dd{t}$, $\gamma_1$ is a constant, and $\mcM (\bx)$ is the smeared mass density operator defined as follows
\begin{equations}\label{eq:SmearedMassDensityOperator}
	&\textrm{Fields:}\quad
	\mcM (\bx) = \sum_{k} m_k \int \dd[3]{\by} g_{r_C} (\by-\bx) \ad_k(\by) a_k(\by), 
	\\
	&\textrm{Particles:}\quad
	\mcM (\bx) = \sum_{k} m_k g_{r_C} (\bx-\hbq_k), 
\end{equations}
where $g_{r_C}(\bx) = (2 \pi r_C^2)^{-3/2}\exp[-\bx^2/(2 r_C^2)]$ and $r_C$ is the \enquote{collapse radius}. When dealing with fields, $m_k$ is the mass of the $k$-th kind of particle, $a_k^\dg (\by)$ is the particle creation operator of particle-kind $k$ at $\by$, and $a_k (\by)$ is the corresponding annihilation operator. When dealing with particles, $\hbq_k$ is the position operator of the $k$-th particle and $m_k$ is its mass. The stochastic equation gives the actual dynamics followed by the quantum state $\ket{\psi_t}$ for a given realization of the noise, while the master equation gives the dynamics averaged over the noise, thus providing the empirical predictions of the model. It is important to recall that the stochastic equation always leads (with extremely high probability) to a fast localization of macroscopic objects~\cite{Bassi2003Dynamical,Bassi2013Models}, thus solving the measurement problem.

In the models presented in Ref.~\cite{Piccione2023Collapse}, the dynamics are governed by the following stochastic and master equations~\cite{Piccione2023Collapse}:
\begin{equations}\label{eq:PSLDynamicsEquations}
	\dd{\ket{\psi_t}}
	&= \prtqB{-\frac{i}{\hbar} \hat{H}\dd{t} -\int \dd[3]{\bx} \prt{1 + \frac{i \hL(\bx)}{\sqrt{\ev{\hL^2 (\bx)}}}}\dd{N_t} (\bx)\\
		&\qquad - \frac{1}{2}\int \dd[3]{\bx} \prt{\hL^2(\bx)-\ev{\hL^2 (\bx)}}\dd{t}
	} \ket{\psi_t},\\
	\dv{t} \rho_t 
	&= -\frac{i}{\hbar}\comm{\hat{H}}{\rho_t} - \frac{1}{2}\int \dd[3]{\bx} \comm{\hL (\bx)}{\comm{\hL (\bx)}{\rho_t}},
\end{equations}
where $\hL (\bx)$ is a generic Hermitian operator and we have the Poisson processes $\dd{N}_t (\bx)$ such that $\dd{N}_t (\bx) \dd{t} = 0$, $\dd{N}_t (\bx)\dd{N}_t (\by) = \delta (\bx-\by)\dd{N}_t (\bx)$, and $\mbE\prtq{\dd{N}_t (\bx)}= \ev{\hL^2 (\bx)}{\psi_t} \dd{t}$. These dynamics can be visualized as the ones obtained by having spacetime randomly filled with detectors which click with a probability proportional to $\ev{\hL^2 (\bx)}{\psi_t}$~\cite{Piccione2023Collapse}. The happening of such an event is usually called a \enquote{flash}, following standard terminology in the foundations of physics~\cite{Book_Tumulka2022Foundations}. For this reason, it makes sense to assume that $\hL (\bx) = \sqrt{\gamma_{1/2} \mcM (\bx)/m_0}$, with $\gamma_{1/2} >0$ being a coupling constant fulfilling the same role of $\gamma_{1}$ in the CSL model. With this choice of $\hL (\bx)$, the probability of a jumping event at time $t$ is related to the amount of mass around point $\bx$ at time $t$, loosely speaking. Hereafter, we will refer to the model corresponding to this choice of $\hL (\bx)$ as Poissonian Spontaneous Localization (PSL).

One could wonder why not to consider an operator such as 
\begin{equation}
	\hL (\bx) = \sqrt{\frac{\gamma_{1/2}}{m_0}}\sum_k \sqrt{m_k g_{r_C} (\bx-\hbq_k)}
\end{equation}
instead, in the case of particles.
A reason is that this does not lead to a flash probability proportional to the mass around $\bx$ in the case of a rigid body of constant density.
In fact, let us consider a body composed of $n_T$ types of atoms and/or molecules whose distance among each other is much less than the collapse radius $r_C$. Let us also consider a flash well inside the volume of space occupied by the body and let us denote by $\mcN_T$ the (constant within the body) density of objects of type $T$. Then, we can estimate the action of the operator
\begin{equation}\label{eq:WrongAlternativeSmearedMassDensity}
	\mcM_{\alpha} (\bx) := \sum_k \prt{m_k g_{r_C} (\bx-\hbq_k)}^\alpha,    
\end{equation}
as
\begin{multline}
	\mcM_{\alpha} (\bx) 
	= \sum_{T=1}^{n_T} \mcN_T m_T^\alpha \int \dd[3]{\by}g_{r_C}^\alpha (\bx-\hbq_k)
	=\\
	= \alpha^{-3/2}\sum_{T=1}^{n_T} \mcN_T m_T^\alpha
	= \sum_{T=1}^{n_T} \mu_T m_T^{\alpha-1},
\end{multline}
where $\mu_T$ is the body's constant mass density due to objects of type $T$. Taking $\alpha=1/2$, we see that this is not the desired results which is instead that $\ev{\hL^2 (\bx)} \propto \mu_0$, where $\mu_0 = \sum_T \mu_T$ is the body's constant density.
Additionally, this choice of $\hL (\bx)$ does not lead to a model with compoundation invariance~(see Sec.~\ref{subsec:CompoundationInvariance}).


For the sake of generality, we will consider the following master equation
\begin{equation}\label{eq:GeneralMasterEquation}
	\dv{t} \rho_t \!
	=\! -\frac{i}{\hbar}\comm{\hat{H}}{\rho_t} - \frac{\gamma_f}{2}\!\int \dd[3]{\bx} \comm{f\prt{\mcM (\bx)}}{\comm{f\prt{\mcM (\bx)}}{\rho_t}}, 
\end{equation}
where the constant $\gamma_f$ depends on the choice of the function $f$. The above master equation admits an unraveling both in terms of generalized CSL and PSL models. The CSL unraveling is obtained by substituting $\mcM (\bx)$ with $f\prt{\mcM (\bx)}$ in Eq.~\eqref{eq:CSLDynamicsEquations} while the PSL unraveling by setting $\hL (\bx)=\sqrt{\gamma_f} f\prt{\mcM (\bx)}$ in Eq.~\eqref{eq:PSLDynamicsEquations}. It is worth mentioning that Eq.~\eqref{eq:GeneralMasterEquation} (as the standard CSL model~\cite{Fu1997SpontaneousRadiation}) also admits a unitary unraveling (see Appendix~\ref{APPSec:StochasticPotential}):
\begin{equations}\label{eq:UnitaryUnraveling}
	i \hbar \dv{t} \ket{\psi_t} &= \prt{\hat{H}+ \hV(t)}\ket{\psi_t},\\
	\hV(t) &= -\hbar \sqrt{\gamma_f} \int \dd[3]{\bx} f\prt{\mcM (\bx)}w(\bx,t),
\end{equations}
where $w(\bx,t)$ is a white noise such that 
\begin{equation}
	\mbE\prtq{w(\bx,t)}=0
	\qquad
	\mbE \prtq{w(\bx,t)w(\by,s)}=\delta(t-s)\delta^{(3)}(\bx-\by).
\end{equation}
This unitary unraveling proves useful to simplify calculations and it also helps getting an intuitive explanation of some of them. One reason for this is that the stochastic potential $\hV(t)$ can also be seen as implementing a stochastic force on a system with momentum operator $\hbp$ equal to $\hbF = (i/\hbar)\comm{\hV(t)}{\hbp}$ which is the sum of many stochastic and uncorrelated forces\footnote{Because $V(t)$ is given by an integral of operators.}. It follows that the net effect of these forces on the average of $\hbp$ is null. However, for example, the effect is not null for $\ev{\hbp^2}$ so that we can immediately understand why, for any system, $\ev{\hbq}_t$ and $\ev{\hbp}_t$ evolve as if the spontaneous collapses were not present, but this is not the case for $\ev{\hbq^2}_t$ and $\ev{\hbp^2}_t$.

In the following sections, we will show that some important properties of the CSL model are retained in the generalized CSL and PSL models for reasonable functions $f$. Then we will focus on the phenomenology arising from the master equation with $f\prt{\mcM (\bx)} = \mcM^\alpha (\bx)/m_0^{\alpha}$, paying particular attention to the CSL and PSL cases, i.e., the cases $\alpha=1/2$ and $\alpha=1$. Before continuing, however, let us comment on a further reason for why the coefficients $\alpha=1/2$ and $\alpha=1$ are special. These are the only values that allow for the implementation of Newtonian gravity according to the Tilloy-Diósi prescription~\cite{Tilloy2016CSLGravity,Tilloy2017LeastDecoherence,Tilloy2018GRWGravity,Piccione2025NewtonianPSL}. This prescription consists of treating the spontaneous collapses as a spontaneous measurement process whose results source the gravitational, classical, feedback. In particular, $\alpha=1$ allows for a gravitational Markovian feedback based on continuous spontaneous collapse models such as the CSL and the Diósi-Penrose models~\cite{Tilloy2016CSLGravity,Tilloy2017LeastDecoherence,GaonaReyes2021GravitationalFeedback} while $\alpha=1/2$ allows for a gravitational Markovian feedback based on jumping events of the wavefunction~\cite{Piccione2023Collapse,Piccione2025NewtonianPSL}. All other values of $\alpha$ would require a non-Markovian gravitational feedback because Markovian feedback necessitates a linear dependence on the measured quantity~\cite{Wiseman2002BayesianFeedback,Tilloy2024HybridDynamics}. This suggests that other values of $\alpha$ make the connection of spontaneous collapse models and classical gravity more difficult.

\section{Two general features retained in the generalized CSL and PSL models\label{Sec:GeneralFeaturesModels}}

Here we show how two important properties of the CSL model are retained in the generalized CSL and PSL models. These are compoundation invariance~\cite{Rimini1997CompoundObjects} and the dynamical decoupling of a rigid body center of mass~\cite{Bassi2003Dynamical}.

\subsection{Compoundation invariance\label{subsec:CompoundationInvariance}}

An important feature of the standard CSL model is that of compoundation~\cite{Rimini1997CompoundObjects}. This property consists of the possibility, within a given theory and under certain conditions, to treat an object composed of two or more subsystems as an elementary object itself. A typical example is how, within classical mechanics and Newtonian gravity, we can treat planets as point particles at the solar system scale. In Ref.~\cite{Rimini1997CompoundObjects}, compoundation is shown for standard quantum mechanics and the mass-proportional CSL under the condition that there is a unique noise field acting on quantum matter. Notice that this condition has to hold at the level of the stochastic \Schr equation and not at the master equation level. Here we will show that this compoundation property holds in the generalized versions of the CSL and PSL models.

The scenario considered in Ref.~\cite{Rimini1997CompoundObjects} is the following:
\begin{itemize}
	\item There are a certain number $n_T$ of type $T$ of microscopic objects, labeled by the Greek index $\omega$. Objects of the same type are indistinguishable (e.g. two hydrogen atoms of the same isotope).
	\item There are particles of kind $k$ within each of those objects. The number of particle of kind $k$ within an object of type $T$ is $n_{T,k}$ while the total number of particle of kind $k$ is $n_k$.
\end{itemize}
A model has the compoundation property if its dynamical equation can be written at the level of the objects alone, while maintaining the same form of that at the elementary particles level. Regarding the standard CSL model, Ref.~\cite{Rimini1997CompoundObjects} shows that if each kind of elementary particle couples to a different noise, this is not possible. If they all couple to the same noise, compoundation holds and a natural way to obtain it is to make the interaction strength mass-proportional. 

To generalize the result of Ref.~\cite{Rimini1997CompoundObjects} it is sufficient to analyze the action of the smeared mass density operator.
First we rewrite it in terms of center-of-mass coordinates $\hbQ_{T,\omega}$ of the $\omega$-th object of type $T$ and relative coordinates $\hbr_{T,\omega,k,j}$ of the $j$-th particle of kind $k$ within that object:
\begin{multline}
	\mcM(\bx) 
	= \sum_k m_k \sum_{j=1}^{n_k}  g_{r_C}(\hbq_{k,j}-\bx)
	=\\
	=\sum_k m_k \sum_T \sum_{\omega=1}^{n_T} \sum_{j=1}^{n_{T,k}}  g_{r_C}(\hbQ_{T,\omega} + \hbr_{T,\omega,k,j}-\bx).
\end{multline}
Then, since the objects are microscopic we have that $\abs{\hbr_{T,\omega,k,j}} \ll r_C$, allowing us to make the following approximation:
\begin{multline}
	\mcM(\bx) 
	= \sum_k m_k \sum_T \sum_{\omega=1}^{n_T} \sum_{j=1}^{n_{T,k}}  g_{r_C}(\hbQ_{T,\omega} + \hbr_{T,\omega,k,j}-\bx)
	\simeq\\
	\simeq
	\sum_k m_k \sum_T \sum_{\omega=1}^{n_T} \sum_{j=1}^{n_{T,k}}  g_{r_C}(\hbQ_{T,\omega} -\bx).
\end{multline}
At this point, the sum over $j$ gives a factor $n_{T,k}$ because $g_{r_C}(\hbQ_{T,\omega} -\bx)$ is independent of $j$ (and also of $k$). Finally, since the mass of an object of type $T$ is $m_{T} = \sum_k n_{T,k} m_k$, we have that
\begin{equation}
	\label{eq:CompoundationInvarianceSmearedMassDensity}
	\mcM(\bx) \simeq \sum_T m_T \sum_{\omega=1}^{n_T}  g_{r_C}(\hbQ_{T,\omega} -\bx).
\end{equation}
The final result is that the smeared mass density operator acts at the level of objects in the same way as at the elementary constituents level. Therefore, these objects can be treated within the stochastic equations as elementary ones as long as the function $f$ is not too bizarre\footnote{To be more precise, denoting by $\delta \mcM$ the error induced by employing Eq.~\eqref{eq:CompoundationInvarianceSmearedMassDensity}, we assume that we can write $f\prt{\mcM+\delta \mcM} \simeq f\prt{\mcM}+f'\prt{\mcM}\delta \mcM$. Since $\delta \mcM \ll \mcM$, compoundation invariance is lost when $f'\prt{\mcM}\gtrsim f\prt{\mcM}/\delta \mcM$ for relevant values of $\mcM$.}. Notice that for the compoundation property to hold also the Hamiltonian has to possess it independently of the spontaneous collapse part of the dynamics. For example, neutral atoms cannot be treated as elementary neutral particles for the purpose of calculating the spontaneous radiation in the $X$-ray domain because (despite being much smaller than $r_C$) the average distance between electrons among themselves and with the nucleus is comparable to or much higher than the wavelength of the emitted radiation~\cite{Donadi2014RadiationEmission,Donadi2015Radiation,Donadi2021NovelCSLBounds,Donadi2021UndergroundTest}.

As anticipated in Sec.~\ref{Sec:Models_CSL_PSL}, we now show that using an operator such as $\mcM_{\alpha} (\bx)$ of Eq.~\eqref{eq:WrongAlternativeSmearedMassDensity} instead of $f(\mcM (\bx))$ does not allow to maintain compoundation invariance, further justifying our generalization of the CSL model dynamics with respect to another one which is, seemingly, equally natural. 
In fact, repeating the calculation done above with $\mcM_{\alpha} (\bx)$, we have that:
\begin{multline}
	\mcM_{\alpha} (\bx) 
	= \sum_k m_k^\alpha \sum_{j=1}^{n_k}  g_{r_C}^\alpha (\hbq_{k,j}-\bx)
	=\\
	= \sum_k m_k^\alpha \sum_T \sum_{\omega=1}^{n_T} \sum_{j=1}^{n_{T,k}}  g_{r_C}^\alpha (\hbQ_{T,\omega} + \hbr_{T,\omega,k,j}-\bx)
	\simeq\\ 
	\simeq \sum_k m_k^\alpha \sum_T \sum_{\omega=1}^{n_T} \sum_{j=1}^{n_{T,k}}  g_{r_C}^\alpha (\hbQ_{T,\omega} -\bx)
	=\\
	=\sum_k m_k^\alpha \sum_T \sum_{\omega=1}^{n_T} n_{T,k} g_{r_C}^\alpha (\hbQ_{T,\omega} -\bx)
	\neq\\ 
	\neq 
	\sum_T m_T^\alpha \sum_{\omega=1}^{n_T}  g_{r_C}^\alpha(\hbQ_{T,\omega} -\bx),
\end{multline}
where in the last line the equality holds only for $\alpha=1$ since $m_T = \sum_k n_{T,k} m_k$.

\subsection{Decoupling of a rigid body center of mass}

An important property of collapse models such as the GRW and CSL models is that the dynamics of the center of mass of a rigid body decouple from their internal one. Generalizing this result is quite straightforward as the decoupling is guaranteed by the fact that the action of $\mcM (\bx)$ can be approximated as an action on the center of mass degree of freedom. Then, any sufficiently regular function of $\mcM (\bx)$ maintains the validity of this approximation. This decoupling of the action of $\mcM (\bx)$ is shown, for example, in Sec. 8.2 of Ref.~\cite{Bassi2003Dynamical}. However, for commodity of the reader, we repeat here the argument. 

We start by considering the smeared mass density operator for $N$ particles in position representation. Denoting by $q$ the coordinates of all particles, we get that [cf. Eq.~\eqref{eq:SmearedMassDensityOperator}]
\begin{equation}
	\mel{q}{f(\mcM(\bx))}{\psi}
	=
	f\prt{\sum_k m_k g_{r_C} (\bx-\bq_k)}\psi(q),
\end{equation}
where $\psi(q)$ is the wavefunction in spatial representation, $m_k$ is the mass of the $k$-th particle, and $\bq_k$ is the position of the $k$-th particle. We can then make the substitution $\bq_k = \bQ + \tilde{\bq}_k (r)$, where $r$ denotes internal variables of the rigid body, $\bQ = \sum_k (m_k/M) \bq_k$ is the center of mass position, and $M=\sum_k m_k$ is the total mass of the body. We then assume that the wavefunction of the rigid body can be decomposed as $\psi(q) = \Psi (\bQ)\xi(r)$, where $\xi(r)$ is sharply peaked (with respect to $r_C$) around the coordinate $r_0$, which represents the equilibrium position of these internal variables. It follows that one can assume that the structure of the rigid body is basically unaffected by the spontaneous collapses so that, in the collapse operator, $r$ can be substituted with $r_0$. This implies that
\begin{equations}\label{eq:SmearedMassDensityDecoupling}
	\mel{q}{f(\mcM(\bx))}{\psi}
	&=
	f\prt{\mcM_{\rm CM}(\bQ-\bx)}\Psi(\bQ)\xi(r),\\
	\mcM_{\rm CM}(\bQ-\bx)
	&\equiv 
	\sum_k m_k g_{r_C} (\bQ+\tl{\bq}_k(r_0)-\bx).
\end{equations}
Indeed, the above expression for $\mcM (\bx) \simeq \mcM_{\rm CM} (\hbQ-\bx)$ only acts on the Hilbert space of the center of mass.
In general, the Hamiltonian of the rigid body can be written as the sum of a Hamiltonian for the center of mass and a Hamiltonian for the internal structure of the rigid body. Since the collapse operators only act on the center of mass part of the wavefunction, the dynamics of the center of mass is decoupled from that of the rigid body internal structure and the spontaneous collapses only affect the center of mass dynamics in this approximation. Notice that, under these approximations, this holds true at the wavefunction level, not just at the density matrix one. Moreover, notice that, up to now, we did not make assumptions on the dimensions or the structure of the rigid body.

\begin{figure}[t]
	\centering
	\includegraphics[width=0.48\textwidth]{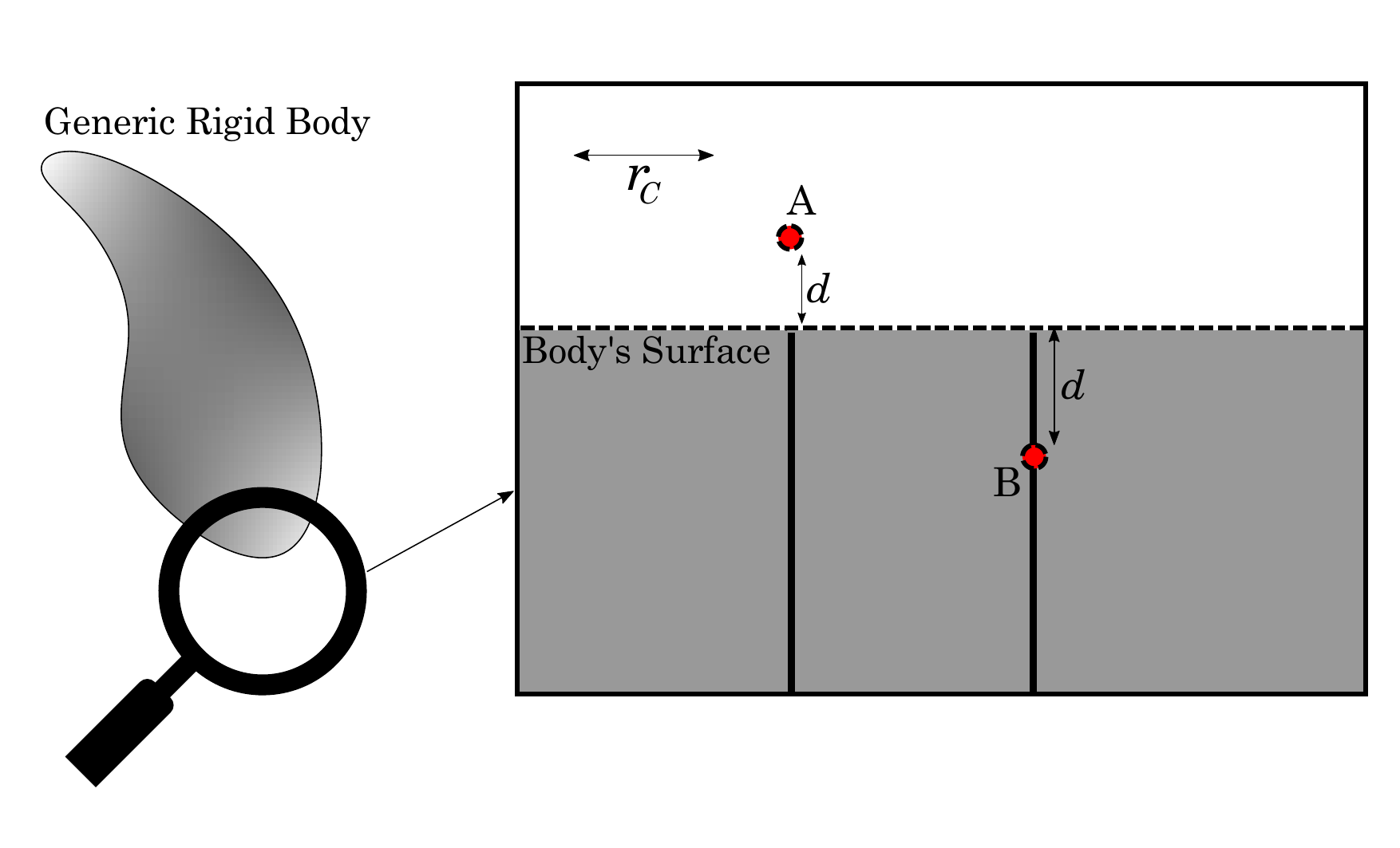}
	\caption{Pictorial explanation of the reasoning behind Eq.~\eqref{eq:ConstantDensitySmearedMassDensityOperator}. When computing $\mcM (\bx)$ with $\bx$ at a distance of order $r_C$ from the surface, we obtain what depicted in the picture: the surface appears constant on the $r_C$ lengthscale. The thick vertical lines represent the Gaussian integration on the direction perpendicular to the rigid body's surface. So, if $\bx$ is outside the body (like point $A$) the Gaussian integration has to stop at $+\abs{d}$, while if $\bx$ is inside the body (like point $B$) the Gaussian integration has to stop at $-\abs{d}$.}
	\label{fig:SmearedDensityOperatorConstantDensityExplanation}
\end{figure}

Let us now consider a rigid body much larger than the collapse radius $r_C$ in all directions. Then, we can employ the so-called \enquote{macroscopic density approximation}~\cite{Bassi2003Dynamical}\footnote{See Secs. 8.3 and 8.4 of Ref.~\cite{Bassi2003Dynamical}.} to write $\mcM_{\rm CM} (\bx) = \int \dd[3]{\by} \mu (\by) g_{r_C} (\bx-\by)$, where $\mu (\bx)$ is the body's mass density. In many cases of interest, we consider rigid bodies with a constant density $\mu_0$, i.e., $\mu (\bx) = \mu_0 \chi_V (\bx)$, where $\chi_V (\bx)$ is the indicator function associated to the body's volume and shape with respect to its center of mass. In this case, a closed formula for $\mcM_{\rm CM} (\bx)$ can be found. To do this, we assume the body's surface to be locally flat on the lengthscale $r_C$. For $r_C \sim 10^{-7} {\rm m}$ this appears reasonable as the surface irregularities due to the atomicity of matter will be on a scale of $10^{-10}-10^{-9} {\rm m}$ and thus invisible to the collapse operators. We define a function $d(\bx)$ that quantifies the distance between a spatial point $\bx$ and the rigid body's surface $\partial V$ and such that $d(\bx)<0$ if $\bx$ is within the body and $d(\bx)>0$ if $\bx$ is outside of the body. 
Clearly, when $d(\bx) \ll -1$ one has that $\mcM_{\rm CM} (\bx) \simeq \mu_0/m_0$ while $d(\bx) \gg 1$ implies $\mcM_{\rm CM} (\bx) \simeq 0$. Then, it remains to understand how to compute $\mcM_{\rm CM} (\bx)$ for all the spatial points whose distance from the rigid body's surface is comparable with $r_C$. Because of our assumptions, at this scale the surface appears as a plane (see Fig.~\ref{fig:SmearedDensityOperatorConstantDensityExplanation}) so that for each point we can choose a coordinate system in which a variable (say, $y_3$) lies on the line connecting $\bx$ to $\partial V$. Formally, this coordinate system is reached by first making a translation so that $\bx=0$ and then a rotation $\mcR$ which leaves $g_{r_C}(\by)$ invariant. Then, the Gaussian integration along the orthogonal directions gives one and we get
\begin{multline}\label{eq:ConstantDensitySmearedMassDensityOperator}
	\mcM_{\rm CM} (\bx) 
	\simeq
	\int \dd[3]{\by} \mu_0 (\by) g_{r_C} (\bx-\by)
	=\\
	= \mu_0 \int \dd[3]{\by} \chi_V \prt{\mcR(\by+\bx)} g_{r_C} (\by)
	=\\
	= \mu_0 \intmp \dd{y_1}\dd{y_2} \frac{e^{-(y_1^2+y_2^2)/2 r_C^2}}{(2 \pi r_C^2)}\int_{-\infty}^{-d} \frac{e^{-y_3^2/2 r_C^2}}{(2 \pi r_C^2)^{1/2}} \dd{y_3}
	=\\
	= \mu_0 \Phi \prt{\frac{-d(\bx)}{r_C}},
\end{multline}
where we recall that $\Phi (z) = [1+\erf(z/\sqrt{2})]/2$ is the cumulative distribution function of the standard normal distribution\footnote{It is straightforward to check that if we compute the above quantity near a surface separating two rigid bodies with densities $\mu_1$ and $\mu_2$ we get $\mcM_{\rm CM} (\bx) = (\mu_1/m_0)\Phi(-d(\bx)/r_C) + (\mu_2/m_0)\Phi(+d(\bx)/r_C)$, with $d(\bx)<0$ if $\bx$ is within the body with density $\mu_1$.}. By taking the limit $r_C \rightarrow 0$ one obtains the result named \enquote{sharp scanning approximation} in Ref.~\cite{Bassi2003Dynamical}. 
Moreover, we can use Eq.~\eqref{eq:ConstantDensitySmearedMassDensityOperator} as a general formula for any $\bx$ as it correctly gives $\mcM_{\rm CM} (\bx)\simeq 0$ when $d(\bx) \gg 1$ and $\mcM_{\rm CM} (\bx)\simeq \mu_0$ when $d(\bx) \ll -1$. This novel way of writing an approximate version of $\mcM_{\rm CM} (\bx)$ will be used in Sec.~\ref{subsec:RigidBodyDynamics} to compute the effects of the spontaneous collapses on the dynamics of a rigid body.


\section{Phenomenology of the generalized models\label{Sec:Phenomenology}}

In this section we explore the phenomenology of the generalized CSL and PSL models. In order to do so, hereafter we specialize to the case in which $f\prt{\mcM(\bx)} = \mcM^\alpha (\bx)/m_0^\alpha$. Moreover, we will also write $\gamma_\alpha$ instead of $\gamma_f$ to denote the coupling constant so to remember that now we are considering only powers of the smeared mass density operator.

\subsection{Single particle dynamics\label{subsec:SingleParticleDynamics}}

Let us start by considering a single particle of mass $m$.
Neglecting the standard Hamiltonian of the particle, the density matrix in position representation evolves as follows:
\begin{equations}\label{eq:SingleParticleDecoherence}
	&\dv{t} \mel{\bq'}{\rho_t}{\bq''} 
	= 
	- \Gamma_\alpha (\bd) \mel{\bq'}{\rho_t}{\bq''},
	\qquad
	\bd := \bq'-\bq'',
	\\
	&\Gamma_\alpha (\bd) 
	= \frac{\gamma_\alpha}{2} \prt{\frac{m}{m_0}}^{2\alpha}\int \dd[3]{\bx} \prtq{g_{r_C}^\alpha (\bx) - g_{r_C}^\alpha(\bx - \bd)}^2,\\
	&\hskip 9mm = \lambda_\alpha \prt{\frac{m}{m_0}}^{2\alpha} \prtq{1-e^{-\alpha \bd^2/(4 r_C^2)}},
	\\
	&\mcM(\bx) = m g_{r_C}\prt{\hbq - \bx},
	\qquad
	\lambda_\alpha := \gamma_\alpha \frac{\prt{\pi r_C^2/\alpha}^{3/2}}{\prt{2 \pi r_C^2}^{3\alpha}}.
\end{equations}
where $\hbq$ denotes the vector position operator for the particle.
We can see that, with just one particle, the dynamics are functionally (but not physically) equivalent for any value of $\alpha$. The most important difference when varying $\alpha$ is the variation of the functional dependence of the decoherence rate on the particle's mass. The mass dependence cannot be eliminated in any way and makes the models behave differently in practice, even if they appear formally equivalent.

The standard ($\alpha=1$) CSL collapse rate~\cite{Bassi2013Models} amounts to $\gamma_1 \sim 10^{-36} \rm{m}^{3} \rm{s}^{-1}$ and corresponds to $\lambda_1 \sim 2.2 \times 10^{-17} \rm{s}^{-1}$ when choosing $r_C \sim 10^{-7} \rm{m}$, which could be taken as the reference value for the collapse radius~\cite{Bassi2013Models}. The collapse rate $\lambda_1$ corresponds to the decoherence rate $\Gamma_1 (\bd)$ of Eq.~\eqref{eq:SingleParticleDecoherence} when $\bd \gg r_C$ and $m=m_0$, i.e., when we are studying the dynamics of a nucleon. Taking this as a reference, we can get the same decoherence rate for $\alpha=1/2$ by simply choosing $\lambda_{1/2}=\gamma_{1/2} \sim 2.2 \times 10^{-17} \rm{s}^{-1}$. Notice that $\alpha=1/2$ is the only value of $\alpha$ for which, once $\gamma_\alpha$ is chosen, $\lambda_\alpha$ is independent of $r_C$.

The decoherence rate of Eq.~\eqref{eq:SingleParticleDecoherence} is also equivalent to the GRW dynamics of Ref.~\cite{Ghirardi1986Unified}, and, in fact, when considering a particle with Hamiltonian $H= \hbp^2/2m$, we can directly read the evolution of the density matrix by looking at Eqs.~(3.6) and~(3.7) [pag. 34] of Ref.~\cite{Ghirardi1986Unified}:
\begin{multline}
	\rho_t (\bq',\bq'') 
	= \frac{1}{(2\pi)^3} \int \dd{\bk} \dd[3]{\bx} 
	e^{-i \bk \cdot \bx} \times
	\\
	\times F_t(\bk,\bq'-\bq'') \rho^{\rm Sch}_t (\bq'+\bx,\bq''+\bx),
\end{multline}
where
\begin{multline}
	F_t(\bk,\bd) = \exp \prtgB{-\lambda_\alpha \prtB{\frac{m}{m_0}}^{2\alpha} \prtqB{\\
			t - \int_0^t \exp[-\frac{\alpha}{4 r_C^2}\prt{\bd-s\frac{\hbar}{m}\bk}^2] \dd{s}}},
\end{multline}
and $\rho^{\rm Sch}_t$ denotes the density matrix which evolved uniquely under the standard Schr\"{o}dinger's evolution.
From the above equation, following again Ref.~\cite{Ghirardi1986Unified}, one can get the time evolution for averages of observables.
The average values of position and momentum are not affected by the spontaneous collapses. However, their squares are:
\begin{equations}\label{eq:QuadraticObservablesSingleParticle}
	\ev{\hq_j^2} &= \ev{\hq_j^2}_{\rm Sch} + \frac{\alpha \lambda_\alpha}{r_C^2} \prt{\frac{m}{m_0}}^{2\alpha}\frac{\hbar^2}{6 m^2} t^3,
	\\
	\ev{\hp_j^2} &= \ev{\hp_j^2}_{\rm Sch} + \frac{\alpha \lambda_\alpha}{r_C^2} \prt{\frac{m}{m_0}}^{2\alpha}\frac{\hbar^2}{2} t,
\end{equations}
where the subscript $j=1,2,3$ denotes an arbitrary direction and $\ev*{\hat{O}}_{\rm Sch}$ denotes the value the operator would have in absence of spontaneous collapses. We see that different values of $\alpha$ lead to different dynamics because of the different dependence on the mass, which is indeed of the same kind as for the decoherence rate.

All of the previous equations have been derived assuming a completely isolated particle. If this is not the case, the dynamics of the single particle depends on the surrounding masses when $\alpha \neq 1$ and this change is appreciable if these other masses are located within a distance comparable to $r_C$ of the particle under scrutiny.
We can see this by exploiting the unitary unraveling of Eq.~\eqref{eq:UnitaryUnraveling}. The stochastic force acting on the particle becomes
\begin{multline}\label{eq:ForceNotIsolatedSingleParticle}
	\hbF = (i/\hbar)\comm{\hV(t)}{\hbp} 
	=\\
	= \frac{\alpha \hbar\sqrt{\gamma_\alpha}}{m_0^{\alpha-1}} \prt{\frac{m}{m_0}}\!\int \dd[3]{\bx} \mcM^{\alpha-1} (\bx) [\nabla g_{r_C} (\hbq-\bx)]w(\bx,t),
\end{multline}
where we see that only for $\alpha=1$ the force remains the same. For $\alpha>1$, every additional particle within the radius $r_C$ increases the force felt by the particle under consideration [cf.~\eqref{eq:SmearedMassDensityOperator}]. For $\alpha<1$, the presence of additional masses lowers the magnitude of the force exerted on the particle under consideration. In particular, considering $N+1$ particles and focusing on the decoherence of the $(N+1)$-th one (labeled by the subscript $p$), we get
\begin{widetext}
	\begin{multline}
		\dv{t} \rho_t^{(p)}(\bx_p,\by_p) = 
		-\frac{\gamma_\alpha}{2m_0^{2\alpha}} \int \dd[3]{\bz}\dd[3]{\bz_1}\dots\dd[3]{\bz_N} 
		\rho_t (\bx_p,\bz_1,\dots,\bz_N ;\by_p,\bz_1,\dots,\bz_N)
		\times \\ \times
		\prtqB{
			\prt{m_p g_{r_C} (\bx_p-\bz) + \sum_k m_k g_{r_C} (\bz_k-\bz)}^\alpha
			-\prt{m_p g_{r_C} (\by_p-\bz) + \sum_k m_k g_{r_C} (\bz_k-\bz)}^\alpha}^2.
	\end{multline}
	If we assume that $\rho_t = \rho^{(p)}_t \otimes \rho_t^{(N)}$ we get that $\dot{\rho}^{(p)}_t (\bx_p,\by_p) = -\Gamma_\alpha (\bx_p,\by_p,\rho_t^{(N)}) \rho_t^{(p)}$, where
	\begin{multline}
		\Gamma_\alpha (\bx_p,\by_p,\rho_t^{(N)}) = 
		\frac{\gamma_\alpha}{2m_0^{2\alpha}} \int \dd[3]{\bz}\dd[3]{\bz_1}\dots\dd[3]{\bz_N} 
		\rho_t^{(N)} (\bz_1,\dots,\bz_N;\bz_1,\dots,\bz_N)
		\times \\ \times
		\prtqB{
			\prt{m_p g_{r_C} (\bx_p-\bz) + \sum_k m_k g_{r_C} (\bz_k-\bz)}^\alpha
			-\prt{m_p g_{r_C} (\by_p-\bz) + \sum_k m_k g_{r_C} (\bz_k-\bz)}^\alpha}^2.
	\end{multline}
\end{widetext}
Focusing on any $m_k$, we can see\footnote{To see this mathematically, one can take the derivative of the term inside the square brackets with respect to $m_k$. The result would be $2 \alpha g_{r_C} (\bz_k-\bz)[(\dots)_\bx^\alpha-(\dots)_{\by}^\alpha][(\dots)_\bx^{\alpha-1}-(\dots)_{\by}^{\alpha-1}]$, with implicit meaning of symbols. For $\alpha>1$, the two quantities within square brackets have the same sign while, for $\alpha<1$, they have opposite sign.} how substituting it with $m'_k > m_k$ would increase $\Gamma_\alpha (\bx_p,\by_p,\rho_t^{(N)})$ if $\alpha>1$, leave it unvaried for $\alpha=1$, and decrease it for $\alpha<1$. In the PSL model, the intuition behind this behavior stems from the fact that the probability of a collapse event is related to the amount of total mass around the point at which the event takes place. If very heavy masses are nearby (on the $r_C$ lengthscale) two locations at which a small-mass particle is spatially superposed, the collapse events are mainly due to the heavy masses and so the small-mass particle is not localized.

It is worth noting that our generalized CSL model does not satisfy the \enquote{innocent bystander} assumption introduced in Ref.~\cite{Nimmrichter2013Macroscopicity}. This assumption requires additional non-interacting particles to not affect the decoherence behavior of the particle under consideration. In our model, this condition is generally violated for $\alpha\neq 1$. However, the violation is local in space and substantially occurs only within distances comparable to the collapse length scale $r_C$.

\subsection{Rigid body dynamics\label{subsec:RigidBodyDynamics}}

We now focus on the dynamics of a rigid body. We start by analyzing their decoherence rate, which is given by [see Eq.~\eqref{eq:SmearedMassDensityDecoupling} and cf. Eq.~\eqref{eq:SingleParticleDecoherence}]
\begin{equation}\label{eq:CenterOfMassDecoherenceRate}
	\Gamma^{\rm{CM}}_\alpha (\bD) = \frac{\gamma_\alpha}{2m_0^{2\alpha}} \int \dd[3]{\bx} \prtq{\mcM^\alpha_{\rm CM}(\bx) - \mcM^\alpha_{\rm CM}(\bx-\bD)}^2.
\end{equation}
Generally speaking, $\mcM_{\rm CM} (\bx)$ is hard to evaluate exactly. However, it can be easily estimated when considering rigid bodies with dimensions much larger than $r_C$ in all directions and whose coarse-grained density $\mu(\bx)$ varies on scale larger than $r_C$. Then, the Gaussian convolution mostly has no effect on the density so that we can write $\mcM_{\rm CM}(\bx) \simeq \mu(\bx)$\footnote{This is called \emph{sharp scanning approximation} in Refs.~\cite{Ghirardi1990_CSL,Bassi2003Dynamical}.}. Substituting this back in Eq.~\eqref{eq:CenterOfMassDecoherenceRate}, we have:
\begin{equation}\label{eq:RigidBodyDecoherenceRateGeneralFormula}
	\Gamma^{\rm{CM}}_\alpha (\bD) 
	\simeq 
	\frac{\gamma_\alpha}{m_0^{2\alpha}} \int \dd[3]{\bx} \prtg{\mu^{2\alpha} (\bx) - \prtq{\mu(\bx) \mu (\bx-\bD)}^\alpha}.
\end{equation}
When $\bD$ is larger than the body dimensions (i.e., $\mu(\bx)\mu(\bx-\bD)=0\ \forall \bx$), the decoherence rate reduces to $\Gamma^{\rm{CM}}_\alpha = (\gamma_\alpha/m_0)\int \dd[3]{\bx} \mu^{2\alpha} (\bx)$, which corresponds to the single particle case [cf. Eq.~\eqref{eq:SingleParticleDecoherence}] only for $\alpha=1/2$. Moreover, for $\alpha=1/2$, the terms in Eq.~\eqref{eq:RigidBodyDecoherenceRateGeneralFormula} assume a clearer meaning. The first one corresponds to the total mass $M$ of the body.
The second one corresponds to the integration of the geometric mean between the densities associated to the body's center of mass being in positions $\mathbf{0}$ and $\bD$.

Let us now consider a rigid body whose center of mass is governed by the Hamiltonian $\hbP^2/2M$, where $\hbP$ is the momentum operator of the center of mass. The decoherence rate is given by Eq.~\eqref{eq:CenterOfMassDecoherenceRate} and does not modify the dynamics of $\ev{\hbP}$ and $\ev{\hbQ}$, where $\hbQ$ is the position operator of the center of mass\footnote{See Sec. 8.4 of Ref.~\cite{Bassi2003Dynamical} or Sec. III.B.3 of Ref.~\cite{Ghirardi1990_CSL}.}. However, the spontaneous decoherence does modify the average of the following quantities\footnote{By redoing the calculations, we noticed that the result in Refs.~\cite{Ghirardi1990_CSL} and~\cite{Bassi2003Dynamical} contains an erroneous $1/2$ factor.}
\begin{equations}\label{eq:RigidBodyObservablesDynamics}
	\ev{\hQ_j^2} &= \ev{\hQ_j^2}_{\rm Sch} + \gamma_\alpha \frac{\hbar^2}{3 M^2}t^3 C_j,
	\\
	\ev{\acomm{\hQ_j}{\hP_j}} &= \ev{\acomm{\hQ_j}{\hP_j}}_{\rm Sch} + \gamma_\alpha \frac{\hbar^2}{M^2}t^2 C_j,
	\\
	\ev{\hP_j^2} &= \ev{\hP_j^2}_{\rm Sch} + \gamma_\alpha \hbar^2 t C_j,
\end{equations}
where the subscript $j=1,2,3$ denotes an arbitrary direction, $\ev{\hat{O}}_{\rm Sch}$ denotes the value the operator would have in absence of spontaneous collapses, and $C_j = m_0^{-2\alpha}\int \dd[3]{\by} \prtq{\partial_j \mcM^\alpha_{\rm CM}(\by)}^2$. 

\begin{figure}
	\centering
	\includegraphics[width=0.95\linewidth]{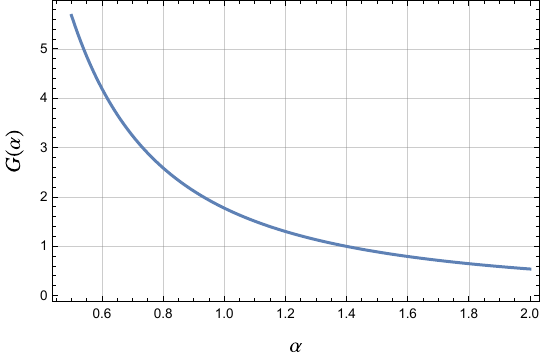}
	\caption{Plot of $G(\alpha)$, defined in Eq.~\eqref{eq:GeneralFormulaCoefficientsRigidBody}.}
	\label{APPfig:funcGplot}
\end{figure}

In Refs.~\cite{Ghirardi1990_CSL,Bassi2003Dynamical}, the quantity $C_j$ is estimated for the standard\footnote{More precisely, the CSL model which is not mass-proportional. However, for bodies with constant density the two models are equivalent up to a numerical factor.} CSL model considering a parallelepiped of constant density $\mu_0$ with edges $L_1,L_2,L_3$. We greatly
generalize this result to any constant density rigid body whose surface can be assumed locally flat on an $r_C$ scale and for a generic exponent $\alpha$. This is done by employing Eq.~\eqref{eq:ConstantDensitySmearedMassDensityOperator} (see Appendix~\ref{APPSec:MomentumSpreadGeneralCase}). The result is
\begin{equations}\label{eq:GeneralFormulaCoefficientsRigidBody}
	C_j 
	&= \alpha^2 \prt{\frac{\mu_0}{m_0}}^{2 \alpha}
	\frac{G(\alpha)}{2 \pi r_C} \int_{\partial V} n^2_j (\bx) \dd{S},
	\\
	G(\alpha) &:= \int \dd{z} \Phi^{2\alpha-2}(z)e^{-z^2},
\end{equations}
where $\partial V$ is the rigid body's surface, $n_j (\bx)$ the $j$-th component of the unit normal vector $\bn (\bx)$ to the surface, and $\Phi (z) = [1+\erf(z/\sqrt{2})]/2$ is the cumulative distribution function of the standard normal distribution. The function $G(\alpha)$ can be computed exactly for $\alpha=1$ and $\alpha=3/2$ giving $G(1)=\sqrt{\pi}$ and $G(3/2)=\sqrt{\pi}/2$. Indeed, the value for $\alpha=1$ coincides with the one given in Ref.~\cite{Ghirardi1990_CSL}. In general, we did not find other special cases but $G(\alpha)$ is easy to compute numerically. We report its graph in Figure~\ref{APPfig:funcGplot}. The numerical value for $\alpha=1/2$ is $G(1/2) \simeq 5.675$.

From Eq.~\eqref{eq:GeneralFormulaCoefficientsRigidBody}, using the fact that $\sum_j n^2_j (\bx)=1$ and $M=\mu_0 V$, we can write the following formulas:
\begin{equations}\label{eq:QuadraticObservablesConstantDensityRigidBody}
	\ev{\hbQ^2} &= \ev{\hbQ^2}_{\rm Sch} +  \frac{\gamma_\alpha}{3 \mu_0^2} \prt{\frac{\mu_0}{m_0}}^{2 \alpha}\frac{\hbar^2 \alpha^2 G(\alpha)}{2 \pi r_C} 
	\frac{A}{V^2} t^3,
	\\
	\ev{\hbP^2} &= \ev{\hbP^2}_{\rm Sch} + \gamma_\alpha \prt{\frac{\mu_0}{m_0}}^{2 \alpha}
	\frac{\hbar^2 \alpha^2 G(\alpha)}{2 \pi r_C} A t,
\end{equations}
where $A$ is the surface area of the rigid body. In the above formulas, the only extensive properties of the rigid body are its area and its volume. So, we see that by increasing the body's dimensions the impact on the momentum dispersion increases but that on its position dispersion decreases. Moreover, also the impact on the rigid body's energy decreases with the dimensions of the rigid body as we have
\begin{equation}\label{eq:EnergyIncreaseConstantDensityRigidBody}
	\dv{t} \ev{\frac{\hbP^2}{2M}} =  
	\frac{\gamma_\alpha \hbar^2 \alpha^2 G(\alpha)}{4 \pi r_C m_0} \prt{\frac{\mu_0}{m_0}}^{2 \alpha-1} \frac{A}{V}.
\end{equation}

Two comments on Eqs.~\eqref{eq:QuadraticObservablesConstantDensityRigidBody} and~\eqref{eq:EnergyIncreaseConstantDensityRigidBody} are in order.
First, a comparison with the single particle case [cf. Eq.~\eqref{eq:QuadraticObservablesSingleParticle}] shows that in the constant density rigid body's case the role of the particle's mass $m$ is taken by the rigid body's density and not by its total mass $M$. Second, the dependence on $A$ and $V$ hints at how the effects of the spontaneous collapses can be maximized so to obtain more stringent experimental bounds on the free parameters of the spontaneous collapse theories. For example, if we were interested in maximizing the energy increase we should search for a body shape which maximizes its surface area for a given volume, keeping in mind that we still need it to always be large with respect to $r_C$ for the formulae to hold. In other words, its thickness should be much larger than $r_C$ everywhere.
This idea has already been explored for the standard CSL model in Refs.~\cite{Vinante2020UltracoldLayeredForce,Adler2021LayeringEffect,Diosi2021SurfaceTensors}.

As a final note, we think that the appearance of the rigid body's surface area in Eq.~\eqref{eq:QuadraticObservablesConstantDensityRigidBody} can be intuitively understood as follows. Let us look at Eq.~\eqref{eq:UnitaryUnraveling} with $\mcM(\bx)\simeq \mcM_{\rm CM} (\bx)$ [Eqs.\eqref{eq:SmearedMassDensityDecoupling} and~\eqref{eq:ConstantDensitySmearedMassDensityOperator}]. Well inside the rigid body, the collapse operator becomes proportional to the identity while well outside the body it is vanishing. So, the stochastic force acting on the center of mass is applied on the rigid body's surface and this explains why the momentum dispersion changes proportionally to the area of the rigid body. Moreover, the proportionality of this force to $(\mu_0/m_0)^\alpha$ is again clear from Eqs.~\eqref{eq:UnitaryUnraveling} and~\eqref{eq:ConstantDensitySmearedMassDensityOperator}.

\subsection{Spontaneous Radiation\label{subsec:SpontaneousRadiation}}

One of the most striking consequence of spontaneous collapse theories is that they entail the emission of spontaneous radiation from charged particles~\cite{Fu1997SpontaneousRadiation}. This is easily understood by unraveling the master equation of Eq.~\eqref{eq:GeneralMasterEquation} by means of a stochastic potential, which accelerates charged particles in random directions. This unraveling is discussed in Appendix~\ref{APPSec:StochasticPotential}. In addition, this peculiar phenomenon can be used to put some of the most stringent bounds on the parameters of the standard CSL model~\cite{Donadi2021UndergroundTest,Donadi2021NovelCSLBounds,MAJORANA2022WaveFunctionCollapse}.

Let us start by considering a single charged particle with mass $m$ and charge $e$ (the electron charge). Exploiting the unitary unraveling of Eq.~\eqref{eq:UnitaryUnraveling}, we can see how the particle is subjected to the force (see Appendix~\ref{APPSec:LarmorFormulaEmissionRate})
\begin{multline}
	\hbF = (i/\hbar)\comm{\hV(t)}{\hbp} 
	=\\
	= \hbar\sqrt{\gamma_\alpha} \prt{\frac{m}{m_0}}^{\alpha}\int \dd[3]{\bx} [\nabla g_{r_C}^\alpha (\hbq-\bx)]w(\bx,t),
\end{multline}
which accelerates the particle and, therefore, makes it emit radiation by bremsstrahlung. In Appendix~\ref{APPSec:LarmorFormulaEmissionRate}, by employing Larmor's formula\footnote{Even if this calculation is not fully quantum-mechanical, it agrees with the fully quantum-mechanical one, as shown in Refs.~\cite{Adler2007Bounds,Donadi2014RadiationEmission}.}, we derive the radiation emission rate\footnote{This is the number of photons at a given energy, emitted in all directions, per unit time. Notice how, in Eq.~\eqref{eq:EmissionRateSingleFreeParticle}, the term within square parentheses is dimensionless.} $\dd{\Gamma}/\dd{E}$, which turns out to be:
\begin{equations}\label{eq:EmissionRateSingleFreeParticle}
	\dv{\Gamma}{E} 
	&= \prtq{\prt{\frac{m}{m_0}}^{2\alpha-2} \frac{\alpha \hbar e^2}{4 \pi^2 \varepsilon_0 m_0^2 r_C^2 c^3}} \frac{\lambda_\alpha}{E},
	\\
	\lambda_\alpha 
	&= \gamma_\alpha \frac{\prt{\pi r_C^2/\alpha}^{3/2}}{\prt{2 \pi r_C^2}^{3\alpha}},
\end{equations}
where $\varepsilon_0$ is the vacuum permittivity and $E$ is the energy of the emitted radiation\footnote{Eq.~\eqref{eq:EmissionRateSingleFreeParticle} contains an apparent infrared divergence which, citing Ref.~\cite{Fu1997SpontaneousRadiation}, can be treated in a standard way. It poses no physical consequences as the total power is given by $P(t) = \int \dd{E} E [\dd{\Gamma}/\dd{E}]$. On the other hand, this integration leads to an ultraviolet divergence. This has to be ignored because the emission of high energy photons could only be treated within a relativistic spontaneous collapse theory.}. 
We notice that the radiation emission is independent of the particle mass only for $\alpha=1$. In fact, when $\alpha=1$, the stochastic force is proportional to the mass thus giving an acceleration that is independent of it. For $\alpha>1$ we have the peculiar prediction that by increasing the mass of a particle we also increase its spontaneous radiation. On the other hand, for $\alpha<1$ we have that the radiation increases by decreasing the mass of the particles.

Eq.~\eqref{eq:EmissionRateSingleFreeParticle} has been derived under the assumption that the radiating particle is isolated. Intuitively, this holds approximately true as long as there are no other masses within a distance $r_C$ from the particle under consideration. If there are other masses within this radius, the collapse operator changes and, therefore, also the stochastic force on the particle. In particular, the force acting on the particle becomes that of Eq.~\eqref{eq:ForceNotIsolatedSingleParticle}, where we see that only for $\alpha=1$ the force remains the same. For $\alpha>1$, every additional particle within the radius $r_C$ increases the force felt by the particle under consideration [cf.~\eqref{eq:SmearedMassDensityOperator}]. For $\alpha<1$, the presence of additional particle decreases the force exerted on the particle under consideration, thus decreasing the amount of emitted radiation.

Now let us consider the radiation from a body, analyzing the case in which this body's dimensions are much larger than $r_C$. One could be tempted to make the rigid body assumption, compute the acceleration of the center of mass due to spontaneous collapses and use that acceleration to compute the emission. This would be wrong, as shown in Appendix~\ref{APPSec:LargeBodyEmissionRateSemiclassicalDerivation}. The physical reason is that one needs to consider the actual (but still semiclassical) force exerted on each particle individually, as that is the force causing the acceleration that makes the particles emit photons. This stochastic force depends, loosely speaking, on the amount of mass in a ball of radius $r_C$ around the particle. Therefore, one qualitatively expects to substitute the mass in the above formulas with the density of the body. Another effect to consider is whether the particles emit coherently or incoherently. If the wavelength $\lambda_{\rm ph}$ of the emitted photons is much larger than the distance $d_p$ between the particles and $d_p \ll r_C$ then this lump of particles behaves as a single particle of charge $Q$, where $Q$ is its total charge, because of the compoundation property. In the regimes $d_p \gg \lambda_{\rm ph}$ or $d_p \gg r_C$ we instead expect an incoherent emission. In the former case due to the fact that the electromagnetic wave can resolve the individual position of the emitters and in the latter to the fact that their accelerations will be uncorrelated. In our case, we are interested in X-rays experiments, which offer the most useful data, so that we can consider the protons in nuclei to emit coherently but the electrons to emit incoherently. By applying the above considerations to a body of constant density made of $N_P$ atoms, we get (detailed calculations in Appendix~\ref{APPSec:LargeBodyEmissionRateSemiclassicalDerivation}):
\begin{equation}\label{eq:RadiationSmallRC}
	\dv{\Gamma}{E}
	\!=\!\!
	\prtq{\alpha^{7/2}\! \prt{\frac{\mu_0 \prt{\sqrt{2\pi}r_C}^3}{m_0}}^{\!\! 2\alpha-2} 
		\!\!
		\frac{\hbar N_P\prtq{Q_N^2 + N_e e^2}}{4 \pi^2 \varepsilon_0 m_0^2 r_C^2 c^3}}
	\!\!
	\frac{\lambda_\alpha}{E},
\end{equation}
where $\varepsilon_0$ is the vacuum permittivity, $\mu_0$ is the constant density of the emitting body, $Q_N = Z e$ is the nucleus charge and $N_e$ the number of electrons (in each atom) participating in the radiative process. If, instead, we consider $r_C$ to be much larger than the emitting body's dimensions, the stochastic force exerted by the spontaneous collapses will be the same for all the atoms in the body. Thus, we just have to use the single particle formula [Eq.~\eqref{eq:EmissionRateSingleFreeParticle}] for the emission rate but with charge $N_P\prtq{Q_N^2 + N_e e^2}$:
\begin{equation}\label{eq:RadiationLargeRC}
	\dv{\Gamma}{E} 
	= \prtq{\alpha \prt{\frac{M}{m_0}}^{2\alpha-2} \frac{\hbar N_P\prtq{Q_N^2 + N_e e^2}}{4 \pi^2 \varepsilon_0 m_0^2 r_C^2 c^3}} \frac{\lambda_\alpha}{E},
\end{equation}
where $M$ is the total mass of the body. For $\alpha=1$, Eqs.~\eqref{eq:RadiationSmallRC} and~\eqref{eq:RadiationLargeRC} are equivalent and correspond to Eq.~(5) of Ref.~\cite{Donadi2021NovelCSLBounds}. This is again a consequence of the stochastic force being proportional to the mass. Notice, moreover, that the above equation is derived under the assumption that there are no additional masses within a radius $r_C$ of the body. If this is not the case, the same considerations expressed under Eq.~\eqref{eq:ForceNotIsolatedSingleParticle} hold: additional masses would increase the amount of emitted radiation for $\alpha>1$ and decrease it for $\alpha<1$.

As a final comment, let us notice how Eq.~\eqref{eq:RadiationSmallRC} on the one hand and Eqs.~\eqref{eq:EmissionRateSingleFreeParticle} and~\eqref{eq:RadiationLargeRC} on the other show us a different behavior with respect to the collapse radius $r_C$. In the single particle case [Eq.~\eqref{eq:EmissionRateSingleFreeParticle}] and when $r_C$ is larger than the emitting body [Eq.~\eqref{eq:RadiationLargeRC}], increasing $r_C$ lowers the spontaneously emitted radiation, as expected from the fact that $r_C \rightarrow \infty$ corresponds to standard quantum mechanics. However, for $r_C$ smaller than the emitting body's dimensions [Eq.~\eqref{eq:RadiationSmallRC}], this behavior changes whether $\alpha <4/3$ or $\alpha>4/3$. For $\alpha>4/3$ we have the counterintuitive effect that by increasing $r_C$ we also increase the amount of emitted radiation. On the contrary, for both the PSL ($\alpha=1/2$) and CSL ($\alpha=1$) models, increasing $r_C$ decreases the amount of emitted radiation in both regimes (small and large $r_C$).

\section{Theoretical and experimental bounds on the generalized models\label{Sec:ExperimentalBounds}}

Given the results of the previous section, we now discuss the lower and upper bounds on the parameters $\lambda_\alpha$ and $r_C$ of the generalized CSL and PSL models. In Secs.~\ref{subsec:LocalizationThinDisk}, \ref{subsec:UpperBoundsRadiation}, \ref{subsec:UpperBoundsLigoLisa}, we explain how these bounds are obtained and refer the reader to the appendices for the detailed calculations. Then, in Sec.~\ref{subsec:ExclusionPlots}, we compare the models with different values of $\alpha$.

\subsection{Theoretical lower bounds based on localization of a thin graphene disk\label{subsec:LocalizationThinDisk}}

Spontaneous collapse models aim at explaining the emergence of the macroscopic world. This implies that the impact of the collapses on the standard unitary dynamics cannot be arbitrarily small. So, there have to be lower bounds dictated by this necessity. Here, we follow the requirement proposed in Ref.~\cite{Toros2018BoundsCalculations}, even if other choices are possible~\cite{Adler2007Bounds,Bassi2010HumanPerception,Feldmann2012ParameterDiagramsCSL}. This bound, while somewhat debatable, is widely used in the literature and provides a reasonable estimate of collapse strength necessary to prevent macroscopic superpositions.

For the standard CSL model, the theoretical lower bound is obtained by requiring that a single-layered graphene disk of radius $r_D = 10^{-5} \rm{m}$ is localized within $\tau_D = 10^{-2} \rm{s}$ when starting in a spatial superposition of two locations separated by the same distance $r_D$. These values of $r_D$ and $\tau_D$ are chosen as an estimation of the spatial and time resolution of the human eye~\cite{Toros2018BoundsCalculations}. The graphene disk is made of atoms occupying an area $\pi r_a^2$ where $r_a = 10^{-10} \rm{m}$ and each of them has mass $m_a = 12 m_0$. Assuming they fill the space of the disk, the disk contains $n_a = (r_D/r_a)^2 = 10^{10}$ atoms. In Appendix~\ref{APPSec:ThinDiskLocalization}, we obtain a generalization of Adler's formula~\cite{Toros2018BoundsCalculations}, first obtained in Ref.~\cite{Adler2007Bounds}. 
This formula quantifies the decoherence rate of the graphene disk's center of mass.
Its generalized version is
\begin{equations}\label{eq:AdlerEffectiveDecoherenceRate}
	\Gamma^{\rm{CM}}_\alpha (\bD) 
	&\simeq
	\Lambda_\alpha \prtq{1-e^{-\alpha \bD^2/(4 r_C^2)}},
	\\
	\Lambda_\alpha &= \frac{n_a}{n(r_C)} \prt{\frac{m_a n(r_C)}{m_0}}^{2\alpha} \lambda_\alpha,
	\\
	&n(r_C) = 
	\begin{cases}
		1, &\quad r_C < r_a,\\
		(r_C/r_a)^2, &\quad r_a \leq r_C \leq r_D,\\
		n_a, &\quad r_D < r_C,
	\end{cases}
\end{equations}
where $\bD$ is the distance between the two positions of the center of mass on the plane of the graphene disk. The formula is not exact, but a more sophisticated analysis gives the same qualitative results. Since the theoretical lower bounds already contain a noticeable degree of arbitrariness, the error of the generalized Adler's formula is completely negligible (see Appendix~\ref{APPSec:ThinDiskLocalization} for more details).
It is worthwhile noticing that Eq.~\eqref{eq:AdlerEffectiveDecoherenceRate} is different from Eq.~\eqref{eq:RigidBodyDecoherenceRateGeneralFormula}, which was derived under the assumption of a rigid body whose extension is much larger than $r_C$ in all directions. Here, the graphene disk is instead considered to be basically two-dimensional, hence the difference. A particular case of interest is obtained for $\alpha=1/2$, which gives $\Lambda_{1/2} = (M/m_0) \lambda_{1/2}$ independently of $r_C$, where $M$ is the total mass of the graphene disk. In other words, the decoherence rate is exactly the same of a single particle of mass $M$ [cf. Eq~\eqref{eq:SingleParticleDecoherence}]. 

The theoretical lower bound can now be found by requiring that $\Gamma^{\rm{CM}}_\alpha (r_D) \geq 1/\tau_D = 10^{2} \rm{Hz}$:
\begin{equation}\label{eq:MinimumValueLocalizationRate}
	\frac{1}{\lambda_{\alpha} (r_C)} \leq \tau_D \frac{n_a}{n(r_C)} \prt{\frac{m_a n(r_C)}{m_0}}^{2\alpha}  \prt{1-e^{-(\alpha/4) r_D^2/r_C^2}}.
\end{equation}
Notice that this bound has been derived assuming that, for any $r_C$, there are no other masses nearby the graphene disk in that radius. This makes no difference as long as $r_C \ll r_D$ but it can drastically change the decoherence rate for $r_C \gg r_D$ since, in that regime, surrounding masses contribute significantly to the local collapse rate when $\alpha \neq 1$ [see the discussion below Eq.~\eqref{eq:ForceNotIsolatedSingleParticle}]. When $\alpha>1$, the presence of other masses increases the decoherence while for $\alpha<1$ it decreases it. Therefore, the theoretical lower bound obtained in Sec.~\ref{subsec:ExclusionPlots} is valid in empty space and could actually be much different when considering more realistic situations.

\subsection{Upper bounds from spontaneous radiation\label{subsec:UpperBoundsRadiation}}

One of the strongest bounds on the standard CSL noise comes from the spontaneous radiation of free or quasi-free particles~\cite{Donadi2021UndergroundTest,Donadi2021NovelCSLBounds,MAJORANA2022WaveFunctionCollapse}, the strongest being the one presented in Ref.~\cite{MAJORANA2022WaveFunctionCollapse}. 

In the experiment of Ref.~\cite{MAJORANA2022WaveFunctionCollapse}, the radiation from 44.1 kg of Germanium is measured in the range $[19-100\ {\rm keV}]$. The atomic radius of Germanium is\footnote{See \url{https://en.wikipedia.org/wiki/Germanium}. We point out that, in the experiment of Ref.~\cite{MAJORANA2022WaveFunctionCollapse}, a large part of the Germanium is enriched. Therefore, its atomic weight and other properties are slightly different than those used here for our estimations. These differences are hardly appreciable at the level of resolution of the exclusion plots of Sec.~\ref{subsec:ExclusionPlots}. Therefore, this discrepancy does not change the conclusions that we draw at the end of this section.} $1.22 \times 10^{-10} {\rm m}$, its lattice constant is $5.66 \times 10^{-10} {\rm m}$, and its mass density is $\mu_0 = 5327\ {\rm kg}\ {\rm m}^{-3}$. 
For ease of notation, let use introduce $[\dd{\Gamma}/\dd{E}]\vert_{\alpha}$ and $[\dd{\Gamma}/\dd{E}]\vert_{\rm Exp}$, which are, respectively, the theoretical prediction on spontaneous radiation and the (total) measured radiation. The experimental upper bound is obtained by imposing $[\dd{\Gamma}/\dd{E}]\vert_{\alpha} \leq [\dd{\Gamma}/\dd{E}]\vert_{\rm Exp}$ because if this were not true, the predicted spontaneous radiation should be higher than the detected one, thus experimentally refuting the model. If some of the sources of the measured radiation are characterized, their radiation can be subtracted from $[\dd{\Gamma}/\dd{E}]\vert_{\rm Exp}$ in order to attain even stronger bounds on the collapse parameters. 

An analysis of the kind above is conducted in Ref.~\cite{MAJORANA2022WaveFunctionCollapse}. Since we are interested in bounding $\lambda_\alpha$ as a function of $r_C$ it is convenient to write $[\dd{\Gamma}/\dd{E}]\vert_{\alpha}=\lambda_\alpha K(\alpha)$. 
The result of Ref.~\cite{MAJORANA2022WaveFunctionCollapse} is that $\lambda_1 \leq r_C^2 \times 4.79 \times 10^{-1} {\rm s}^{-1} {\rm m}^{-2}$, which we use to get \begin{multline}\label{eq:UpperBoundRadiation_Part1}
	\lambda_\alpha 
	\leq \frac{\dd{\Gamma}/\dd{E}\vert_{\rm Exp}}{K(\alpha)}
	= \frac{\dd{\Gamma}/\dd{E}\vert_{\rm Exp}}{K(1)}\frac{K(1)}{K(\alpha)}
	=\\
	= r_C^2 \frac{K(1)}{K(\alpha)}\times 4.79 \times 10^{-1} {\rm s}^{-1} {\rm m}^{-2},
\end{multline}
where, denoting by $D$ the typical dimensions of the emitting body, [cf. Eqs.~\eqref{eq:RadiationSmallRC} and~\eqref{eq:RadiationLargeRC}]
\begin{equations}\label{eq:UpperBoundRadiation_Part2}
	\rvt{\frac{K(1)}{K(\alpha)}}_{r_C \ll D} &= \alpha^{-7/2}\prtq{ \frac{\mu_0 \prt{\sqrt{2\pi}r_C}^3}{m_0} }^{2-2\alpha},
	\\
	\rvt{\frac{K(1)}{K(\alpha)}}_{r_C \gg D} &= \alpha^{-1}\prt{\frac{M}{m_0}}^{2-2\alpha}.
\end{equations}
These formulae allow to compute the upper bounds, despite not having a formula for the regime $r_C \sim D$. This is achieved by analyzing where the bounds coming from the regimes of small and large $r_C$ intersect and by joining the two curves (see Sec.~\ref{subsec:ExclusionPlots}). We also point out that the above formula for the regime $r_C \ll D$ is valid for $r_C \gg 5.66 \times 10^{-10} \rm{m}$, the lattice constant of the Germanium crystal. This is not a problem because $r_C \leq 10^{-9} \rm{m}$ is not allowed (see Sec.~\ref{subsec:ExclusionPlots}).

In Eq.~\eqref{eq:UpperBoundRadiation_Part2}, for $r_C \gg D$, we used $M = 44.1\ \rm{kg}$. This value does not keep into account the exact geometry of the experiment and, most importantly, does not keep into account the presence of other masses around the Germanium emitters. Following the discussion below Eq.~\eqref{eq:ForceNotIsolatedSingleParticle}, what this implies is that the estimated bounds that we derive are conservative for $\alpha>1$ and stronger than the actual ones for $\alpha<1$. For $\alpha=3/2$ or $\alpha=2$ (the values used for Fig.~\ref{fig:exclusionPlots}), this is not a problem as we will see that already with this value of $M$ the conclusion to be drawn is that the models with $\alpha>1$ have to be discarded. For $\alpha=1/2$, the upper bound due to spontaneous radiation when $r_C \gg D$ is much higher than those coming from gravitational wave detectors (see Sec.~\ref{subsec:UpperBoundsLigoLisa}).

\subsection{Upper bounds from gravitational wave detectors\label{subsec:UpperBoundsLigoLisa}}

The gravitational wave detectors Advanced LIGO and LISA Pathfinder are characterized by an extremely small noise. As such, their calibration can be used to put bounds on spontaneous collapse models~\cite{Carlesso2016ExperimentalBounds}. Here we generalize the analysis performed in Ref.~\cite{Carlesso2016ExperimentalBounds} to case of the generalized CSL and PSL models. 

\begin{table}[]
	\centering
	\resizebox{0.48\textwidth}{!}{%
		\begin{tabular}{l|l|l|}
			\cline{2-3}
			& LIGO            & LISA     \\ \hline
			\multicolumn{1}{|l|}{Mass {[}kg{]}} & $40$            & $1.928$  \\ \hline
			\multicolumn{1}{|l|}{a {[}m{]}}     & $4 \times 10^3$ & $0.376$  \\ \hline
			\multicolumn{1}{|l|}{L {[}m{]}}     & $0.02$          & $0.046$  \\ \hline
			\multicolumn{1}{|l|}{R {[}m{]}}     & $0.017$         & N/A      \\ \hline
			\multicolumn{1}{|l|}{$S_{\rm Exp}$ {[}$\rm{kg}^2 \rm{m}^2 \rm{s}^{-3}${]}} & $9.025 \times 10^{27}$ & $2.5091 \times 10^{-29}$ \\ \hline
		\end{tabular}%
	}
	\caption{This table shows the parameters used in Ref.~\cite{Carlesso2016ExperimentalBounds} to compute the standard CSL bounds. Notice that the experimental force noise spectral density of LIGO has to be divided by $4$ while that of LISA by 2 when comparing them with Eq.~\eqref{eq:ForceNoiseDensity}.}
	\label{tab:GravWaveParameters}
\end{table}

Both LIGO and LISA monitor the distance $a$ between pairs of masses. In the former case, there are perpendicular two arms in the detector and in both of them two cylindrical masses of radius $R$ and length $L$ are separated by a distance $a$ along the arm. In the latter case, there is only one arm and the masses are cubes of side $L$. These quantities are reported in Table~\ref{tab:GravWaveParameters}. 
The derivation of the upper bounds, in this context, is based on the rigid body assumption and on approximations allowed by the fact that the displacement of each particle in the apparatus is small with respect to $r_C$. In particular, one exploits the unitary unraveling of Eq.~\eqref{eq:GeneralMasterEquation} to analyze the stochastic force acting on the centers of mass of the bodies composing the apparatuses in the frequency domain, so to obtain the so-called force noise spectral density $S$\footnote{In Ref.~\cite{Carlesso2016ExperimentalBounds}, this is denoted as $S_{FF} (\omega)$. However, we will see it is actually independent of $\omega$ as we are dealing with a white noise force.} which has to be compared with the experimentally measured one $S_{\rm Exp}$, reported in Table~\ref{tab:GravWaveParameters}\footnote{In Ref.~\cite{Carlesso2016ExperimentalBounds}, the data reported are the square-root noise spectrum for LIGO and the acceleration noise for LISA. In the table, we report the minimum of the force noise spectral density coming from those data.}. The theoretical force noise density, computed in Appendix~\ref{APPSec:LigoLisaCalculations}, is equal to
\begin{equations}\label{eq:ForceNoiseDensity}
	S\vert_{r_C\ll L,R} &= \lambda_\alpha \prtq{\frac{\mu_0 \prt{\sqrt{2\pi}r_C}^3}{m_0}}^{2\alpha} \frac{\alpha^{7/2} G(\alpha) \hbar^2}{2 \pi^{5/2} r_C^4} A_P,
	\\
	S\vert_{r_C\gg L,R} &= \frac{\alpha^{5/2} \hbar^2 \lambda_\alpha}{4\sqrt{\pi} r_C^2}\prt{\frac{M}{m_0}}^{2\alpha} f_S \prt{\frac{a}{r_C},\alpha},
\end{equations}
where $\mu_0$ is the bodies density, $A_P$ the area of their cross-section\footnote{$A_P= L^2$ for the cube and $A_P = \pi R^2$ for the cylinder.}, and $f(x,\alpha)$ is a function reported in Eq.~\eqref{APPeq:FunctionfS} of Appendix~\ref{APPSec:LigoLisaCalculations}.
Contrary to what has been done in Ref.~\cite{Carlesso2016ExperimentalBounds} we did not find expressions of $S$ valid for any value of $r_C$. The method employed there could not be used in this more general case due to the higher mathematical complexity of the models with $\alpha \neq 1$. However, as already discussed for the radiation bounds, we can plot the two curves corresponding to the regimes of small and large $r_C$ and join them where they intersect. The error made in the region where is neither much smaller or much bigger than both $L$ and $R$ does not change the qualitative results of Sec.~\ref{subsec:ExclusionPlots}. As a final note, we mention that the experimental noise of LIGO has to be divided by $4$ because there are two arms in the experiment and the noise is measured with a one sided-spectrum~\cite{Carlesso2016ExperimentalBounds}. Similarly, for LISA, we have to divide the experimental noise by $2$ because of the one-sided spectrum used in measurements~\cite{Carlesso2016ExperimentalBounds}.

\subsection{Comparison of the models and exclusion plots\label{subsec:ExclusionPlots}}

\begin{figure*}
	\centering
	\includegraphics[width=0.9\textwidth]{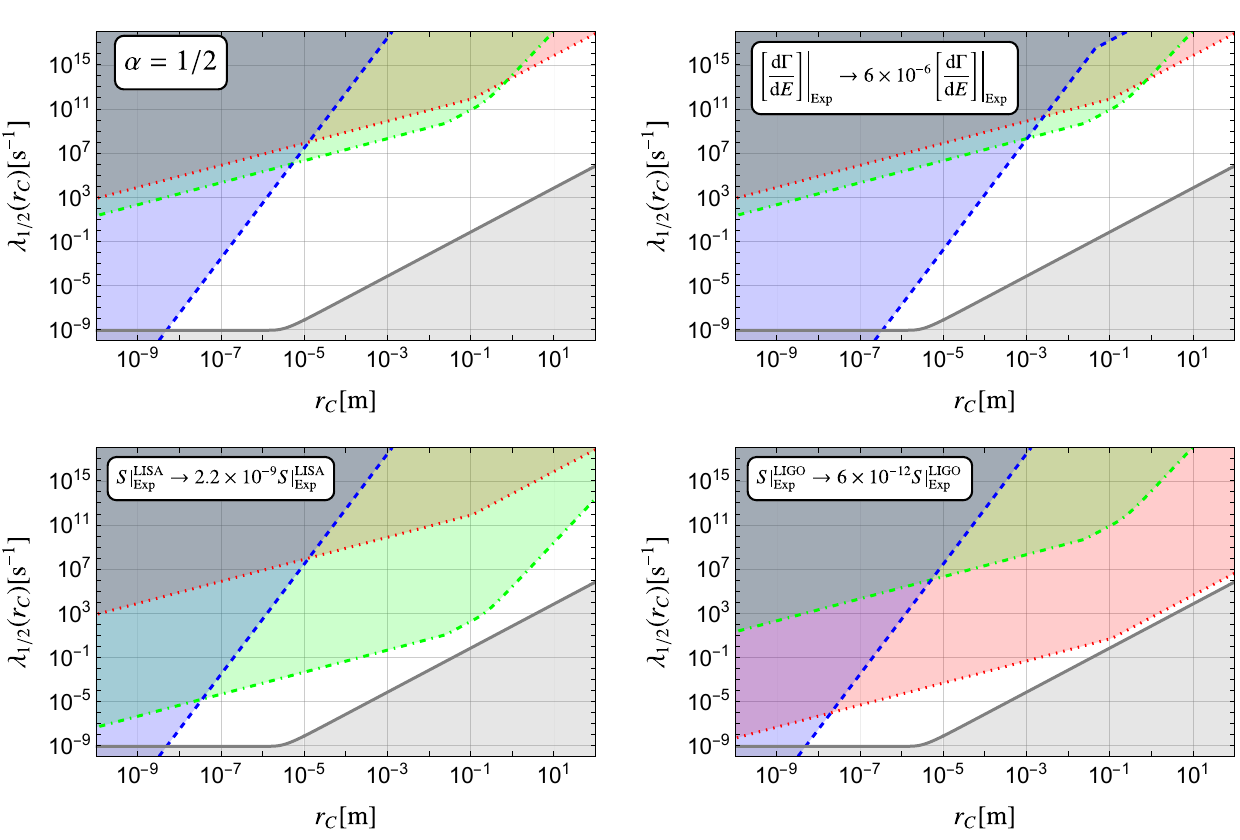}
	\caption{Exclusion plots of the PSL model for different modifications of the experimental data such that the CSL model would be experimentally ruled out. 
		In each plot, the experimentally allowed parameters ($\lambda$,$r_C$) have to reside within the white region lower bounded by the continuous gray line and upper bounded by the other lines. The lower bounds (continuous gray lines) are computed by means of Eq.~\eqref{eq:MinimumValueLocalizationRate}, the radiation bounds (dashed blue lines) by employing Eqs.~\eqref{eq:UpperBoundRadiation_Part1} and~\eqref{eq:UpperBoundRadiation_Part2}, and the LIGO and LISA (respectively the dotted red line and the dot-dashed green line) bounds by employing Eq.~\eqref{eq:ForceNoiseDensity} and Table~\ref{tab:GravWaveParameters}. The radiation bounds are artificially lowered in the top-right plot, LISA's bounds in the bottom-left one, and LIGO's ones in the bottom-right plot.}
	\label{fig:exclusionPlotsComparisonModExpValuesOneHalf}
\end{figure*}

Having collected all the bounds for the experiments considered in Ref.~\cite{Carlesso2016ExperimentalBounds} and Refs.~\cite{Donadi2021NovelCSLBounds,MAJORANA2022WaveFunctionCollapse}, and having computed the theoretical lower bounds following Ref.~\cite{Toros2018BoundsCalculations}, we can compare the different models. The results are presented in Fig.~\ref{fig:exclusionPlots}. 

Taking for granted the theoretical lower bounds, we see that the model with $\alpha=2$ is experimentally excluded; in particular, the radiation bounds alone are sufficient to exclude it. The model with $\alpha=3/2$, instead, is almost excluded but the GRW value $r_C \sim 10^{-7} {\rm m}$ is nevertheless ruled out. For $\alpha=1$, corresponding to the standard CSL model, we have the exclusion plot reported in \cite{Altamura2024NonInterfCSL}. The errors due to using approximate formula with respect to Ref.~\cite{Carlesso2016ExperimentalBounds} have no consequences on the shape and dimensions of the allowed parameter region. Finally, the PSL model ($\alpha=1/2$) turns out to be the most resilient to the experiments considered, showing that no upper bound is currently imposed. Focusing on the region around $r_C \sim 10^{-7} {\rm m}$ we see that the allowed values of $\lambda_{1/2}$ comprise an interval that is different orders of magnitude higher than for $\lambda_1$. Indeed, experiments that we have not considered because they were not providing additional bounds for $\alpha=1$ could instead be the source of significant bounds for $\alpha=1/2$.

As first discussed at the end of Sec.~\ref{subsec:SingleParticleDynamics} and then in Sec.~\ref{subsec:LocalizationThinDisk}, the bounds strongly depend on the assumption of isolated systems when $r_C$ is greater than the typical dimension of the system under consideration and $\alpha \neq 1$. This is so because when $\alpha \neq 1$ the system spontaneous collapse dynamics changes due to the presence of other masses within a length $r_C$. For example, for $\alpha=1/2$ and $r_C \gtrsim 10^{-5}\ \rm{m}$, the thin graphene disk localization would decelerate due to the presence of other (non-interacting) masses; to localize the disk within the required time, the collapse rate $\lambda_{1/2}$ should be higher.
At the same time, in realistic situations it is questionable to assume that the environment (at least through  the small but unavoidable gravitational interactions) does not accelerate the thin disk decoherence.
In conclusion, the bounds displayed in Fig.~\ref{fig:exclusionPlots} are  reliable when the collapse radius is smaller then the dimensions of the object under investigation,  otherwise they are less reliable, when also the object is  not isolated.

Due to the higher falsification resiliency of PSL ($\alpha=1/2$), future experiments could rule out CSL ($\alpha=1$) while still leaving a region of allowed parameters for PSL. In Fig.~\ref{fig:exclusionPlotsComparisonModExpValuesOneHalf}, we show how the exclusion plot of the PSL model changes when we artificially modify the experimental data so that they would exclude the CSL model. We can see that in all three cases considered, a complete refutation of the CSL model would not rule out the PSL model. We stress again, however, that the theoretical lower bounds are quite debatable.

\section{Conclusions\label{Sec:Conclusions}}

We introduced a generalization of the Continuous Spontaneous Localization (CSL) model by allowing the collapse operator to depend on an arbitrary function of the smeared mass density. This extension was partially motivated by the Poissonian Spontaneous Localization (PSL) model, according to which the collapse scales linearly with mass rather than quadratically as it does for CSL. By exploring both structural and phenomenological aspects of the generalized dynamics, we investigated whether different forms of mass dependence are internally consistent and compatible with existing experimental data.

Our analysis showed that two fundamental properties of CSL, namely compoundation invariance and the decoupling of internal and center-of-mass motion, remain intact for a broad class of functional dependencies. This suggests that the CSL framework is structurally more robust than previously thought and can accommodate a wider range of theoretical possibilities.

Specializing to power-law mass dependencies, we derived analytical results for the decoherence rates of single particles and rigid bodies, as well as the rate of spontaneous radiation emission from charged particles. We compared these predictions with current experimental bounds: the analysis revealed that models with $\alpha \gtrsim 3/2$, corresponding to a cubic or higher mass dependence, are severely constrained, while those with weaker scaling (such as PSL and CSL) remain viable across a broader region of parameter space. In particular, PSL ($\alpha=1/2$) appears significantly more robust to experimental falsification than standard CSL ($\alpha=1$).

Overall, our results suggest that models with sub-quadratic mass density scaling deserve greater attention in both theoretical and experimental investigations of spontaneous collapse models. 



\section*{Acknowledgements}
N. P. thanks S. Donadi, M. Carlesso, and D. G. A. Altamura for useful discussions about their works. 
The authors acknowledge support from the PNRR MUR project PE0000023-NQSTI, INFN, and the University of Trieste.
N. P. acknowledges support also from the European Union Horizon’s 2023 research and innovation programme [HORIZON-MSCA-2023-PF-01] under the Marie Sklodowska Curie Grant Agreement No. 101150889 (CPQM).
A. B. acknowledges support also from the EU EIC Pathfinder project QuCoM (Grant Agreement No. 101046973).

\section*{Data availability}

The data that support the finding of this article are openly available at \url{https://github.com/Knomes02/arXiv-2501.17637}.


%

\onecolumngrid
\appendix

\clearpage
\section{Unraveling by means of a Stochastic Potential\label{APPSec:StochasticPotential}}

Here we show that the master equation~\eqref{eq:GeneralMasterEquation} admits a unitary unraveling, a standard result used in the context of collapse models since Ref.~\cite{Fu1997SpontaneousRadiation}.
More in general, let us consider a master equation of the following kind:
\begin{equation}\label{APPeq:SuperGeneralMasterEquationContinuousNoise}
	\dv{t} \rho_t =  
	-\frac{i}{\hbar}\comm{\hat{H}}{\rho_t} 
	- \frac{1}{2} \int \dd[3]{\bx}\dd[3]{\by} \mcD (\bx,\by) \comm{\hL (\bx)}{\comm{\hL (\by)}{\rho_t}},
\end{equation}
which corresponds to Eq.~\eqref{eq:GeneralMasterEquation} when $\mcD (\bx,\by)= \gamma \delta (\bx-\by)$.
The unitary unraveling is given by adding the following stochastic potential to the Schr\"{o}dinger equation:
\begin{equation}
	\hV(t) = \hbar \int \dd[3]{\bx} \hL (\bx) w(\bx,t),
\end{equation}
where $w(\bx,t)$ is a white-in-time noise characterized by 
\begin{gather}
	\mbE\prtq{w(\bx,t)}=0,
	\qquad
	\mbE \prtq{w(\bx,t)w(\by,s)}=\delta\prt{t-s} \mcD(\bx,\by),
\end{gather}
where $\mcD(\bx,\by) = \mcD (\abs{\bx-\by})$.

To prove the validity of the unraveling one starts with a density matrix $\rho_t$ and evolves in time to $t+\delta t$, up to second order in $\delta t$, according to a certain realization of the noise with
\begin{equation}
	U(t+\delta t,t) = \mcT \exp{-\frac{i}{\hbar}\int_t^{t+\delta t} H'(s) \dd{s}}
	\simeq
	1 - \frac{i}{\hbar}\int_t^{t+\delta t} H'(s) \dd{s} -\frac{1}{\hbar^2}\int_t^{t+\delta t}\int_t^{t_1}  H'(t_1)H'(t_2) \dd{t_1}\dd{t_2},
\end{equation}
where $H'(t) = \hat{H} + \hV (t)$.
The average density matrix is obtained by taking the average over the noise of $\rho_{t+\delta t} = U(t+\delta t,t)\rho_t U^\dg(t+\delta t,t)$ and keeping only terms up to first order in $\delta t$. The heuristic argument\footnote{This connected to the idea of writing $w(\bx,t) = \dd{W_t (\bx)}/\dd{t}$, where $\dd{W_t} (\bx)$ is the generalized Wiener increment such that $\dd{W_t} (\bx) \dd{W_t} (\by) = \mcD (\bx,\by) \dd{t}$. This kind of generalized Wiener increment can be obtained by convoluting the usual Wiener increment. See page 22 (492) of Ref.~\cite{Bassi2013Models}.} to do so is to consider $\prtq{w(\bx,t)}\sim \delta t^{-1/2}$ so that, for example, a term like $\int \hat{H} \hV (t_2)\dd{t_1}\dd{t_2}$ and $\int \hV (t_1)\hV(t_2)\hV(t_3) \dd{t_1}\dd{t_2}\dd{t_3}$ are both of higher order and have to be neglected. Then, one must use equalities such as $\mbE \prtq{\int_t^{t+\delta t} \hV(s)\dd{s}}=0$,
\begin{equation}
	\mbE\prtq{\int_t^{t+\delta t}\int_t^{t+\delta t} \dd{t_1} \dd{t_2} \hV(t_1)\rho_t \hV (t_2)}
	= \hbar^2 \delta t \int \dd[3]{\bx}\dd[3]{\by} \mcD(\bx,\by) \hL(\bx)\rho_t \hL(\by),
\end{equation}
and
\begin{multline}
	\mbE\prtq{\int_t^{t+\delta t}\int_t^{t_1} \dd{t_1} \dd{t_2} \hV(t_1)\hV(t_2)\rho_t}
	=\\
	= \hbar^2 \prtq{\int_t^{t+\delta t}\int_t^{t_1} \delta(t_1-t_2)\dd{t_1}\dd{t_2}} \int \dd[3]{\bx}\dd[3]{\by} \mcD(\bx,\by)\hL(\bx)\hL (\by)\rho_t
	= \frac{\delta t}{2}\hbar^2 \int \dd[3]{\bx}\dd[3]{\by} \mcD(\bx,\by)\hL(\bx)\hL (\by)\rho_t.
\end{multline}
The master equation is finally obtained by writing $\dot{\rho}_t = \lim_{\delta t \rightarrow 0} (\rho_{t+\delta t} - \rho_t)/\delta t$.

Since we provided a unitary unraveling with a white noise potential, one may wonder if the same can be obtained with a Poissonian potential. We did not find any such unraveling. However, by slightly generalizing the Poisson noise, we can find one which is unitary albeit non-linear in the case when $\mcD (\bx,\by)= \gamma \delta (\bx-\by)$. We use a Poisson-like function $N(x,t)$ which can take positive and negative values, i.e., it can assume only the values $-1$, $0$, and $+1$ with $\pm 1$ being equiprobable and their probability infinitesimal. So we have $N^2 (\bx,t)=\abs{N(\bx,t)}$, $\mbE\prtq{N(\bx,t)}=0$, $\mbE\prtq{\abs{N(\bx,t)}}= \ev{\hL^2 (\bx)}$, and $\mbE \prtq{N(\bx,t)N(\by,s)}=\ev{\hL^2 (\bx)}\delta(t-s)\delta^{(3)}(\bx-\by)$.
We can now define our stochastic Hamiltonian as $H'(t) = H - \hbar \sqrt{\gamma}\int \dd[3]{\bx}\hL(\bx)\ev{\hL^2 (\bx)}^{-1/2} N(\bx,t)$. The same procedure as before leads to the correct master equation with $\mcD (\bx,\by)= \gamma \delta (\bx-\by)$.

\clearpage
\section{Calculation of momentum spread for a general rigid body\label{APPSec:MomentumSpreadGeneralCase}}

We compute the quantity $C_j = m_0^{-2\alpha}\int \dd[3]{\bx} \prtq{\partial_j \mcM_{\rm CM}^\alpha(\bx)}^2$ appearing in Eqs.~\eqref{eq:RigidBodyObservablesDynamics} and~\eqref{eq:GeneralFormulaCoefficientsRigidBody}. To do this, we use Eq.~\eqref{eq:ConstantDensitySmearedMassDensityOperator}. Then, a simple substitution gives
\begin{equation}
	C_j 
	= m_0^{-2\alpha}\int \dd[3]{\bx} \prtq{\partial_j \mcM_{\rm CM}^\alpha (\bx)}^2
	= \alpha^2 \prt{\frac{\mu_0}{m_0}}^{2 \alpha}
	\int \Phi^{2\alpha-2} \prt{\frac{-d(\bx)}{r_C}} \frac{e^{-d^2(\bx)/r_C^2}}{2\pi r_C^2} \prtq{\partial_j d(\bx)}^2 \dd[3]{\bx}.
\end{equation}

To proceed with the calculation we have to consider that all non-negligible contributions to the above integral come from spatial points near the surface.
In particular, we can evaluate $\partial_j d(\bx)$ by considering that, for all relevant points (like those depicted in Fig.~\ref{fig:SmearedDensityOperatorConstantDensityExplanation}), a small displacement $\bv$ changes the distance from the surface by $\bn(\bx) \cdot \bv$, where $\bn(\bx)$ is the unit normal vector to the surface $\partial V$ on the closest point to $\bx$. Then, one gets that $\partial_j d(\bx) = \bn(\bx) \cdot \be_j = n_j (\bx)$, where $\be_j$ is the unit vector in the $j$-th direction. Let us now choose a point on the surface and a local orthogonal coordinate system such that two coordinates vary on the locally flat surface and the third one, which we denote by $z_p$, varies perpendicularly to it. With these coordinates, we get that $\partial_{z_p} \bn (\bx) = 0$ and $d(\bx)=z_p$ so that we can integrate along the $z_p$ coordinate\footnote{In the integration, changing $\Phi(-d)$ with $\Phi(d)$ does not change the result.}:
\begin{equation}
	\intmp \Phi^{2\alpha-2} \prt{\frac{z_p}{r_C}} \frac{e^{-z_p^2/r_C^2}}{2\pi r_C^2} \dd{z_p}
	=
	\frac{1}{2 \pi r_C} \intmp \Phi^{2\alpha-2} \prt{z} e^{-z^2} \dd{z}
	=
	\frac{G(\alpha)}{2 \pi r_C}.
\end{equation}
The function $G(\alpha)$ has to be computed numerically. A plot of it is given in Fig.~\ref{APPfig:funcGplot}.
The remaining integration to do is a surface integral of the scalar function $n^2_j (\bx)$, i.e.,
\begin{equation}
	C_j 
	= \alpha^2 \prt{\frac{M}{m_0}\frac{1}{V}}^{2 \alpha}
	\frac{G(\alpha)}{2 \pi r_C} \int_{\partial V} n^2_j (\bx) \dd{S}.
\end{equation}
For a parallelepiped, for example, we have that $n^2_3 (\bx) = 1$ on the faces perpendicular to the third direction and zero otherwise. The integral then gives $\int_{\partial V} n^2_j (\bx) \dd{S} = 2 L_1 L_2$, thus giving back the result obtained in Ref.~\cite{Ghirardi1990_CSL,Bassi2003Dynamical}.

\clearpage
\section{Semiclassical derivation of the emission rates from Larmor's formula\label{APPSec:LarmorFormulaEmissionRate}}

We compute the emission rate of a free particle by means of a semiclassical derivation based on Larmor's formula. This approach can be found in Refs.~\cite{Adler2007Bounds,Donadi2014RadiationEmission,Donadi2021NovelCSLBounds}.

The power of radiation emitted by a point particle with charge $q$ is
\begin{equations}\label{APPeq:LarmorFormula}
	P(t) 
	&= \frac{q^2}{6 \pi \varepsilon_0 c^3} a^2 (t)
	= \frac{q^2}{6 \pi \varepsilon_0 c^3} \sum_{j=1}^{3} \frac{1}{2\pi} \int \dd{\omega}\dd{\nu} e^{-i(\nu+\omega)t} \tl{a}_j (\nu) \tl{a}_j (\omega),
	\\
	P(t) &= \int_0^{\infty} \dd{\omega} \hbar \omega \dv{\Gamma}{\omega}, 
\end{equations}
where the first equation is Larmor's formula and $\tl{a}_j (\omega)$ is the Fourier transform of $a_j (t)$. The second equation is the integration over all frequencies of the rate of photons emitted at a given frequency times their energy. 
According to our unitary stochastic unraveling [see Appendix~\ref{APPSec:StochasticPotential}], the free particle is subjected to the following acceleration:
\begin{multline}
	a_j(t) 
	= \frac{F_j(t)}{m} 
	= \frac{i}{\hbar m}\comm{\hV(t)}{\hp_j}
	= -\frac{i\sqrt{\gamma_\alpha}}{m} \prt{\frac{m}{m_0}}^{\alpha}\int \dd[3]{\bx} \comm{g_{r_C}^\alpha (\hbq-\bx)}{\hp_j} w(\bx,t)
	=\\= \frac{\hbar\sqrt{\gamma_\alpha}}{m} \prt{\frac{m}{m_0}}^{\alpha}\int \dd[3]{\bx} [\partial_j g_{r_C}^\alpha (\hbq-\bx)]w(\bx,t)
	= \frac{\hbar\sqrt{\gamma_\alpha}}{m} \frac{1}{\sqrt{2\pi}}\prt{\frac{m}{m_0}}^{\alpha}\int \dd[3]{\bx} [\partial_j g_{r_C}^\alpha (\hbq-\bx)] \int \dd{\omega}\tl{w}(\bx,\omega) e^{-i \omega t}.
\end{multline}
Thus, we have
\begin{equation}
	\tl{a}_j (\omega) = \frac{\hbar\sqrt{\gamma_\alpha}}{m} \prt{\frac{m}{m_0}}^{\alpha}\int \dd[3]{\bx} [\partial_j g_{r_C}^\alpha (\hbq-\bx)] \tl{w}(\bx,\omega).
\end{equation}
We can now compute the average power in frequency. Using 
\begin{equation}
	\mbE\prtq{w(\bx,\omega)w(\by,\omega')} = \delta(\omega+\omega')\delta^{(3)}(\bx-\by),
\end{equation}
we get
\begin{multline}
	\mbE[P(t)] =\\
	=\frac{q^2 \hbar^2 \gamma_\alpha}{12 \pi^2 \varepsilon_0 c^3 m^2} \prt{\frac{m}{m_0}}^{2\alpha} \sum_{j=1}^{3} \int \dd[3]{\bx}\dd[3]{\by} \dd{\omega}\dd{\nu} 
	\prtgB{
		\ev{[\partial_j g_{r_C}^\alpha (\hbq-\bx)] [\partial_j g_{r_C}^\alpha (\hbq-\by)]} \mbE[\tl{w}(\bx,\nu)\tl{w}(\bx,\omega)]e^{-i(\nu+\omega)t}}
	=\\=
	\frac{q^2 \hbar^2 \gamma_\alpha}{12 \pi^2 \varepsilon_0 c^3 m^2} \prt{\frac{m}{m_0}}^{2\alpha} \sum_{j=1}^{3} \int \dd[3]{\bx}\dd{\omega} [\partial_j g_{r_C}^\alpha (\bx)]^2
	=
	\intmp \dd{\omega} \prtg{\frac{q^2 \hbar^2 \gamma_\alpha}{12 \pi^2 \varepsilon_0 c^3 m^2} \prt{\frac{m}{m_0}}^{2\alpha} \sum_{j=1}^{3} \int \dd[3]{\bx} [\partial_j g_{r_C}^\alpha (\bx)]^2}.
\end{multline}
So, by comparison with Eq.~\eqref{APPeq:LarmorFormula} we get that
\begin{equation}
	\dv{\Gamma (t)}{\omega} = \frac{q^2 \hbar \gamma_\alpha}{6 \pi^2 \varepsilon_0 c^3 m^2 \omega} \prt{\frac{m}{m_0}}^{2\alpha} \sum_{j=1}^{3} \int \dd[3]{\bx} [\partial_j g_{r_C}^\alpha (\bx)]^2.
\end{equation}
For a point particle we can analytically solve the integrals and we get
\begin{equation}
	\int \dd[3]{\bx} [\partial_j g_{r_C}^\alpha (\bx)]^2
	=
	\frac{\pi^{3/2} r_C}{2 \sqrt{\alpha}} (2 \pi r_C^2)^{-3\alpha}
	\implies
	\dv{\Gamma}{\omega} = \frac{\alpha \lambda_\alpha q^2 \hbar}{4 \pi^2 \varepsilon_0 c^3 m^2 r_C^2 \omega} \prt{\frac{m}{m_0}}^{2\alpha},
\end{equation}
which is Eq.~\eqref{eq:EmissionRateSingleFreeParticle}.

\clearpage
\section{Semiclassical derivation of the emission rate of a large body\label{APPSec:LargeBodyEmissionRateSemiclassicalDerivation}}

In this Appendix, we derive Eq.~\eqref{eq:RadiationSmallRC} and Eq.~\eqref{eq:RadiationLargeRC}. First, we do it by considering the emission of each particle independently and using heuristic arguments on whether the emissions sum coherently or not. Then, we follow Ref.~\cite{Donadi2021NovelCSLBounds} and make a more sophisticated calculation. Finally, we show how using blindly the rigid body assumption leads to an incorrect result.

\subsection{Derivation with heuristic arguments}

Starting from the actual formula for $\mcM(\bx)$ we get that the acceleration of a single particle is
\begin{equation}
	\mcM(\bx) = \sum_k m_k g_{r_C}(\hbq_k-\bx),
	\implies
	\tl{a}_j (\omega) 
	= \frac{F_j (\omega)}{m}
	= \alpha \frac{\hbar \sqrt{\gamma}}{m_0^\alpha} \int \dd[3]{\br} \mcM^{\alpha-1} (\br) [\partial_j g_{r_C}(\br)] \tl{w}(\br,\omega),
\end{equation}
where $m$ is the particle of the mass under consideration. 
This expression does not depend on the type of particle because we are assuming that the nuclei and the electrons occupy basically the same place on a scale of $r_C$ and almost all emitting particles are well-within the body. When $r_C$ is much smaller than the body's dimensions, we can employ the macroscopic density approximation and the sharp scanning approximation to get
\begin{equation}
	\tl{a}_j (\omega) = \alpha \frac{\hbar \sqrt{\gamma_\alpha}}{m_0^\alpha} \mu_0^{\alpha-1}\int \dd[3]{\br} [\partial_j g_{r_C}(\br)] \tl{w}(\br,\omega),
\end{equation}
which is the same acceleration as that of a single particle with the replacements $m \rightarrow m_0^\alpha/\alpha$,  $(m/m_0)^\alpha\rightarrow \mu_0^{\alpha-1}$, and $g_{r_C}^\alpha \rightarrow g_{r_C}$. Then, summing the charges according the whether the emission is expected to be coherent or incoherent according to the discussion in Sec.~\ref{subsec:SpontaneousRadiation}, we can immediately write down the final result:
\begin{multline}\label{APPeq:SpontaneousRadiationSmallRCHeuristicDerivation}
	\dv{\Gamma (t)}{\omega} 
	= N_P\prtq{Q_N^2 + N_e e^2}\frac{\alpha^2 \hbar \gamma_\alpha}{6 \pi^2 \varepsilon_0 c^3 m_0^{2\alpha} \omega} \mu_0^{2\alpha-2} \sum_{j=1}^{3} \int \dd[3]{\bx} [\partial_j g (\bx)]^2
	=\\= 
	N_P\prtq{Q_N^2 + N_e e^2}\frac{\alpha^2 \hbar \mu_0^{2\alpha-2} \gamma_\alpha}{32 \pi^{7/2} \varepsilon_0 c^3 m_0^{2\alpha} r_C^5 \omega}
	= N_P\prtq{Q_N^2 + N_e e^2}\frac{\alpha^{7/2} \lambda_\alpha \hbar (2\pi r_C^2)^{3\alpha} \mu_0^{2\alpha-2}}{32 \pi^{5} \varepsilon_0 c^3 m_0^{2\alpha} r_C^8 \omega}.
\end{multline}

The formula for $r_C$ much larger than the body's dimensions is simply the formula for a single emitting particle with $N_P\prtq{Q_N^2 + N_e e^2}$ in place of $q^2$.

\subsection{More rigorous calculation}

Here we adapt the calculation presented in the Appendix of Ref.~\cite{Donadi2021NovelCSLBounds} to our generalized CSL models. The first part of the appendix is devoted to finding approximate formulas for the power of the emitted radiation at long distances from the source. This is independent of the spontaneous collapse model we use and will be our starting point. We have that\footnote{Contrary to the rest of this work, for ease of comparison with Ref.~\cite{Donadi2021NovelCSLBounds}, we will use their conventions for Fourier transforms.}
\begin{equation}
	P(t) = \frac{1}{64 \pi^4 \varepsilon_0 c^3} \intmp \dd{\omega} \intmp \dd{\nu} e^{i(\omega+\nu)(t-R_{sp}/c)} \sum_{i,j} q_i q_j J_{i,j} (\omega,\nu),
\end{equation}
where $R_{sp}$ is the radius of the sphere over which we measure the emitted radiation, $\bn$ is the unit normal vector on the sphere surface and pointing outward, and
\begin{equation}
	J_{i,j} (\omega,\nu) =
	4 \pi \prtq{ \ddot{\br}_i (\omega) \cdot \ddot{\br}_j (\nu) \frac{(b^2-1)\sin(b) + b\cos(b)}{b^3} - \ddot{\br}^z_i (\omega) \ddot{\br}^z_j (\nu) \frac{(b^2-3)\sin(b) + 3b\cos(b)}{b^3}},
\end{equation}
where $b = \abs{\omega \ev{\br_i} + \nu \ev{\br_j}}/c$, with $\br_j$ the position of the $j$-th particle, and we are using the convention that upper indices denote the spatial direction. To get to the above formulas, one has to heavily rely on the idea that $R_{sp}$ and $\abs{\br}$ are both much larger than $\ev{\br_k}$, so that we can use almost everywhere this average in place of $\br_k$, the actual position of the particle. This average does not change in time.

The acceleration of the particles in frequency domain is
\begin{equation}
	\ddot{\br}^{j}_k (\omega) = \alpha \frac{\hbar \sqrt{\gamma}}{m_0^\alpha} \int \dd[3]{\br} \mcM^{\alpha-1} (\br) [\partial^j g_{r_C}(\hbq_k - \br)] \tl{w}(\br,\omega),
	\qq{where}
	\tl{w}(\br,\omega) = \int \dd{t} e^{-i \omega t} w(\br,t).
\end{equation}
With the conventions we are using, we have $\mbE[\tl{w}(\br,\omega)\tl{w}(\br',\nu)] = 2\pi \delta^{(3)}(\br-\br')\delta(\omega+\nu)$. Again we exploit $\abs{\br}\gg \ev{\hbq_k}$ to make the substitution $\hbq_k \rightarrow \ev{\br_k}$ both in $g_{r_C}(\bx)$ and $\mcM(\bx)$.
So now we can compute the averages
\begin{equations}
	\mbE[\ddot{\br}_{k} (\omega) \cdot \ddot{\br}_{k'} (\nu)]
	&=
	2\pi \alpha^2  \frac{\hbar^2 \gamma}{m_0^{2\alpha}} \sum_j \int \dd[3]{\br} \mcM^{2\alpha-2} (\br) [\partial^j g_{r_C}(\ev{\br_k} - \br)][\partial^j g_{r_C}(\ev{\br_{k'}} - \br)] \delta(\omega + \nu),
	\\
	\mbE[\ddot{\br}_{k}^z (\omega) \cdot \ddot{\br}_{k'}^z (\nu)]
	&=
	2\pi \alpha^2  \frac{\hbar^2 \gamma}{m_0^{2\alpha}} \int \dd[3]{\br} \mcM^{2\alpha-2} (\br) [\partial^z g_{r_C}(\ev{\br_k} - \br)][\partial^z g_{r_C}(\ev{\br_{k'}} - \br)] \delta(\omega + \nu).
\end{equations}
We can assume that, for the vast majority of particles within the emitting body, all directions look the same so that we can write $\mbE[\ddot{\br}_{k}^j (\omega) \cdot \ddot{\br}_{k'}^j (\nu)] = (1/3) \mbE[\ddot{\br}_{k} (\omega) \cdot \ddot{\br}_{k'} (\nu)]$, for all directions $j$. This leads to
\begin{equation}
	\mbE[J_{i,j} (\omega,\nu)] 
	= \frac{8 \pi \sin(b)}{3 b}\mbE[\ddot{\br}_{k} (\omega) \cdot \ddot{\br}_{k'} (\nu)]
	= \alpha^2  \frac{\hbar^2 \gamma}{m_0^{2\alpha}} \frac{16 \pi^2 \sin(b)}{3 b}\delta(\omega + \nu) f_{k,k'}.
\end{equation}
where
\begin{equation}
	f_{k,k'} := \sum_j \int \dd[3]{\br} \mcM^{2\alpha-2} (\br) [\partial^j g_{r_C}(\ev{\br_k} - \br)][\partial^j g_{r_C}(\ev{\br_{k'}} - \br)].
\end{equation}
Substituting this in the expression for the power, we get
\begin{equation}
	P(t) = \frac{\alpha^2 \hbar^2 \gamma}{12 \pi^2 \varepsilon_0 c^3 m_0^{2\alpha}} \sum_{i,j} q_i q_j f_{i,j} \intmp \dd{\omega} \frac{\sin(b_{i,j} (\omega,-\omega))}{b_{i,j} (\omega,-\omega)},
\end{equation}
where $b_{i,j} (\omega,-\omega) = (\omega/c)\abs{\ev{\br_i}-\ev{\br_j}}$. Comparing this result with the decomposition of $P(t)$ as an integral over the emission rate one gets
\begin{equation}
	\dv{\Gamma_t}{\omega} = \frac{\alpha^2 \hbar \gamma}{6 \pi^2 \varepsilon_0 c^3 m_0^{2\alpha} \omega} \sum_{i,j} q_i q_j f_{i,j} \frac{\sin(b_{i,j} (\omega,-\omega))}{b_{i,j} (\omega,-\omega)}.
\end{equation}

Now we consider two relevant regimes:
\begin{equations}
	\abs{\ev{\br_i}-\ev{\br_j}} \gg \lambda_\omega
	&\implies
	\dv{\Gamma_t}{\omega} = \frac{\alpha^2 \hbar \gamma}{6 \pi^2 \varepsilon_0 c^3 m_0^{2\alpha} \omega} \sum_{k} q_k^2 f_{k,k},
	\\
	\abs{\ev{\br_i}-\ev{\br_j}} \ll \lambda_\omega
	&\implies
	\dv{\Gamma_t}{\omega} = \frac{\alpha^2 \hbar \gamma}{6 \pi^2 \varepsilon_0 c^3 m_0^{2\alpha} \omega} \sum_{i,j} q_i q_j f_{i,j},
\end{equations}
where $\lambda_\omega = 2\pi c/\omega$ is the wavelength of the emitted radiation. Since the case we are interested in is that of a solid composed of only one type of atoms, we can simplify the calculation by exploiting the compoundation property for the nuclei. We treat each nucleus as a single particle of mass $M_N$ and charge $Q_N$. Electrons are instead always taken as being at distances much larger than $\lambda_\omega$ among themselves and with the nuclei. However, since an atom has dimensions much smaller than $r_C$, their position can be taken to always coincide with that of the nucleus. The contribution of each atom is the same so we have just to compute it for one of them and multiply by $N_P$, the number of atoms in the emitting body. Moreover, we notice that $f_{k,k}$ is the same for both nuclei and electrons once one makes the approximations that the electrons average position is at their nucleus position. Therefore we get
\begin{equation}
	\dv{\Gamma_t}{\omega} = N_P\prtq{Q_N^2 + N_e e^2} \frac{\alpha^2 \hbar \gamma}{6 \pi^2 \varepsilon_0 c^3 m_0^{2\alpha} \omega} f_{k,k},
\end{equation}
where $N_e$ is the number of electrons participating in the emission. Using the 
macroscopic density approximation and the sharp scanning approximation, we can substitute $\mcM(\bx)$ with $\mu_0$, the density of the emitting body. Thus we get
\begin{equation}
	f_{k,k} 
	=3 \mu_0^{2\alpha-2} \int \dd[3]{\br} [\partial^j g_{r_C}(\br)]^2
	=\frac{3 \mu_0^{2\alpha-2}}{16 \pi^{3/2} r_C^5}.
\end{equation}
The final formula is then
\begin{equation}
	\dv{\Gamma_t}{\omega} 
	= N_P\prtq{Q_N^2 + N_e e^2} \frac{\alpha^2 \hbar \gamma}{32 \pi^2 \varepsilon_0 c^3 m_0^{2\alpha} \omega} \frac{\mu_0^{2\alpha-2}}{\pi^{3/2} r_C^5}
	= N_P\prtq{Q_N^2 + N_e e^2} \frac{\alpha^{7/2} \lambda_\alpha \hbar (2 \pi r_C^2)^{3\alpha} \mu_0^{2\alpha-2}}{32 \pi^5 \varepsilon_0 c^3 m_0^{2\alpha} r_C^8 \omega},
\end{equation}
which is the same as Eq.~\eqref{APPeq:SpontaneousRadiationSmallRCHeuristicDerivation}.

\subsection{Getting the wrong result by blindly using the rigid body assumption}

When treating an ideal rigid body, we assume that all particles undergo the same exact acceleration, i.e., that of the center of mass. Based on their distances compared to the wavelength of the emitted radiation they can emit coherently or not. As explained in the main text, we consider that protons in nuclei emit coherently while electrons and different nuclei do not. Charges emitting coherently are first summed and then squared in Larmor's formula while for incoherent emission one just adds the square of the charges.
Repeating the calculation of the single particle case with $\mcM_{\rm CM}^\alpha (\bx)$ would get us to
\begin{equation}
	\dv{\Gamma (t)}{\omega} = N_P\prtq{Q_N^2 + N_e e^2}\frac{\hbar \gamma_\alpha}{6 \pi^2 \varepsilon_0 c^3 m^2 m_0^{2\alpha} \omega} \sum_{j=1}^{3} \int \dd[3]{\bx} [\partial_j \mcM^\alpha (\bx)]^2,
\end{equation}
where the integral seems exactly as the one we computed in Appendix~\ref{APPSec:MomentumSpreadGeneralCase}. If we were to treat as in there, we would get
\begin{equation}
	\dv{\Gamma (t)}{\omega} = N_P\prtq{Q_N^2 + N_e e^2}\frac{\alpha^2 \hbar \gamma_\alpha}{6 \pi^2 \varepsilon_0 c^3 M^2 \omega}  \prt{\frac{M}{m_0}\frac{1}{V}}^{2 \alpha}
	\frac{G(\alpha)}{2 \pi r_C} A,
\end{equation}
where $A$ is the surface area of the rigid body. Indeed, the above equation does not coincide with Eq.~\eqref{eq:RadiationSmallRC}. As predicted, the complete rigid body assumption is not appropriate for this problem.

\clearpage
\section{Localization of a thin disk\label{APPSec:ThinDiskLocalization}}

\begin{figure}
	\centering
	\includegraphics[width=1.0\textwidth]{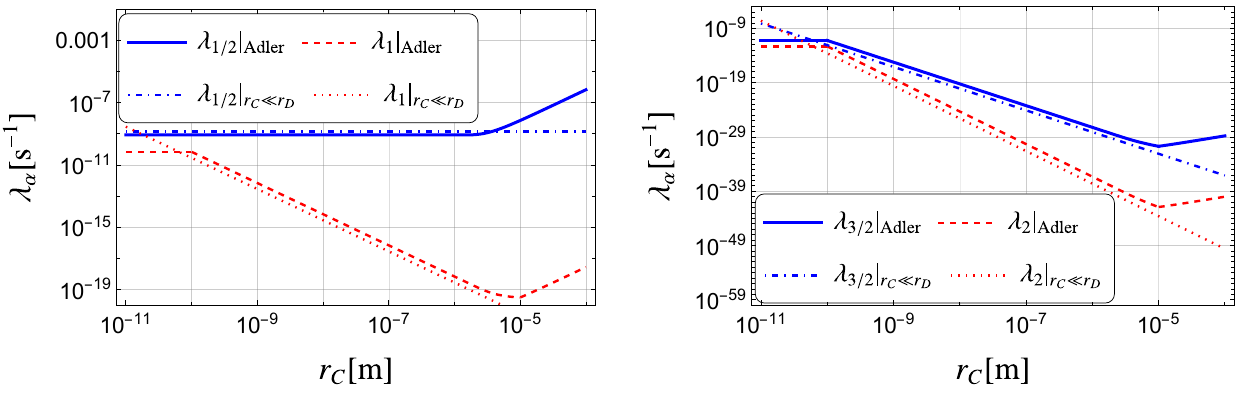}
	\caption{Theoretical lower bound for the localization rate $\lambda_\alpha$ as a function of $r_C$. The curves obtained using Eq.~\eqref{eq:MinimumValueLocalizationRate} are denoted by the subscript \enquote{Adler} in the plot legends while those obtained using Eq.~\eqref{APPeq:MinimumValueLocalizationRateSophisticatedSmallRc} are denoted with the subscript $r_C \ll r_D$. The parameters used are $r_a=10^{-10} \rm{m}$, $r_D = 10^{-5} \rm{m}$, $m_a = 12 m_0$, $n_A = 10^{10}$, $\tau=10^{-2} \rm{s}$, $\abs{\bD} = r_D$. There is a good agreement in the region $10^{-10} \rm{m} \leq r_C \leq 10^{-5} \rm{m}$, as expected. For $r_C\geq 10^{-5} \rm{m}$ there is no need to make comparisons because Eq.~\eqref{APPeq:LargeCollapseRadiusDecoherenceRateThinDisk} and Eq.~\eqref{eq:AdlerEffectiveDecoherenceRate} give the same results.}
	\label{APPfig:TheoryBoundPlots}
\end{figure}

We explain how to obtain Eq.~\eqref{eq:AdlerEffectiveDecoherenceRate} of the main text in Sec.~\ref{APPsubsec:AdlerFormulaGeneralization}. We also get a more sophisticated formula in Sec~\ref{APPsubsec:HomogeneousThinDisk}. The two results are compared in Fig.~\ref{APPfig:TheoryBoundPlots}.

\subsection{Generalized Adler's formula\label{APPsubsec:AdlerFormulaGeneralization}}

As explained in Sec.~\ref{subsec:LocalizationThinDisk}, we consider every atom (of mass $m_a$) of the graphene disk to occupy an area $\pi r_a^2$, while the disk has radius $r_D$. 

If $r_C \gg r_D$, because of compoundation (see Sec.~\ref{subsec:CompoundationInvariance}), we get that
\begin{equation}
	\Gamma^{\rm{CM}}_\alpha (\bD) = \Lambda_\alpha \prtq{1-e^{-\alpha \bD^2/(4 r_C^2)}},
	\qq{where}
	\Lambda_\alpha = \prt{\frac{M_D}{m_0}}^{2\alpha} \lambda_\alpha,
\end{equation}
where $M_D$ is the total mass of the disk. 

In the opposite regime, $r_C \ll r_a$, we can write
\begin{equation}
	\mcM(\bx) 
	=
	m_a \sum_k g_{r_C} (\hbq_k - \bx)
	\simeq
	m_a g_{r_C} (\hbQ + \bq_c (\bx-\hbQ) - \bx)
\end{equation}
where $\bq_c (\bx-\hbQ)$ is the position of the closest atom to the spatial point $\bx$, given a certain position of the center of mass of the disk. If there is more than one atom minimizing this distance, than we are sufficiently far away from all of them that $\mcM(\bx) \sim 0$, therefore it does not matter that there are more contributions from equidistant atoms, since all of them are negligible. In this case we get [cf.~\eqref{eq:CenterOfMassDecoherenceRate}]
\begin{equation}
	\Gamma^{\rm{CM}}_\alpha (\bD) 
	\simeq 
	\frac{\gamma_\alpha}{2} \prt{\frac{m_a}{m_0}}^{2\alpha} \int \dd[3]{\bx} \prtq{g_{r_C}^\alpha (\bq_c (\bx) - \bx) - g_{r_C}^\alpha (\bq_c (\bx-\bD) - \bx - \bD)}^2
	\simeq
	n_a \gamma_\alpha \prt{\frac{m_a}{m_0}}^{2\alpha} \int \dd[3]{\bx} g_{r_C}^{2\alpha} (\bx).
\end{equation}
We have made the approximations to neglect the cross terms because for most values of $\bD$ the lattices correspondent to the two center of mass superpositions will not overlap over a coarse grain of $r_C \ll r_a$. Performing the calculation, we get
\begin{equation}
	\Gamma^{\rm{CM}}_\alpha (\bD) 
	\simeq 
	2\sqrt{2} n_a \prt{\frac{m_a}{m_0}}^{2\alpha} \lambda_\alpha 
	\sim 
	\Lambda_\alpha \prtq{1-e^{-\alpha \bD^2/(4 r_C^2)}},
	\qq{where}
	\Lambda_\alpha = n_a \prt{\frac{m_a}{m_0}}^{2\alpha} \lambda_\alpha.
\end{equation}
The last substitution has been made in order to obtain the same result of Adler's formula for $r_C \ll r_a$~\cite{Toros2018BoundsCalculations}. 

Finally, let us consider $r_a \ll r_C \ll r_D$. In this case, we write
\begin{equation}
	\mcM(\bx) 
	=
	m_a \sum_k g_{r_C} (\hbq_k - \bx)
	\simeq
	m_a n(r_C) g_{r_C} (\hbQ - \bx),
\end{equation}
where $n(r_C)=(r_C/r_a)^2$ estimates the number of atoms giving a meaningful contribution to the action of $\mcM (\bx)$ on the center of mass. The number of circles of radius $r_C$ covering the disk can be estimated as $(r_D/r_C)^2 \sim n_a/n(r_C)$. Then, since $\gamma_\alpha \int \dd[3]{\bx} g_{r_C}^{2\alpha} \sim \lambda_\alpha$ [cf. Eq.~\eqref{eq:SingleParticleDecoherence}], we can estimate the total decoherence rate as
\begin{equation}
	\Gamma^{\rm{CM}}_\alpha (\bD) \sim \Lambda_\alpha \prtq{1-e^{-\alpha \bD^2/(4 r_C^2)}},
	\qq{where}
	\Lambda_\alpha = \frac{n_a}{n(r_C)} \prt{\frac{m_a n(r_C)}{m_0}}^{2\alpha} \lambda_\alpha.
\end{equation}

The three formula obtained above can be put together as follows
\begin{equation}\label{APPeq:GeneralizedAdlerFormula}
	\Lambda_\alpha = \frac{n_a}{n(r_C)} \prt{\frac{m_a n(r_C)}{m_0}}^{2\alpha} \lambda_\alpha,
	\qq{where}
	n(r_C) = 
	\begin{cases}
		1, &\quad r_C < r_a,\\
		(r_C/r_a)^2, &\quad r_a \leq r_C \leq r_D,\\
		n_a, &\quad r_D < r_C.
	\end{cases}
\end{equation}
This is the generalized Adler's formula.

To be sure of the validity of our generalized Adler's formula, we deal with this same problem in a more sophisticated way in the next subsection.

\subsection{Homogeneous thin disk calculation\label{APPsubsec:HomogeneousThinDisk}}

Here we compute the decoherence rate of the disk generalizing the calculations detailed in section 4 of Ref.~\cite{Toros2018BoundsCalculations}. Unfortunately, the generalization cannot be done by following exactly the same steps due to the higher mathematical complexity of the generalized CSL and PSL models. We stress that these calculations are valid for $r_C \gg r_a$. 

In this regime, we can assume that the coarse-grained density is constant and we write it as follows:
\begin{equation}
	\mu (\bx) = \frac{M_D}{\pi r_D^2 \varepsilon} \chi_{D} (\bx),
	\implies
	\mcM_{\rm CM} (\bx) = \frac{M_D}{(\pi r_D^2 \varepsilon)(2 \pi r_C^2)^{3/2}}  \int \dd[3]{\by} \chi_D (\by) \exp(-\frac{(\bx - \by)^2}{2 r_C^2}),
\end{equation}
where $M_D$ is the total mass of the graphene disk, $r_D$ is its radius and $\chi_D$ is an indicator function for a disk of radius $r_D$ and depth $\varepsilon$. To compute the convolution we exploit the convolution theorem\footnote{Notice that because of the convention we are using for the Fourier transforms, the convolution theorem gets the form $\mcF \prtg{f \star g} = (2\pi)^{3/2} \mcF \prtg{f} \mcF\prtg{g}$.} and first compute the Fourier transform of the Gaussian and of the indicator function\footnote{Notice that we take the disk to be planar on the $x$-$y$ plane.}
\begin{equations}
	\tilde{g} (\bk) &= \frac{1}{(2\pi)^{3/2}} \int \dd[3]{\by} \exp(-\frac{\by^2}{2 r_C^2})e^{i \bk \cdot \by}
	= r_C^3 e^{-r_C^2 \bk^2/2},
	\\
	\tilde{\chi}_D (\bk) &= \frac{1}{(2\pi)^{3/2}} \int \dd[3]{\by} \chi_D (\by) e^{i \bk \cdot \by}
	= \frac{r_D \varepsilon}{\sqrt{2\pi} \abs{\bk_D}} J_1 (r_D \abs{\bk_D}){\rm sinc} \prt{\frac{k_z \varepsilon}{2}},
\end{equations}
where $\bk_D$ denotes the projection of $\bk$ on the $k_x$-$k_y$ plane and $J_1$ denotes the Bessel function of the first kind\footnote{See \url{https://reference.wolfram.com/language/ref/BesselJ.html?q=BesselJ}.}. Considering that the graphene is a single layer we have that $\varepsilon \ll r_C$ so we can take $\varepsilon \rightarrow 0$ to simplify our calculations. We get
\begin{equations}
	\mcM_{\rm CM} (\bx) &= \frac{M_D}{(\pi r_D^2)(2 \pi r_C^2)^{3/2}}  \frac{ r_C^3 r_D}{\sqrt{2 \pi}}
	\int \dd[3]{\bk} \frac{J_1 (r_D \abs{\bk_D})}{\abs{\bk_D}} e^{-r_C^2 \bk^2/2} e^{- i \bx \cdot \bk}
	= M_D\frac{2 e^{-\frac{x_3^2}{2 r_C^2}}}{(2 \pi)^{3/2} r_C r_D} G_2(r,r_D,r_C),
	\\
	G_2(r,r_D,r_C) &= \int_0^\infty \dd{k_D} J_0 (r k_D) J_1 (r_D k_D) e^{-k_D^2 r_C^2/2},
\end{equations}
where $x_3$ is the $z$-component of $\bx$ and $r$ is the radial one on the $x$-$y$ plane. It does not seem possible to derive an explicit (non-integral) expression for $G_2(r,r_D,r_C)$ except for the following special cases:
\begin{equation}
	G_2(0,r_D,r_C) = \frac{1-e^{-r_D^2/2 r_C^2}}{r_D},
	\qquad
	G_2(r_D,r_D,r_C) = \frac{1-e^{-r_D^2/r_C^2}I_0(r_D^2/r_C^2)}{2 r_D},
\end{equation}
where $I_0$ denotes the modified Bessel function of the first kind\footnote{See \url{https://reference.wolfram.com/language/ref/BesselI.html}.}. These two cases give the value of $\mcM_{\rm CM} (\bx)$ at the center of the disk and at its border. We can also find approximate values in the two regimes $r_C \ll r_D $ and $r_C \gg r_D$. When $r_C \ll r_D$, we can set $r_C =0$ while for the other regime we must expand $J_1 (r_D k_D)$ to first order. We get
\begin{equation}
	G_2(r,r_D,0) = \frac{1}{r_D}\Theta_{\rm{H}} (r_D-r),
	\qquad
	G_2(r,r_D,r_C)\vert_{r_D \ll r_C} \simeq \frac{r_D}{2 r_C^2}e^{-r^2/2 r_C^2}.
\end{equation}
$G_2(r,r_D,0)$ corresponds to the two exact cases calculated before when setting $\Theta_{\rm{H}} (0)=1/2$ and taking the limit $r_C \rightarrow 0$. On the other hand $G_2(r,r_D,r_C)\vert_{r_D \ll r_C}$ corresponds to both exact cases when they are expanded to first order in $r_D$.

The general formula for $\Gamma_{\alpha}^{\rm CM}$ is now given by 
\begin{equation}\label{APPeq:IntegralThinDiskPlanarDisplacement}
	\Gamma^{\rm{CM}}_\alpha \prt{\bD} =
	\frac{\gamma_\alpha}{2} \prt{\frac{M_D}{m_0}}^{2\alpha}
	\prt{\frac{2}{(2\pi)^{3/2}r_C r_D}}^{2\alpha} \frac{\sqrt{\pi}r_C}{\sqrt{\alpha}}
	\int \dd[2]{\bx} \prtq{G_2^\alpha (\bx) - G_2^\alpha (\bx - \bD)}^{2},
\end{equation}
where we used $\int \exp(-\alpha z^2/r_C^2) \dd{z} = \sqrt{\pi}r_C/\sqrt{\alpha}$ and the fact that $\bD$ is assumed to lie in the $x$-$y$ plane. 

Let us start by considering the case $G_2 (\bx,r_D,0)$. Then, the integral in Eq.~\eqref{APPeq:IntegralThinDiskPlanarDisplacement} evaluates to $1/r_D^{2\alpha}$ times the area of the non-overlapping regions of two circles of radius $r_D$ and distance $\abs{\bD}$. This area is $2 \pi r_D^2$ if $d=\abs{\bD} \geq 2 r_D$, while it is given by
\begin{equation}
	A(d) = 2 \pi r_D^2 - 4 r_D^2\prtg{\arcsin(\sqrt{1-\prt{\frac{d}{2 r_D}}^2}) - \frac{d}{2r_D}\sqrt{1-\prt{\frac{d}{2 r_D}}^2}}.
\end{equation}
for $0\leq d \leq 2 r_D$. By defining $\tilde{d}_D=d/(2 r_D)$, we get
\begin{equation}\label{APPeq:DecoherenceRateThinDiskSophisticated}
	\Gamma^{\rm{CM}}_\alpha \prt{\bD} =
	\frac{\gamma_\alpha}{2} \prt{\frac{M_D}{m_0}}^{2\alpha}
	\prt{\frac{2}{(2\pi)^{3/2}r_C r_D}}^{2\alpha} \frac{\sqrt{\pi}r_C}{\sqrt{\alpha}} \frac{2 r_D^2}{r_D^{2\alpha}} \prtg{\pi - 2\prtq{\arcsin(\sqrt{1-\tilde{d}_D^2}) - \tilde{d}_D\sqrt{1-\tilde{d}_D^2}}},
\end{equation}
from which we can compute, in analogy to Eq.~\eqref{eq:MinimumValueLocalizationRate}\footnote{See also Eq.~\eqref{eq:SingleParticleDecoherence} for the definition of $\lambda_\alpha$.},
\begin{equation}\label{APPeq:MinimumValueLocalizationRateSophisticatedSmallRc}
	\lambda_\alpha (r_C) \geq \prtq{\tau_D \frac{\alpha}{\pi} \prt{\frac{2 M_D}{m_0}}^{2\alpha} \prt{\frac{r_C}{r_D}}^{4\alpha-2} \prt{\pi - 2\prtq{\arcsin(\sqrt{1-\tilde{d}_D^2}) - \tilde{d}_D\sqrt{1-\tilde{d}_D^2}}}}^{-1}.
\end{equation}
Indeed we have $\Gamma^{\rm{CM}}_\alpha \prt{0}=0$. Some special cases of interest are
\begin{equations}\label{APPeq:SmallCollapseRadiusDecoherenceRateThinDisk}
	\Gamma^{\rm{CM}}_\alpha \prt{r_D} 
	&= \frac{\gamma_\alpha}{2} \prt{\frac{M_D}{m_0}}^{2\alpha}
	\prt{\frac{2}{(2\pi)^{3/2}r_C r_D}}^{2\alpha} \frac{\sqrt{\pi}r_C}{\sqrt{\alpha}} \frac{2 r_D^2}{r_D^{2\alpha}} \prt{\frac{\pi}{3}+\frac{\sqrt{3}}{4}},\\
	\Gamma^{\rm{CM}}_{1/2} \prt{r_D} 
	&\simeq 6.1 \times 10^{-1} \times \gamma_{1/2} \frac{M_D}{m_0}\\
	\Gamma^{\rm{CM}}_1 \prt{r_D} 
	&= 5.5 \times 10^{-2} \times \gamma_1 \prt{\frac{M_D}{m_0}}^2 \frac{1}{r_C r_D^2}
	= 5.5 \times 10^{-2} \times \lambda_1 (4 \pi r_C^2)^{3/2} \prt{\frac{M_D}{m_0}}^2 \frac{1}{r_C r_D^2}.
\end{equations}

Let us now consider the case $G_2(\bx,r_D,r_C)\vert_{r_D \ll r_C}$. We get
\begin{equation}
	\int \dd[2]{\bx} \prtq{G_2^\alpha (\bx) - G_2^\alpha (\bx - \bD)}^{2}
	=2 \pi \frac{r_C^2}{\alpha}\prt{\frac{r_D}{2 r_C^2}}^{2\alpha}\prtq{1 - e^{-\alpha d^2/4 r_C^2}},
\end{equation}
so that
\begin{equation}\label{APPeq:LargeCollapseRadiusDecoherenceRateThinDisk}
	\Gamma^{\rm{CM}}_\alpha \prt{\bD} = \lambda_\alpha \prt{\frac{M_D}{m_0}}^{2\alpha} \prtq{1 - e^{-\alpha d^2/4 r_C^2}},
\end{equation}
as expected from compoundation. Indeed, this is the same result of the generalized Adler's formula [Eq.~\eqref{eq:AdlerEffectiveDecoherenceRate}] for $r_C \gg r_D$.

\clearpage
\section{Analysis of the spontaneous collapse noise for LIGO and LISA Pathfinder\label{APPSec:LigoLisaCalculations}}

The goal of this section is to derive Eq.~\eqref{eq:ForceNoiseDensity} of the main text.
We will make a unified analysis valid for both LIGO and LISA Pathfinder in the two regimes $r_C \ll L,R$ and $r_C \gg L,R$ [cf. Table~\ref{tab:GravWaveParameters}].

\subsection{Large collapse radius regime}

Let us start with the regime $r_C \gg L,R$. In this case, every body $B$ in the experiment can be considered as a point particle of mass $M$ and position operator $\hbQ_B$. Moreover, we consider that the displacement $\Delta \hbQ_B = \hbQ_B-\bQ^{(0)}_B$ of each body from its rest position $\bQ^{(0)}_B$ is much smaller than $r_C$. So, we can write:
\begin{equations}
	\mcM^\alpha (\bx) &= M^\alpha \prtq{\sum_B g_{r_C}(\hbQ_B - \bx)}^\alpha
	\simeq
	M^\alpha \prtq{\sum_B g_{r_C}(\bQ^{(0)}_B - \bx) - \sum_B \nabla g_{r_C}(\bQ^{(0)}_B - \bx) \cdot \Delta \hbQ_B}^\alpha,\\
	\simeq
	M^\alpha &\prtq{\sum_B g_{r_C}(\bQ^{(0)}_B - \bx)}^\alpha 
	- \alpha M^\alpha \prtq{\sum_{B'} g_{r_C}(\bQ^{(0)}_{B'} - \bx)}^{\alpha-1} \sum_B \nabla g_{r_C}(\bQ^{(0)}_B - \bx)\cdot \Delta \hbQ_B.
\end{equations}
Due to the unitary unraveling of Eq.~\eqref{eq:UnitaryUnraveling}, we can consider as if on each of them acts a stochastic force given by:
\begin{equation}
	\hbF_B (t) 
	= \frac{i}{\hbar}\comm{\hV(t)}{\hbP_B}
	= 
	\alpha\hbar \sqrt{\gamma_\alpha} \prt{\frac{M}{m_0}}^\alpha \int \dd[3]{\bx} \prtq{\sum_{B'} g_{r_C} (\bQ^{(0)}_{B'} - \bx)}^{\alpha-1} \nabla g_{r_C} (\bQ^{(0)}_B - \bx) w(\bx,t),
\end{equation}
where $w(\bx,t)$ is a white noise with $\mbE[w(\bx,t)]=0$ and $\mbE[w(\bx,t)w(\by,s)]=\delta(t-s)\delta^{(3)}(\bx-\by)$.
The physical quantity discussed in Ref.~\cite{Carlesso2016ExperimentalBounds} is the force noise spectral density\footnote{In Ref.~\cite{Carlesso2016ExperimentalBounds} is denoted as $S_{FF} (\omega)$. However, we will see it is actually independent of $\omega$ as we are taking a white noise force.} 
$S$ which is defined in terms of Fourier transform\footnote{Contrarily to Ref.~\cite{Carlesso2016ExperimentalBounds}, 
	we define the Fourier transform with the prefactor $1/\sqrt{2\pi}$ instead of $1$.} of the stochastic force along a specific direction that we will call $x_1$, as $S (\omega) := (1/2)\int \mbE\prtq{\acomm{\tilde{F} (\omega)}{\tilde{F} (\omega')}}\dd{\omega'}$. The stochastic force has the same form as above with the substitution $w(\bx,t) \rightarrow \tilde{w}(\bx,\omega)$, where $\tilde{w}(\bx,\omega) = \frac{1}{\sqrt{2\pi}}\int w(\bx,t)e^{i\omega t}\dd{t}$. A simple calculation\footnote{For simplicity, let us consider $w(t)$ such that $\mbE[w(t)w(t')]=\delta(t-t')$. Then,\\ $\mbE\prtq{\tilde{w}(\omega)\tilde{w}(\omega')}=(1/2\pi)\int \mbE[w(t)w(t')]e^{i\omega t}e^{i \omega' t'} \dd{t}\dd{t'} = (1/2\pi) \int e^{i(\omega+\omega')t}\dd{t} = \delta(\omega+\omega')$.} 
shows that $\mbE\prtq{w(\bx,\omega)w(\by,\omega')} = \delta(\omega+\omega')\delta^{(3)}(\bx-\by)$. For the LIGO and LISA detectors discussed in Ref.~\cite{Carlesso2016ExperimentalBounds} the considered stochastic force is $F(t)=[F_1(t)-F_2(t)]/2$ so that its force noise spectral density reads\footnote{Hereafter we write $\bQ_B$ instead pf $\bQ^{(0)}_B$ to lighten the notation.}:
\begin{equation}
	S (\omega) =
	\frac{\alpha^2 \hbar^2 \gamma_\alpha}{4}\prt{\frac{M}{m_0}}^{2\alpha}
	\int\prtq{g_{r_C} \prt{\bQ_1 -\bx}+g_{r_C} \prt{\bQ_2 -\bx}}^{2\alpha-2}\prtq{\partial_{x_1} g_{r_C} \prt{\bQ_1 -\bx} - \partial_{x_1} g_{r_C} \prt{\bQ_2 -\bx}}^2 \dd[3]{\bx},
\end{equation}
where we see that, as expected, $S(\omega)$ does not depend on $\omega$ due to the noise being white in time. Since the two bodies are displaced along the $x_1$-axis, we can immediately perform the integration over the other directions. We can also change coordinates for $x_1$ so that $Q_1=0$ and $Q_2=-a$, obtaining:
\begin{equation}
	S = \frac{\alpha \hbar^2 \gamma_\alpha}{4^{\alpha+1}\pi^{2\alpha-1}r_C^{4\alpha-2}}\prt{\frac{M}{m_0}}^{2\alpha}
	\int\prtq{g_{r_C} \prt{x_1}+g_{r_C} \prt{x_1+a}}^{2\alpha-2}\prtq{\frac{x1 g_{r_C} \prt{x1} - (x_1+a) g_{r_C} \prt{x1+a}}{r_C^2}}^2 \dd{x_1},
\end{equation}
where $g_{r_C} (x_1) = [\sqrt{2\pi}r_C]^{-1/2}\exp[-x_1^2/2 r_C^2]$. By making the substitution $\tl{a}=a/r_C$ and $\tl{x}=x_1/r_C$ one gets
\begin{equation}
	S = \frac{\alpha \hbar^2 \gamma_\alpha}{4^{\alpha+1}\pi^{2\alpha-1}r_C^{6\alpha-1}}\prt{\frac{M}{m_0}}^{2\alpha} f_S \prt{\frac{a}{r_C},\alpha}
	= \frac{\alpha^{5/2} \hbar^2 \lambda_\alpha}{4\sqrt{\pi} r_C^2}\prt{\frac{M}{m_0}}^{2\alpha} f_S \prt{\frac{a}{r_C},\alpha},
\end{equation}
where
\begin{equation}\label{APPeq:FunctionfS}
	f_S (\tl{a},\alpha) = \int\prtq{e^{-\tl{x}^2/2}+e^{-\prt{\tl{x}+\tl{a}}^2/2}}^{2\alpha-2}\prtq{\tl{x} e^{-\tl{x}^2/2} - (\tl{x}+\tl{a})e^{-\prt{\tl{x}+\tl{a}}^2/2} }^2 \dd{\tl{x}}.
\end{equation}
We have that
\begin{equation}
	f(a \ll 1,\alpha) = a^2 \frac{\sqrt{\pi}(3+4\alpha(\alpha-1))}{\alpha^{5/2} 4^{\alpha-2}},
	\quad
	f(a\gg 1,\alpha) = \frac{\sqrt{\pi}}{\alpha^{3/2}},
\end{equation}
and
\begin{equations}
	f(a,1) &= \sqrt{\pi} \prtq{1+\prt{\frac{1}{2}a^2 -1}e^{-a^2/4}},
	\\
	f(a,3/2) &= \frac{2}{3}\sqrt{\frac{2\pi}{3}} \prtq{1+\prt{\frac{5}{3}a^2 -1}e^{-a^2/3}},
	\\
	f(a,2) &= \frac{1}{2}\sqrt{\frac{\pi}{2}} \prtq{1+\prt{3a^2 -1}e^{-a^2/2} +2 a^2 e^{-(3/8)a^2}}.
\end{equations}
Unfortunately, we could not analytically solve the integral for $\alpha=1/2$. However, a simple analytical function sufficiently similar (for our purposes) to it is
\begin{equation}
	f(a,1/2) \simeq 2 \sqrt{2\pi} \prt{1-e^{-a^2/2}}.
\end{equation}
On can check that this has the correct limiting behavior for $\tl{a}\ll 1$ and $\tl{a} \gg 1$.

\subsection{Small collapse radius regime}

In the regime $r_C \ll L,R$ we have to consider the constituents of the rigid bodies. However, we will still consider values of $r_C$ much larger than the distance between the atoms and molecules composing the rigid bodies. 

We denote each particle with two indexes: $B$ and $k$. Index $B$ denotes to which macroscopic body the particle belongs to, while index $k$ enumerates them within the $B$-th body. In this case, we can write the smeared mass density operator as follows
\begin{equation}
	\mcM(\bx) = \sum_B \mcM_B (\bx),
	\qquad
	\mcM_B (\bx) = \sum_k m_{B,k} g \prt{\hbq_{B,k} - \bx},
\end{equation}
where $\hbq_{B,k}$ denotes the position operator of the $k$-th particle within the $B$-th body. As in the previous subsection, we consider rigid bodies oscillating around their equilibrium positions. We write $\hbq_{B,k} = \bq^{(0)}_{B,k} + \Delta \hbq_{B,k} + \hbQ_B$, where $\bq^{(0)}_{B,k}$ is the equilibrium position of the $k$-th particle in the $B$-th body, $\hbQ_B$ is the displacement operator of the $B$-th rigid body center of mass\footnote{With respect to an equilibrium position $\bQ^{(0)}_B$.}, and $\Delta \hbq_{B,k}$ is the remaining displacement operator at the level of single particle and not accounted for by $\hbQ_B$. However, under the rigid body assumption, this last operator can be neglected as all particles are assumed to move together. Moreover, assuming that the displacement of each rigid body is much smaller than $r_C$, we can expand at first order in $\hbQ/r_C$ and write\footnote{Notice that $\nabla$ denotes the gradient with respect to the position vector $\bx$.}
\begin{equation}
	\mcM_B (\bx) 
	\simeq \sum_k m_{B,k} \prtq{ g\prt{\bq_{B,k}^{(0)} - \bx} - \nabla g\prt{\bq_{B,k}^{(0)} - \bx}\cdot \hbQ_B}
	=
	\mcM_B^{(0)} (\bx) - \nabla \mcM_B^{(0)} (\bx) \cdot \hbQ_B,
\end{equation}
where $\mcM_B^{(0)} (\bx) = \sum_k m_{B,k} g\prt{\bq_{B,k}^{(0)} - \bx}$. We also define $\mcM^{(0)} (\bx) = \sum_B \mcM_B^{(0)} (\bx)$. Since we are interested in the generalized case with collapse operator $\mcM^\alpha (\bx)$, we write
\begin{equations}
	\mcM^\alpha (\bx)
	&\simeq 
	\prt{\mcM^{(0)} (\bx) - \sum_B \nabla \mcM_B^{(0)} (\bx) \cdot \hbQ_B}^\alpha,\\
	&\simeq 
	\prt{\mcM^{(0)} (\bx)}^\alpha
	- \alpha \prt{\mcM^{(0)} (\bx)}^{\alpha-1} \sum_B \nabla \mcM_B^{(0)} (\bx) \cdot \hbQ_B,\\
	&\simeq
	\prt{\mcM^{(0)} (\bx)}^\alpha
	- \alpha \sum_B \prt{\mcM_B^{(0)} (\bx)}^{\alpha-1} \nabla \mcM_B^{(0)} (\bx) \cdot \hbQ_B,\\
	&=
	\prt{\mcM^{(0)} (\bx)}^\alpha
	- \sum_B \nabla\prt{\mcM_B^{(0)} (\bx)}^{\alpha} \cdot \hbQ_B,
\end{equations}
where, for the last line, we have exploited the fact that different bodies occupy different positions\footnote{This approximation is debatable when $r_C \geq a$. However, in this regime we always have $r_C \ll L \lesssim a$.}. 

We need the stochastic force acting on a body in the direction $x_1$:
\begin{equation}
	\tilde{F}_B (\omega) 
	= \hbar \frac{\sqrt{\gamma_\alpha}}{m_0^\alpha}\int \dd[3]{\bx} \partial_{x_1}\prt{\mcM_B^{(0)}(\bx)}^{\alpha} \tilde{w}(\bx,\omega).
\end{equation}
Then, as before, we compute the spectral density $S$ related to $F(t)=[F_1(t)-F_2(t)]/2$:
\begin{equation}
	S = \frac{\gamma_\alpha \hbar^2}{4 m_0^{2\alpha}} 
	\int \dd[3]{\bx} \prtg{\prtq{\partial_{x_1} \prt{\mcM_1^{(0)} (\bx)}^\alpha}^2
		+\prtq{\partial_{x_1} \prt{\mcM_2^{(0)} (\bx)}^\alpha}^2
		-2\prtq{\partial_{x_1} \prt{\mcM_1^{(0)} (\bx)}^\alpha}\prtq{\partial_{x_1} \prt{\mcM_2^{(0)} (\bx)}^\alpha}}.
\end{equation}
Since the second body is identical to first but displaced along the $x_1$-axis by a length $a$, we can write $\mcM_2^{(0)} (\bx) = \mcM_1^{(0)} (\bx - a \be_1)$, with $\be_1$ the unit normal vector on the $x$-axis. It follows that
\begin{equation}\label{APPeq:ForceNoiseSpectralDensityTwoBodies}
	S
	= \frac{\gamma \hbar^2}{2 m_0^{2\alpha}} \int \dd[3]{\bx} \prtg{\prtq{\partial_{x_1} \prt{\mcM_1^{(0)} (\bx)}^{2\alpha}}
		-\prtq{\partial_{x_1} \prt{\mcM_1^{(0)} (\bx)}^\alpha}\prtq{\partial_{x_1} \prt{\mcM_1^{(0)} (\bx-a\be_1)}^\alpha}}.
\end{equation}

We recall that we are considering objects of volume $V = A_P L$, where $A_P$ is the (constant) cross-section perpendicular to the $x_1$-axis and $L$ is their length across the $x_1$ axis. Their centers are separated by a distance $a$ along this same axis. Then, for each them we can write
\begin{equation}
	\mcM_B^{(0)} (\bx) = \mu_0 s_1 (x_1) \chi_P (\bx_P),
	\implies
	\prt{\mcM_B^{(0)} (\bx)}^\alpha = \mu_0^\alpha s_1^\alpha (x_1) \chi_P (\bx_P), 
\end{equation}
where $\bx_P = (x_2,x_3)$ are the coordinates on the plane perpendicular to the $x_1$-axis and $\chi_P (\bx_P)$ denotes the indicator function for the constant cross-section of the rigid body under consideration. Indeed, this cross section must be characterized by a radius $R \gg r_C$ for our approximations to work. The function $s_1(x_1)$ is given by
\begin{equations}
	s_1 (x) &= \int_{-L/2}^{+L/2} \dd{y} \frac{e^{-(x-y)^2/2 r_C^2}}{(2 \pi r_C^2)^{1/2}}
	=
	\frac{1}{2}\prtq{\erf\prt{\frac{L+2x}{2\sqrt{2}r_C}}+\erf\prt{\frac{L - 2 x}{2\sqrt{2} r_C}}},
	\\
	s'_1 (x) &= \frac{e^{-(L+2x)^2/8 r_C^2}-e^{-(L-2x)^2/8 r_C^2}}{\sqrt{2\pi}r_C}.
\end{equations}
which in turn implies $\partial_{x_1} \prt{\mcM_B^{(0)} (\bx)}^\alpha = \alpha \mu_0^\alpha \chi_P (\bx_P) s_1^{\alpha-1} (x_1) s'_1 (x_1)$.
To compute the integral in Eq.~\eqref{APPeq:ForceNoiseSpectralDensityTwoBodies} we can neglect the second term because $r_C \ll a$. We get
\begin{multline}
	S \simeq \frac{\gamma_\alpha \hbar^2}{2 m_0^{2\alpha}} \int \dd[3]{\bx}\prtq{\partial_{x_1} \prt{\mcM_B^{(0)} (\bx)}^\alpha}^2
	= \alpha^2 A_P\frac{\gamma_\alpha \hbar^2}{2} \prt{\frac{\mu_0}{m_0}}^{2\alpha} \intmp \dd{x_1} \prtq{s_1^{\alpha-1} (x_1) s'_1 (x_1)}^2
	=\\=
	\alpha^2 A_P\frac{\gamma_\alpha \hbar^2}{2} \prt{\frac{\mu_0}{m_0}}^{2\alpha} \frac{G(\alpha)}{\pi r_C}
	=
	\lambda_\alpha \frac{\hbar^2}{2} \prt{\frac{\mu_0}{m_0}}^{2\alpha} \frac{\alpha^{7/2} A_P G(\alpha)\prt{2 \pi r_C^2}^{3\alpha}}{\pi^{5/2} r_C^4}.
\end{multline}

\subsection{Comparison with exact results for standard CSL}

\begin{figure}
	\centering
	\includegraphics[width=0.9\textwidth]{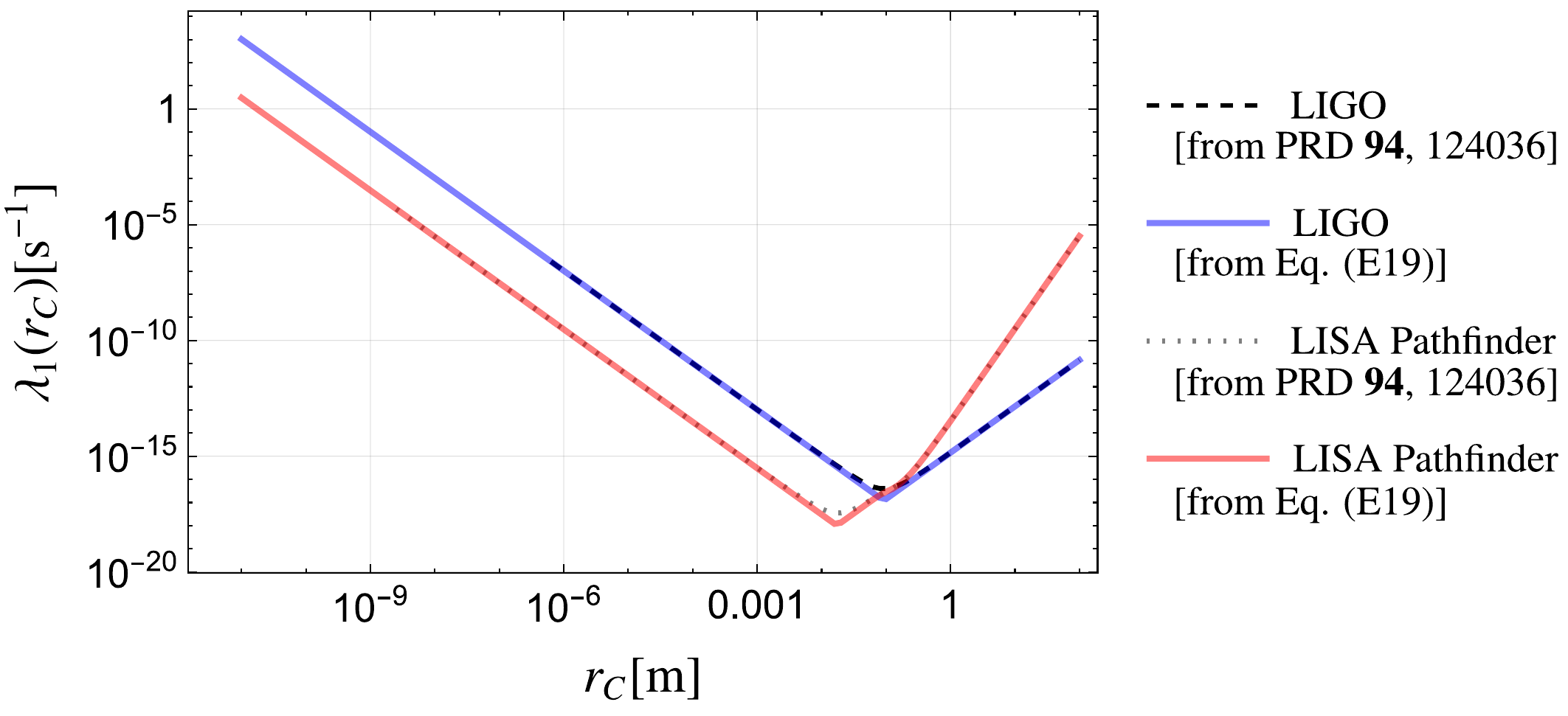}
	\caption{Comparison between the approximated bounds derived from Eq.~\eqref{APPeq:ForceNoiseDensities} for $\alpha=1$ and the more exact result of Ref.~\cite{Carlesso2016ExperimentalBounds}.}
	\label{APPfig:comparisonPlotGravWaves}
\end{figure}

Summing up, the results we obtained in the previous subsections are:
\begin{equation}\label{APPeq:ForceNoiseDensities}
	S\vert_{r_C\ll L} = \lambda_\alpha \frac{\hbar^2}{2} \prt{\frac{M}{m_0}}^{2\alpha} \frac{\alpha^{7/2} A_P G(\alpha)\prt{2 \pi r_C^2}^{3\alpha}}{V^{2\alpha}\pi^{5/2} r_C^4},
	\quad
	S\vert_{r_C\gg L} = \frac{\alpha^{5/2} \hbar^2 \lambda_\alpha}{4\sqrt{\pi} r_C^2}\prt{\frac{M}{m_0}}^{2\alpha} f_S \prt{\frac{a}{r_C},\alpha},
\end{equation}
where $V=L A_P$ is the volume of the rigid body, $L$ its length, and $A_P$ the area of its cross-section\footnote{$A_P= L^2$ for the cube and $A_P = \pi R^2$ for the cylinder.}.
The above formulae, corresponding to Eq.~\eqref{eq:ForceNoiseDensity} of the main text, can be used to set upper bounds on the spontaneous collapse parameters. In Fig.~\ref{APPfig:comparisonPlotGravWaves}, we compare the bounds set by them with the more precise ones obtained in Ref.~\cite{Carlesso2016ExperimentalBounds}. We can see how the two methods give the same result.


\end{document}

%% file: preamble_packages.tex
\usepackage[T1]{fontenc}
\usepackage{graphicx}
\graphicspath{{./Figures/}}
\usepackage{dcolumn}
\usepackage{bm}
\usepackage[dvipsnames]{xcolor}
\usepackage{hyperref}
\hypersetup{colorlinks,linkcolor={MidnightBlue},citecolor={MidnightBlue},urlcolor={MidnightBlue}}  

\usepackage{lipsum}

\usepackage{float}
\usepackage{wrapfig}

\usepackage{tensor}

\usepackage{physics}
\usepackage{amsfonts}
\usepackage{verbatim}
\usepackage{amsmath}
\usepackage{csquotes}
\usepackage{color}
\usepackage{soul}
\usepackage{amsthm}
\usepackage{bm}
\usepackage{graphicx}
\usepackage[normalem]{ulem}
\usepackage{mathtools}

\usepackage{yfonts}

\usepackage{amssymb}
\usepackage{verbatim}


%% file: preamble_commands.tex
\newenvironment{equations}
{\begin{equation}\begin{aligned}}
		{\end{aligned}\end{equation}\ignorespacesafterend}

\newenvironment{equations*}
{\begin{equation*}\begin{aligned}}
		{\end{aligned}\end{equation*}\ignorespacesafterend}

\DeclareFontFamily{U}{mathx}{}
\DeclareFontShape{U}{mathx}{m}{n}{<-> mathx10}{}
\DeclareSymbolFont{mathx}{U}{mathx}{m}{n}
\DeclareMathAccent{\widehat}{0}{mathx}{"70}
\DeclareMathAccent{\widecheck}{0}{mathx}{"71}

\newcommand{\prt}[1]{\left(#1\right)}

\newcommand{\rvt}[1]{\left.#1\right|}

\newcommand{\intmp}{\int_{-\infty}^{+\infty}}

\newcommand{\Schr}{Schr\"{o}dinger\ }

\newcommand{\dg}{\dagger}
\newcommand{\tl}{\tilde}

\newcommand{\prtq}[1]{\left[#1\right]}

\newcommand{\prtg}[1]{\left\{#1\right\}}

\newcommand{\prtB}[1]{\Bigg(#1\Bigg)}

\newcommand{\prtqB}[1]{\Bigg[#1\Bigg]}

\newcommand{\prtgB}[1]{\Bigg\{#1\Bigg\}}

\newcommand{\mcD}{\mathcal{D}}

\newcommand{\mcR}{\mathcal{R}}

\newcommand{\mcT}{\mathcal{T}}

\newcommand{\mcN}{\mathcal{N}}
\newcommand{\mcM}{\mathcal{M}}

\newcommand{\mcF}{\mathcal{F}}

\newcommand{\mbE}{\mathbb{E}}

\newcommand{\ad}{a^\dagger}

\newcommand{\hp}{\hat{p}}
\newcommand{\hq}{\hat{q}}

\newcommand{\hP}{\hat{P}}
\newcommand{\hQ}{\hat{Q}}

\newcommand{\hV}{\hat{V}}

\newcommand{\hL}{\hat{L}}

\newcommand{\bx}{\mathbf{x}}
\newcommand{\by}{\mathbf{y}}
\newcommand{\bz}{\mathbf{z}}

\newcommand{\bq}{\mathbf{q}}
\newcommand{\bk}{\mathbf{k}}

\newcommand{\bd}{\mathbf{d}}

\newcommand{\br}{\mathbf{r}}

\newcommand{\bn}{\mathbf{n}}

\newcommand{\bQ}{\mathbf{Q}}

\newcommand{\bD}{\mathbf{D}}

\newcommand{\hbF}{\hat{\mathbf{F}}}

\newcommand{\be}{\mathbf{e}}

\newcommand{\bv}{\mathbf{v}}

\newcommand{\hbr}{\hat{\mathbf{r}}}

\newcommand{\hbp}{\hat{\mathbf{p}}}
\newcommand{\hbq}{\hat{\mathbf{q}}}

\newcommand{\hbP}{\hat{\mathbf{P}}}
\newcommand{\hbQ}{\hat{\mathbf{Q}}}




%% file: arXiv_ExploringMassDependence_FinalUpdate.bbl
\begin{thebibliography}{42}%
	\makeatletter
	\providecommand \@ifxundefined [1]{%
		\@ifx{#1\undefined}
	}%
	\providecommand \@ifnum [1]{%
		\ifnum #1\expandafter \@firstoftwo
		\else \expandafter \@secondoftwo
		\fi
	}%
	\providecommand \@ifx [1]{%
		\ifx #1\expandafter \@firstoftwo
		\else \expandafter \@secondoftwo
		\fi
	}%
	\providecommand \natexlab [1]{#1}%
	\providecommand \enquote  [1]{``#1''}%
	\providecommand \bibnamefont  [1]{#1}%
	\providecommand \bibfnamefont [1]{#1}%
	\providecommand \citenamefont [1]{#1}%
	\providecommand \href@noop [0]{\@secondoftwo}%
	\providecommand \href [0]{\begingroup \@sanitize@url \@href}%
	\providecommand \@href[1]{\@@startlink{#1}\@@href}%
	\providecommand \@@href[1]{\endgroup#1\@@endlink}%
	\providecommand \@sanitize@url [0]{\catcode `\\12\catcode `\$12\catcode
		`\&12\catcode `\#12\catcode `\^12\catcode `\_12\catcode `\%12\relax}%
	\providecommand \@@startlink[1]{}%
	\providecommand \@@endlink[0]{}%
	\providecommand \url  [0]{\begingroup\@sanitize@url \@url }%
	\providecommand \@url [1]{\endgroup\@href {#1}{\urlprefix }}%
	\providecommand \urlprefix  [0]{URL }%
	\providecommand \Eprint [0]{\href }%
	\providecommand \doibase [0]{https://doi.org/}%
	\providecommand \selectlanguage [0]{\@gobble}%
	\providecommand \bibinfo  [0]{\@secondoftwo}%
	\providecommand \bibfield  [0]{\@secondoftwo}%
	\providecommand \translation [1]{[#1]}%
	\providecommand \BibitemOpen [0]{}%
	\providecommand \bibitemStop [0]{}%
	\providecommand \bibitemNoStop [0]{.\EOS\space}%
	\providecommand \EOS [0]{\spacefactor3000\relax}%
	\providecommand \BibitemShut  [1]{\csname bibitem#1\endcsname}%
	\let\auto@bib@innerbib\@empty
	\bibitem [{\citenamefont {Bell}\ and\ \citenamefont
		{Bell}(2004)}]{Book_Bell2004Speakable}%
	\BibitemOpen
	\bibfield  {author} {\bibinfo {author} {\bibfnamefont {J.~S.}\ \bibnamefont
			{Bell}}\ and\ \bibinfo {author} {\bibfnamefont {J.~S.}\ \bibnamefont
			{Bell}},\ }\href {https://doi.org/10.1017/CBO9780511815676} {\emph {\bibinfo
			{title} {Speakable and unspeakable in quantum mechanics: Collected papers on
				quantum philosophy}}}\ (\bibinfo  {publisher} {Cambridge university press},\
	\bibinfo {year} {2004})\BibitemShut {NoStop}%
	\bibitem [{\citenamefont {Norsen}(2017)}]{Book_Norsen2017foundations}%
	\BibitemOpen
	\bibfield  {author} {\bibinfo {author} {\bibfnamefont {T.}~\bibnamefont
			{Norsen}},\ }\href {https://link.springer.com/book/10.1007/978-3-319-65867-4}
	{\emph {\bibinfo {title} {Foundations of quantum mechanics}}}\ (\bibinfo
	{publisher} {Springer},\ \bibinfo {year} {2017})\BibitemShut {NoStop}%
	\bibitem [{\citenamefont {D{\"u}rr}\ and\ \citenamefont
		{Lazarovici}(2020)}]{Book_Durr2020understanding}%
	\BibitemOpen
	\bibfield  {author} {\bibinfo {author} {\bibfnamefont {D.}~\bibnamefont
			{D{\"u}rr}}\ and\ \bibinfo {author} {\bibfnamefont {D.}~\bibnamefont
			{Lazarovici}},\ }\href
	{https://link.springer.com/book/10.1007/978-3-030-40068-2} {\emph {\bibinfo
			{title} {Understanding quantum mechanics}}}\ (\bibinfo  {publisher}
	{Springer},\ \bibinfo {year} {2020})\BibitemShut {NoStop}%
	\bibitem [{\citenamefont {Tumulka}(2022)}]{Book_Tumulka2022Foundations}%
	\BibitemOpen
	\bibfield  {author} {\bibinfo {author} {\bibfnamefont {R.}~\bibnamefont
			{Tumulka}},\ }\href {https://doi.org/10.1007/978-3-031-09548-1} {\emph
		{\bibinfo {title} {Foundations of quantum mechanics}}},\ Vol.\ \bibinfo
	{volume} {1003}\ (\bibinfo  {publisher} {Springer Nature},\ \bibinfo {year}
	{2022})\BibitemShut {NoStop}%
	\bibitem [{\citenamefont {Bassi}\ and\ \citenamefont
		{Ghirardi}(2003)}]{Bassi2003Dynamical}%
	\BibitemOpen
	\bibfield  {author} {\bibinfo {author} {\bibfnamefont {A.}~\bibnamefont
			{Bassi}}\ and\ \bibinfo {author} {\bibfnamefont {G.}~\bibnamefont
			{Ghirardi}},\ }\bibfield  {title} {\bibinfo {title} {Dynamical reduction
			models},\ }\href {https://doi.org/10.1016/S0370-1573(03)00103-0} {\bibfield
		{journal} {\bibinfo  {journal} {Physics Reports}\ }\textbf {\bibinfo {volume}
			{379}},\ \bibinfo {pages} {257} (\bibinfo {year} {2003})}\BibitemShut
	{NoStop}%
	\bibitem [{\citenamefont {Bassi}\ \emph {et~al.}(2013)\citenamefont {Bassi},
		\citenamefont {Lochan}, \citenamefont {Satin}, \citenamefont {Singh},\ and\
		\citenamefont {Ulbricht}}]{Bassi2013Models}%
	\BibitemOpen
	\bibfield  {author} {\bibinfo {author} {\bibfnamefont {A.}~\bibnamefont
			{Bassi}}, \bibinfo {author} {\bibfnamefont {K.}~\bibnamefont {Lochan}},
		\bibinfo {author} {\bibfnamefont {S.}~\bibnamefont {Satin}}, \bibinfo
		{author} {\bibfnamefont {T.~P.}\ \bibnamefont {Singh}},\ and\ \bibinfo
		{author} {\bibfnamefont {H.}~\bibnamefont {Ulbricht}},\ }\bibfield  {title}
	{\bibinfo {title} {Models of wave-function collapse, underlying theories, and
			experimental tests},\ }\href {https://doi.org/10.1103/RevModPhys.85.471}
	{\bibfield  {journal} {\bibinfo  {journal} {Rev. Mod. Phys.}\ }\textbf
		{\bibinfo {volume} {85}},\ \bibinfo {pages} {471} (\bibinfo {year}
		{2013})}\BibitemShut {NoStop}%
	\bibitem [{\citenamefont {Bassi}\ \emph {et~al.}(2023)\citenamefont {Bassi},
		\citenamefont {Dorato},\ and\ \citenamefont
		{Ulbricht}}]{Bassi2023CollapseModels}%
	\BibitemOpen
	\bibfield  {author} {\bibinfo {author} {\bibfnamefont {A.}~\bibnamefont
			{Bassi}}, \bibinfo {author} {\bibfnamefont {M.}~\bibnamefont {Dorato}},\ and\
		\bibinfo {author} {\bibfnamefont {H.}~\bibnamefont {Ulbricht}},\ }\bibfield
	{title} {\bibinfo {title} {Collapse models: A theoretical, experimental and
			philosophical review},\ }\bibfield  {journal} {\bibinfo  {journal} {Entropy}\
	}\textbf {\bibinfo {volume} {25}},\ \href {https://doi.org/10.3390/e25040645}
	{10.3390/e25040645} (\bibinfo {year} {2023})\BibitemShut {NoStop}%
	\bibitem [{\citenamefont {Carlesso}\ \emph {et~al.}(2022)\citenamefont
		{Carlesso}, \citenamefont {Donadi}, \citenamefont {Ferialdi}, \citenamefont
		{Paternostro}, \citenamefont {Ulbricht},\ and\ \citenamefont
		{Bassi}}]{Carlesso2022Present}%
	\BibitemOpen
	\bibfield  {author} {\bibinfo {author} {\bibfnamefont {M.}~\bibnamefont
			{Carlesso}}, \bibinfo {author} {\bibfnamefont {S.}~\bibnamefont {Donadi}},
		\bibinfo {author} {\bibfnamefont {L.}~\bibnamefont {Ferialdi}}, \bibinfo
		{author} {\bibfnamefont {M.}~\bibnamefont {Paternostro}}, \bibinfo {author}
		{\bibfnamefont {H.}~\bibnamefont {Ulbricht}},\ and\ \bibinfo {author}
		{\bibfnamefont {A.}~\bibnamefont {Bassi}},\ }\bibfield  {title} {\bibinfo
		{title} {Present status and future challenges of non-interferometric tests of
			collapse models},\ }\href {https://doi.org/10.1038/s41567-021-01489-5}
	{\bibfield  {journal} {\bibinfo  {journal} {Nature Physics}\ }\textbf
		{\bibinfo {volume} {18}},\ \bibinfo {pages} {243} (\bibinfo {year}
		{2022})}\BibitemShut {NoStop}%
	\bibitem [{\citenamefont {Ghirardi}\ \emph {et~al.}(1990)\citenamefont
		{Ghirardi}, \citenamefont {Pearle},\ and\ \citenamefont
		{Rimini}}]{Ghirardi1990_CSL}%
	\BibitemOpen
	\bibfield  {author} {\bibinfo {author} {\bibfnamefont {G.~C.}\ \bibnamefont
			{Ghirardi}}, \bibinfo {author} {\bibfnamefont {P.}~\bibnamefont {Pearle}},\
		and\ \bibinfo {author} {\bibfnamefont {A.}~\bibnamefont {Rimini}},\
	}\bibfield  {title} {\bibinfo {title} {Markov processes in hilbert space and
			continuous spontaneous localization of systems of identical particles},\
	}\href {https://doi.org/10.1103/PhysRevA.42.78} {\bibfield  {journal}
		{\bibinfo  {journal} {Phys. Rev. A}\ }\textbf {\bibinfo {volume} {42}},\
		\bibinfo {pages} {78} (\bibinfo {year} {1990})}\BibitemShut {NoStop}%
	\bibitem [{\citenamefont {Diósi}(1987)}]{Diosi1987Universal}%
	\BibitemOpen
	\bibfield  {author} {\bibinfo {author} {\bibfnamefont {L.}~\bibnamefont
			{Diósi}},\ }\bibfield  {title} {\bibinfo {title} {A universal master
			equation for the gravitational violation of quantum mechanics},\ }\href
	{https://doi.org/10.1016/0375-9601(87)90681-5} {\bibfield  {journal}
		{\bibinfo  {journal} {Physics Letters A}\ }\textbf {\bibinfo {volume}
			{120}},\ \bibinfo {pages} {377} (\bibinfo {year} {1987})}\BibitemShut
	{NoStop}%
	\bibitem [{\citenamefont {Di\'osi}(1989)}]{Diosi1989Models}%
	\BibitemOpen
	\bibfield  {author} {\bibinfo {author} {\bibfnamefont {L.}~\bibnamefont
			{Di\'osi}},\ }\bibfield  {title} {\bibinfo {title} {Models for universal
			reduction of macroscopic quantum fluctuations},\ }\href
	{https://doi.org/10.1103/PhysRevA.40.1165} {\bibfield  {journal} {\bibinfo
			{journal} {Phys. Rev. A}\ }\textbf {\bibinfo {volume} {40}},\ \bibinfo
		{pages} {1165} (\bibinfo {year} {1989})}\BibitemShut {NoStop}%
	\bibitem [{\citenamefont {Penrose}(1996)}]{Penrose1996gravity}%
	\BibitemOpen
	\bibfield  {author} {\bibinfo {author} {\bibfnamefont {R.}~\bibnamefont
			{Penrose}},\ }\bibfield  {title} {\bibinfo {title} {On gravity's role in
			quantum state reduction},\ }\href {https://doi.org/10.1007/BF02105068}
	{\bibfield  {journal} {\bibinfo  {journal} {General relativity and
				gravitation}\ }\textbf {\bibinfo {volume} {28}},\ \bibinfo {pages} {581}
		(\bibinfo {year} {1996})}\BibitemShut {NoStop}%
	\bibitem [{\citenamefont {Penrose}(2014)}]{Penrose2014Gravitization}%
	\BibitemOpen
	\bibfield  {author} {\bibinfo {author} {\bibfnamefont {R.}~\bibnamefont
			{Penrose}},\ }\bibfield  {title} {\bibinfo {title} {On the gravitization of
			quantum mechanics 1: Quantum state reduction},\ }\href
	{https://doi.org/10.1007/s10701-013-9770-0} {\bibfield  {journal} {\bibinfo
			{journal} {Found. Phys}\ }\textbf {\bibinfo {volume} {44}},\ \bibinfo {pages}
		{557} (\bibinfo {year} {2014})}\BibitemShut {NoStop}%
	\bibitem [{\citenamefont {Carlesso}\ and\ \citenamefont
		{Donadi}(2019)}]{Carlesso2019Collapse}%
	\BibitemOpen
	\bibfield  {author} {\bibinfo {author} {\bibfnamefont {M.}~\bibnamefont
			{Carlesso}}\ and\ \bibinfo {author} {\bibfnamefont {S.}~\bibnamefont
			{Donadi}},\ }\bibfield  {title} {\bibinfo {title} {Collapse models: main
			properties and the state of art of the experimental tests},\ }in\ \href
	{https://doi.org/10.1007/978-3-030-31146-9_1} {\emph {\bibinfo {booktitle}
			{Advances in Open Systems and Fundamental Tests of Quantum Mechanics}}}\
	(\bibinfo  {publisher} {Springer},\ \bibinfo {year} {2019})\ pp.\ \bibinfo
	{pages} {1--13}\BibitemShut {NoStop}%
	\bibitem [{\citenamefont {Pearle}\ and\ \citenamefont
		{Squires}(1994)}]{Pearle1994MassCSL}%
	\BibitemOpen
	\bibfield  {author} {\bibinfo {author} {\bibfnamefont {P.}~\bibnamefont
			{Pearle}}\ and\ \bibinfo {author} {\bibfnamefont {E.}~\bibnamefont
			{Squires}},\ }\bibfield  {title} {\bibinfo {title} {Bound state excitation,
			nucleon decay experiments and models of wave function collapse},\ }\href
	{https://doi.org/10.1103/PhysRevLett.73.1} {\bibfield  {journal} {\bibinfo
			{journal} {Phys. Rev. Lett.}\ }\textbf {\bibinfo {volume} {73}},\ \bibinfo
		{pages} {1} (\bibinfo {year} {1994})}\BibitemShut {NoStop}%
	\bibitem [{\citenamefont {Rimini}(1997)}]{Rimini1997CompoundObjects}%
	\BibitemOpen
	\bibfield  {author} {\bibinfo {author} {\bibfnamefont {A.}~\bibnamefont
			{Rimini}},\ }\bibfield  {title} {\bibinfo {title} {Compound objects as
			particles in quantum mechanics},\ }\href {https://doi.org/10.1007/BF02551445}
	{\bibfield  {journal} {\bibinfo  {journal} {Found. of Phys.}\ }\textbf
		{\bibinfo {volume} {27}},\ \bibinfo {pages} {1689} (\bibinfo {year}
		{1997})}\BibitemShut {NoStop}%
	\bibitem [{\citenamefont {Ghirardi}\ \emph {et~al.}(1986)\citenamefont
		{Ghirardi}, \citenamefont {Rimini},\ and\ \citenamefont
		{Weber}}]{Ghirardi1986Unified}%
	\BibitemOpen
	\bibfield  {author} {\bibinfo {author} {\bibfnamefont {G.~C.}\ \bibnamefont
			{Ghirardi}}, \bibinfo {author} {\bibfnamefont {A.}~\bibnamefont {Rimini}},\
		and\ \bibinfo {author} {\bibfnamefont {T.}~\bibnamefont {Weber}},\ }\bibfield
	{title} {\bibinfo {title} {Unified dynamics for microscopic and macroscopic
			systems},\ }\href {https://doi.org/10.1103/PhysRevD.34.470} {\bibfield
		{journal} {\bibinfo  {journal} {Phys. Rev. D}\ }\textbf {\bibinfo {volume}
			{34}},\ \bibinfo {pages} {470} (\bibinfo {year} {1986})}\BibitemShut
	{NoStop}%
	\bibitem [{\citenamefont {Tumulka}(2006)}]{Tumulka2006spontaneous}%
	\BibitemOpen
	\bibfield  {author} {\bibinfo {author} {\bibfnamefont {R.}~\bibnamefont
			{Tumulka}},\ }\bibfield  {title} {\bibinfo {title} {On spontaneous wave
			function collapse and quantum field theory},\ }\href
	{https://doi.org/10.1098/rspa.2005.1636} {\bibfield  {journal} {\bibinfo
			{journal} {Proceedings of the Royal Society A: Mathematical, Physical and
				Engineering Sciences}\ }\textbf {\bibinfo {volume} {462}},\ \bibinfo {pages}
		{1897} (\bibinfo {year} {2006})}\BibitemShut {NoStop}%
	\bibitem [{\citenamefont {Piccione}(2023)}]{Piccione2023Collapse}%
	\BibitemOpen
	\bibfield  {author} {\bibinfo {author} {\bibfnamefont {N.}~\bibnamefont
			{Piccione}},\ }\bibfield  {title} {\bibinfo {title} {A proposal for a new
			kind of spontaneous collapse model},\ }\href
	{https://doi.org/10.1007/s10701-023-00739-1} {\bibfield  {journal} {\bibinfo
			{journal} {Found. Phys.}\ }\textbf {\bibinfo {volume} {54}},\ \bibinfo
		{pages} {4} (\bibinfo {year} {2023})}\BibitemShut {NoStop}%
	\bibitem [{\citenamefont {Tilloy}\ and\ \citenamefont
		{Di\'osi}(2016)}]{Tilloy2016CSLGravity}%
	\BibitemOpen
	\bibfield  {author} {\bibinfo {author} {\bibfnamefont {A.}~\bibnamefont
			{Tilloy}}\ and\ \bibinfo {author} {\bibfnamefont {L.}~\bibnamefont
			{Di\'osi}},\ }\bibfield  {title} {\bibinfo {title} {Sourcing semiclassical
			gravity from spontaneously localized quantum matter},\ }\href
	{https://doi.org/10.1103/PhysRevD.93.024026} {\bibfield  {journal} {\bibinfo
			{journal} {Phys. Rev. D}\ }\textbf {\bibinfo {volume} {93}},\ \bibinfo
		{pages} {024026} (\bibinfo {year} {2016})}\BibitemShut {NoStop}%
	\bibitem [{\citenamefont {Tilloy}\ and\ \citenamefont
		{Di\'osi}(2017)}]{Tilloy2017LeastDecoherence}%
	\BibitemOpen
	\bibfield  {author} {\bibinfo {author} {\bibfnamefont {A.}~\bibnamefont
			{Tilloy}}\ and\ \bibinfo {author} {\bibfnamefont {L.}~\bibnamefont
			{Di\'osi}},\ }\bibfield  {title} {\bibinfo {title} {Principle of least
			decoherence for newtonian semiclassical gravity},\ }\href
	{https://doi.org/10.1103/PhysRevD.96.104045} {\bibfield  {journal} {\bibinfo
			{journal} {Phys. Rev. D}\ }\textbf {\bibinfo {volume} {96}},\ \bibinfo
		{pages} {104045} (\bibinfo {year} {2017})}\BibitemShut {NoStop}%
	\bibitem [{\citenamefont {Gaona-Reyes}\ \emph {et~al.}(2021)\citenamefont
		{Gaona-Reyes}, \citenamefont {Carlesso},\ and\ \citenamefont
		{Bassi}}]{GaonaReyes2021GravitationalFeedback}%
	\BibitemOpen
	\bibfield  {author} {\bibinfo {author} {\bibfnamefont {J.~L.}\ \bibnamefont
			{Gaona-Reyes}}, \bibinfo {author} {\bibfnamefont {M.}~\bibnamefont
			{Carlesso}},\ and\ \bibinfo {author} {\bibfnamefont {A.}~\bibnamefont
			{Bassi}},\ }\bibfield  {title} {\bibinfo {title} {Gravitational interaction
			through a feedback mechanism},\ }\href
	{https://doi.org/10.1103/PhysRevD.103.056011} {\bibfield  {journal} {\bibinfo
			{journal} {Phys. Rev. D}\ }\textbf {\bibinfo {volume} {103}},\ \bibinfo
		{pages} {056011} (\bibinfo {year} {2021})}\BibitemShut {NoStop}%
	\bibitem [{\citenamefont {Wiseman}\ \emph {et~al.}(2002)\citenamefont
		{Wiseman}, \citenamefont {Mancini},\ and\ \citenamefont
		{Wang}}]{Wiseman2002BayesianFeedback}%
	\BibitemOpen
	\bibfield  {author} {\bibinfo {author} {\bibfnamefont {H.~M.}\ \bibnamefont
			{Wiseman}}, \bibinfo {author} {\bibfnamefont {S.}~\bibnamefont {Mancini}},\
		and\ \bibinfo {author} {\bibfnamefont {J.}~\bibnamefont {Wang}},\ }\bibfield
	{title} {\bibinfo {title} {Bayesian feedback versus markovian feedback in a
			two-level atom},\ }\href {https://doi.org/10.1103/PhysRevA.66.013807}
	{\bibfield  {journal} {\bibinfo  {journal} {Phys. Rev. A}\ }\textbf {\bibinfo
			{volume} {66}},\ \bibinfo {pages} {013807} (\bibinfo {year}
		{2002})}\BibitemShut {NoStop}%
	\bibitem [{\citenamefont {Tilloy}(2024)}]{Tilloy2024HybridDynamics}%
	\BibitemOpen
	\bibfield  {author} {\bibinfo {author} {\bibfnamefont {A.}~\bibnamefont
			{Tilloy}},\ }\bibfield  {title} {\bibinfo {title} {{General quantum-classical
				dynamics as measurement based feedback}},\ }\href
	{https://doi.org/10.21468/SciPostPhys.17.3.083} {\bibfield  {journal}
		{\bibinfo  {journal} {SciPost Phys.}\ }\textbf {\bibinfo {volume} {17}},\
		\bibinfo {pages} {083} (\bibinfo {year} {2024})}\BibitemShut {NoStop}%
	\bibitem [{\citenamefont {Piccione}\ and\ \citenamefont
		{Bassi}(2025)}]{Piccione2025NewtonianPSL}%
	\BibitemOpen
	\bibfield  {author} {\bibinfo {author} {\bibfnamefont {N.}~\bibnamefont
			{Piccione}}\ and\ \bibinfo {author} {\bibfnamefont {A.}~\bibnamefont
			{Bassi}},\ }\href {https://arxiv.org/abs/2502.04996} {\bibinfo {title}
		{Hybrid classical-quantum newtonian gravity with stable vacuum}} (\bibinfo
	{year} {2025}),\ \Eprint {https://arxiv.org/abs/2502.04996} {arXiv:2502.04996
		[quant-ph]} \BibitemShut {NoStop}%
	\bibitem [{\citenamefont {Toroš}\ and\ \citenamefont
		{Bassi}(2018)}]{Toros2018BoundsCalculations}%
	\BibitemOpen
	\bibfield  {author} {\bibinfo {author} {\bibfnamefont {M.}~\bibnamefont
			{Toroš}}\ and\ \bibinfo {author} {\bibfnamefont {A.}~\bibnamefont {Bassi}},\
	}\bibfield  {title} {\bibinfo {title} {Bounds on quantum collapse models from
			matter-wave interferometry: calculational details},\ }\href
	{https://doi.org/10.1088/1751-8121/aaabc6} {\bibfield  {journal} {\bibinfo
			{journal} {Journal of Physics A: Mathematical and Theoretical}\ }\textbf
		{\bibinfo {volume} {51}},\ \bibinfo {pages} {115302} (\bibinfo {year}
		{2018})}\BibitemShut {NoStop}%
	\bibitem [{\citenamefont {Fu}(1997)}]{Fu1997SpontaneousRadiation}%
	\BibitemOpen
	\bibfield  {author} {\bibinfo {author} {\bibfnamefont {Q.}~\bibnamefont
			{Fu}},\ }\bibfield  {title} {\bibinfo {title} {Spontaneous radiation of free
			electrons in a nonrelativistic collapse model},\ }\href
	{https://doi.org/10.1103/PhysRevA.56.1806} {\bibfield  {journal} {\bibinfo
			{journal} {Phys. Rev. A}\ }\textbf {\bibinfo {volume} {56}},\ \bibinfo
		{pages} {1806} (\bibinfo {year} {1997})}\BibitemShut {NoStop}%
	\bibitem [{\citenamefont {Tilloy}(2018)}]{Tilloy2018GRWGravity}%
	\BibitemOpen
	\bibfield  {author} {\bibinfo {author} {\bibfnamefont {A.}~\bibnamefont
			{Tilloy}},\ }\bibfield  {title} {\bibinfo {title} {Ghirardi-rimini-weber
			model with massive flashes},\ }\href
	{https://doi.org/10.1103/PhysRevD.97.021502} {\bibfield  {journal} {\bibinfo
			{journal} {Phys. Rev. D}\ }\textbf {\bibinfo {volume} {97}},\ \bibinfo
		{pages} {021502} (\bibinfo {year} {2018})}\BibitemShut {NoStop}%
	\bibitem [{\citenamefont {Donadi}(2014)}]{Donadi2014RadiationEmission}%
	\BibitemOpen
	\bibfield  {author} {\bibinfo {author} {\bibfnamefont {S.}~\bibnamefont
			{Donadi}},\ }\emph {\bibinfo {title} {Electromagnetic Radiation Emission and
			Flavour Oscillations in Collapse Models}},\ \href
	{https://www.openstarts.units.it/entities/publication/8d2a2d9a-0c80-4a07-818d-ea7b7cbb6973/details}
	{Ph.D. thesis},\ \bibinfo  {school} {University of Trieste} (\bibinfo {year}
	{2014})\BibitemShut {NoStop}%
	\bibitem [{\citenamefont {Donadi}\ and\ \citenamefont
		{Bassi}(2014)}]{Donadi2015Radiation}%
	\BibitemOpen
	\bibfield  {author} {\bibinfo {author} {\bibfnamefont {S.}~\bibnamefont
			{Donadi}}\ and\ \bibinfo {author} {\bibfnamefont {A.}~\bibnamefont {Bassi}},\
	}\bibfield  {title} {\bibinfo {title} {The emission of electromagnetic
			radiation from a quantum system interacting with an external noise: a general
			result},\ }\href {https://doi.org/10.1088/1751-8113/48/3/035305} {\bibfield
		{journal} {\bibinfo  {journal} {Journal of Physics A: Mathematical and
				Theoretical}\ }\textbf {\bibinfo {volume} {48}},\ \bibinfo {pages} {035305}
		(\bibinfo {year} {2014})}\BibitemShut {NoStop}%
	\bibitem [{\citenamefont {Donadi}\ \emph
		{et~al.}(2021{\natexlab{a}})\citenamefont {Donadi}, \citenamefont
		{Piscicchia}, \citenamefont {Del~Grande}, \citenamefont {Curceanu},
		\citenamefont {Laubenstein},\ and\ \citenamefont
		{Bassi}}]{Donadi2021NovelCSLBounds}%
	\BibitemOpen
	\bibfield  {author} {\bibinfo {author} {\bibfnamefont {S.}~\bibnamefont
			{Donadi}}, \bibinfo {author} {\bibfnamefont {K.}~\bibnamefont {Piscicchia}},
		\bibinfo {author} {\bibfnamefont {R.}~\bibnamefont {Del~Grande}}, \bibinfo
		{author} {\bibfnamefont {C.}~\bibnamefont {Curceanu}}, \bibinfo {author}
		{\bibfnamefont {M.}~\bibnamefont {Laubenstein}},\ and\ \bibinfo {author}
		{\bibfnamefont {A.}~\bibnamefont {Bassi}},\ }\bibfield  {title} {\bibinfo
		{title} {Novel csl bounds from the noise-induced radiation emission from
			atoms},\ }\bibfield  {journal} {\bibinfo  {journal} {The European Physical
			Journal C}\ }\textbf {\bibinfo {volume} {81}},\ \href
	{https://doi.org/10.1140/epjc/s10052-021-09556-0}
	{10.1140/epjc/s10052-021-09556-0} (\bibinfo {year}
	{2021}{\natexlab{a}})\BibitemShut {NoStop}%
	\bibitem [{\citenamefont {Donadi}\ \emph
		{et~al.}(2021{\natexlab{b}})\citenamefont {Donadi}, \citenamefont
		{Piscicchia}, \citenamefont {Curceanu}, \citenamefont {Diósi}, \citenamefont
		{Laubenstein},\ and\ \citenamefont {Bassi}}]{Donadi2021UndergroundTest}%
	\BibitemOpen
	\bibfield  {author} {\bibinfo {author} {\bibfnamefont {S.}~\bibnamefont
			{Donadi}}, \bibinfo {author} {\bibfnamefont {K.}~\bibnamefont {Piscicchia}},
		\bibinfo {author} {\bibfnamefont {C.}~\bibnamefont {Curceanu}}, \bibinfo
		{author} {\bibfnamefont {L.}~\bibnamefont {Diósi}}, \bibinfo {author}
		{\bibfnamefont {M.}~\bibnamefont {Laubenstein}},\ and\ \bibinfo {author}
		{\bibfnamefont {A.}~\bibnamefont {Bassi}},\ }\bibfield  {title} {\bibinfo
		{title} {Underground test of gravity-related wave function collapse},\
	}\bibfield  {journal} {\bibinfo  {journal} {Nature Physics}\ }\textbf
	{\bibinfo {volume} {17}},\ \href {https://doi.org/10.1038/s41567-020-1008-4}
	{10.1038/s41567-020-1008-4} (\bibinfo {year}
	{2021}{\natexlab{b}})\BibitemShut {NoStop}%
	\bibitem [{\citenamefont {Nimmrichter}\ and\ \citenamefont
		{Hornberger}(2013)}]{Nimmrichter2013Macroscopicity}%
	\BibitemOpen
	\bibfield  {author} {\bibinfo {author} {\bibfnamefont {S.}~\bibnamefont
			{Nimmrichter}}\ and\ \bibinfo {author} {\bibfnamefont {K.}~\bibnamefont
			{Hornberger}},\ }\bibfield  {title} {\bibinfo {title} {Macroscopicity of
			mechanical quantum superposition states},\ }\href
	{https://doi.org/10.1103/PhysRevLett.110.160403} {\bibfield  {journal}
		{\bibinfo  {journal} {Phys. Rev. Lett.}\ }\textbf {\bibinfo {volume} {110}},\
		\bibinfo {pages} {160403} (\bibinfo {year} {2013})}\BibitemShut {NoStop}%
	\bibitem [{\citenamefont {Vinante}\ \emph {et~al.}(2020)\citenamefont
		{Vinante}, \citenamefont {Carlesso}, \citenamefont {Bassi}, \citenamefont
		{Chiasera}, \citenamefont {Varas}, \citenamefont {Falferi}, \citenamefont
		{Margesin}, \citenamefont {Mezzena},\ and\ \citenamefont
		{Ulbricht}}]{Vinante2020UltracoldLayeredForce}%
	\BibitemOpen
	\bibfield  {author} {\bibinfo {author} {\bibfnamefont {A.}~\bibnamefont
			{Vinante}}, \bibinfo {author} {\bibfnamefont {M.}~\bibnamefont {Carlesso}},
		\bibinfo {author} {\bibfnamefont {A.}~\bibnamefont {Bassi}}, \bibinfo
		{author} {\bibfnamefont {A.}~\bibnamefont {Chiasera}}, \bibinfo {author}
		{\bibfnamefont {S.}~\bibnamefont {Varas}}, \bibinfo {author} {\bibfnamefont
			{P.}~\bibnamefont {Falferi}}, \bibinfo {author} {\bibfnamefont
			{B.}~\bibnamefont {Margesin}}, \bibinfo {author} {\bibfnamefont
			{R.}~\bibnamefont {Mezzena}},\ and\ \bibinfo {author} {\bibfnamefont
			{H.}~\bibnamefont {Ulbricht}},\ }\bibfield  {title} {\bibinfo {title}
		{Narrowing the parameter space of collapse models with ultracold layered
			force sensors},\ }\href {https://doi.org/10.1103/PhysRevLett.125.100404}
	{\bibfield  {journal} {\bibinfo  {journal} {Phys. Rev. Lett.}\ }\textbf
		{\bibinfo {volume} {125}},\ \bibinfo {pages} {100404} (\bibinfo {year}
		{2020})}\BibitemShut {NoStop}%
	\bibitem [{\citenamefont {Adler}\ \emph {et~al.}(2021)\citenamefont {Adler},
		\citenamefont {Bassi},\ and\ \citenamefont
		{Carlesso}}]{Adler2021LayeringEffect}%
	\BibitemOpen
	\bibfield  {author} {\bibinfo {author} {\bibfnamefont {S.~L.}\ \bibnamefont
			{Adler}}, \bibinfo {author} {\bibfnamefont {A.}~\bibnamefont {Bassi}},\ and\
		\bibinfo {author} {\bibfnamefont {M.}~\bibnamefont {Carlesso}},\ }\bibfield
	{title} {\bibinfo {title} {The continuous spontaneous localization layering
			effect from a lattice perspective},\ }\href
	{https://doi.org/10.1088/1751-8121/abdbc8} {\bibfield  {journal} {\bibinfo
			{journal} {Journal of Physics A: Mathematical and Theoretical}\ }\textbf
		{\bibinfo {volume} {54}},\ \bibinfo {pages} {085303} (\bibinfo {year}
		{2021})}\BibitemShut {NoStop}%
	\bibitem [{\citenamefont {Diósi}(2021)}]{Diosi2021SurfaceTensors}%
	\BibitemOpen
	\bibfield  {author} {\bibinfo {author} {\bibfnamefont {L.}~\bibnamefont
			{Diósi}},\ }\bibfield  {title} {\bibinfo {title} {Two invariant
			surface-tensors determine csl of massive body wave function},\ }in\ \href
	{https://doi.org/10.1007/978-3-030-46777-7_17} {\emph {\bibinfo {booktitle}
			{Do Wave Functions Jump?}}}\ (\bibinfo  {publisher} {Springer},\ \bibinfo
	{year} {2021})\ pp.\ \bibinfo {pages} {217--226}\BibitemShut {NoStop}%
	\bibitem [{\citenamefont
		{Collaboration}(2022)}]{MAJORANA2022WaveFunctionCollapse}%
	\BibitemOpen
	\bibfield  {author} {\bibinfo {author} {\bibnamefont {Collaboration}}
		(\bibinfo {collaboration} {Majorana Collaboration}),\ }\bibfield  {title}
	{\bibinfo {title} {Search for spontaneous radiation from wave function
			collapse in the majorana demonstrator},\ }\href
	{https://doi.org/10.1103/PhysRevLett.129.080401} {\bibfield  {journal}
		{\bibinfo  {journal} {Phys. Rev. Lett.}\ }\textbf {\bibinfo {volume} {129}},\
		\bibinfo {pages} {080401} (\bibinfo {year} {2022})}\BibitemShut {NoStop}%
	\bibitem [{\citenamefont {Adler}(2007)}]{Adler2007Bounds}%
	\BibitemOpen
	\bibfield  {author} {\bibinfo {author} {\bibfnamefont {S.~L.}\ \bibnamefont
			{Adler}},\ }\bibfield  {title} {\bibinfo {title} {Lower and upper bounds on
			csl parameters from latent image formation and igm heating},\ }\href
	{https://doi.org/10.1088/1751-8113/40/12/S03} {\bibfield  {journal} {\bibinfo
			{journal} {Journal of Physics A: Mathematical and Theoretical}\ }\textbf
		{\bibinfo {volume} {40}},\ \bibinfo {pages} {2935} (\bibinfo {year}
		{2007})}\BibitemShut {NoStop}%
	\bibitem [{\citenamefont {Bassi}\ \emph {et~al.}(2010)\citenamefont {Bassi},
		\citenamefont {Deckert},\ and\ \citenamefont
		{Ferialdi}}]{Bassi2010HumanPerception}%
	\BibitemOpen
	\bibfield  {author} {\bibinfo {author} {\bibfnamefont {A.}~\bibnamefont
			{Bassi}}, \bibinfo {author} {\bibfnamefont {D.-A.}\ \bibnamefont {Deckert}},\
		and\ \bibinfo {author} {\bibfnamefont {L.}~\bibnamefont {Ferialdi}},\
	}\bibfield  {title} {\bibinfo {title} {Breaking quantum linearity:
			Constraints from human perception and cosmological implications},\ }\href
	{https://doi.org/10.1209/0295-5075/92/50006} {\bibfield  {journal} {\bibinfo
			{journal} {Europhysics Letters}\ }\textbf {\bibinfo {volume} {92}},\ \bibinfo
		{pages} {50006} (\bibinfo {year} {2010})}\BibitemShut {NoStop}%
	\bibitem [{\citenamefont {Feldmann}\ and\ \citenamefont
		{Tumulka}(2012)}]{Feldmann2012ParameterDiagramsCSL}%
	\BibitemOpen
	\bibfield  {author} {\bibinfo {author} {\bibfnamefont {W.}~\bibnamefont
			{Feldmann}}\ and\ \bibinfo {author} {\bibfnamefont {R.}~\bibnamefont
			{Tumulka}},\ }\bibfield  {title} {\bibinfo {title} {Parameter diagrams of the
			grw and csl theories of wavefunction collapse},\ }\href
	{https://doi.org/10.1088/1751-8113/45/6/065304} {\bibfield  {journal}
		{\bibinfo  {journal} {Journal of Physics A: Mathematical and Theoretical}\
		}\textbf {\bibinfo {volume} {45}},\ \bibinfo {pages} {065304} (\bibinfo
		{year} {2012})}\BibitemShut {NoStop}%
	\bibitem [{\citenamefont {Carlesso}\ \emph {et~al.}(2016)\citenamefont
		{Carlesso}, \citenamefont {Bassi}, \citenamefont {Falferi},\ and\
		\citenamefont {Vinante}}]{Carlesso2016ExperimentalBounds}%
	\BibitemOpen
	\bibfield  {author} {\bibinfo {author} {\bibfnamefont {M.}~\bibnamefont
			{Carlesso}}, \bibinfo {author} {\bibfnamefont {A.}~\bibnamefont {Bassi}},
		\bibinfo {author} {\bibfnamefont {P.}~\bibnamefont {Falferi}},\ and\ \bibinfo
		{author} {\bibfnamefont {A.}~\bibnamefont {Vinante}},\ }\bibfield  {title}
	{\bibinfo {title} {Experimental bounds on collapse models from gravitational
			wave detectors},\ }\href {https://doi.org/10.1103/PhysRevD.94.124036}
	{\bibfield  {journal} {\bibinfo  {journal} {Phys. Rev. D}\ }\textbf {\bibinfo
			{volume} {94}},\ \bibinfo {pages} {124036} (\bibinfo {year}
		{2016})}\BibitemShut {NoStop}%
	\bibitem [{\citenamefont {Altamura}\ \emph {et~al.}(2024)\citenamefont
		{Altamura}, \citenamefont {Carlesso}, \citenamefont {Donadi},\ and\
		\citenamefont {Bassi}}]{Altamura2024NonInterfCSL}%
	\BibitemOpen
	\bibfield  {author} {\bibinfo {author} {\bibfnamefont {D.~G.~A.}\
			\bibnamefont {Altamura}}, \bibinfo {author} {\bibfnamefont {M.}~\bibnamefont
			{Carlesso}}, \bibinfo {author} {\bibfnamefont {S.}~\bibnamefont {Donadi}},\
		and\ \bibinfo {author} {\bibfnamefont {A.}~\bibnamefont {Bassi}},\ }\bibfield
	{title} {\bibinfo {title} {Noninterferometric rotational test of the
			continuous spontaneous localization model: Enhancement of the collapse noise
			through shape optimization},\ }\href
	{https://doi.org/10.1103/PhysRevA.109.062212} {\bibfield  {journal} {\bibinfo
			{journal} {Phys. Rev. A}\ }\textbf {\bibinfo {volume} {109}},\ \bibinfo
		{pages} {062212} (\bibinfo {year} {2024})}\BibitemShut {NoStop}%
\end{thebibliography}
